\documentclass[a4paper,11pt]{article}

\usepackage{miama} 
\usepackage[T1]{fontenc}
\usepackage[utf8]{inputenc}
\usepackage{lmodern}
\usepackage{cancel}

\usepackage{tikz} 
\usepackage{xcolor}
\usepackage{graphicx}
\graphicspath{{Figures/}}
\usepackage[left=2.5cm,right=2.5cm,bottom=2.5cm,top=2.5cm]{geometry}
\usepackage[english]{babel}
\usepackage{microtype}
\usepackage{amsmath,amssymb}
\usepackage{float}
\usepackage{bm}
\usepackage{enumitem}
\usepackage[hang,flushmargin]{footmisc}
\usepackage{relsize,exscale}
\usepackage{fancyhdr}
\usepackage{caption}
\usepackage{subcaption}
\usepackage{changepage}
\usepackage[most]{tcolorbox}
\tcbset{colback=yellow!10!white, colframe=red!50!black, 
	highlight math style= {enhanced, %<-- needed for the ’remember’ options
		colframe=red,colback=red!10!white,boxsep=0pt}}
\usepackage{empheq}
\usepackage{gensymb}
\usepackage{dsfont}
\usepackage[only,llbracket,rrbracket]{stmaryrd}
\usepackage{mathtools}
\usepackage{xparse}

\usepackage{slashed}
\usepackage{adjustbox}
\usepackage{cite}
\makeatletter
\let\oldthebibliography\thebibliography
\renewcommand\thebibliography[1]{%
  \oldthebibliography{#1}%
  \setlength{\itemsep}{5pt}% 
  \setlength{\parskip}{0pt}% 
}
\makeatother
\usepackage{hyperref}

\hypersetup{
	colorlinks=true,
	linktoc=section,   
	citecolor=blue, 
	filecolor=blue,
	linkcolor=blue,
	urlcolor=blue,
	pdfborder={0 0 0.5},
	pdfmenubar=false,
	pdftoolbar=false
}
\numberwithin{equation}{section}

\providecommand{\delimsize}{\relax}

\DeclareRobustCommand{\rbig}[1]{\mathopen{\big#1}}
\DeclareRobustCommand{\rBig}[1]{\mathopen{\Big#1}}
\DeclareRobustCommand{\rbigg}[1]{\mathopen{\bigg#1}}

\DeclarePairedDelimiterX{\bra}[1]{\delimsize\langle}{\delimsize\rvert}{#1}
\DeclarePairedDelimiterX{\ket}[1]{\delimsize\lvert}{\delimsize\rangle}{#1}
\DeclarePairedDelimiterX{\makebraket}[1]{\delimsize\langle}{\delimsize\rangle}{#1}

\makeatletter
\newcommand{\braketbar}{%
	\, \delimsize\vert\@ifnextchar|{\!}{\,\!\!\:}%
}
\makeatother
\newcommand{\activatebraketbar}{%
	\begingroup\lccode`~=`|\lowercase{\endgroup\let~}\braketbar
	\mathcode`|="8000
}

\NewDocumentCommand{\braket}{som}{%
	\mathord{%
		\begingroup
		\activatebraketbar
		\IfBooleanTF{#1}
		{\makebraket*{#3}}
		{\IfNoValueTF{#2}{\makebraket{#3}}{\makebraket[#2]{#3}}}%
		\endgroup
	}%
}
\newcommand{\ols}[1]{\mskip.5\thinmuskip\overline{\mskip-.5\thinmuskip {#1} \mskip-.5\thinmuskip}\mskip.5\thinmuskip} % overline short
\newcommand{\olsi}[1]{\,\overline{\!{#1}}} % overline short italic
\makeatletter
\newcommand\closure[1]{
	\tctestifnum{\count@stringtoks{#1}>1} %checks if number of chars in arg > 1 (including '\')
	{\ols{#1}} %if arg is longer than just one char, e.g. \mathbb{Q}, \mathbb{F},...
	{\olsi{#1}} %if arg is just one char, e.g. K, L,...
}
\long\def\count@stringtoks#1{\tc@earg\count@toks{\string#1}}
\long\def\count@toks#1{\the\numexpr-1\count@@toks#1.\tc@endcnt}
\long\def\count@@toks#1#2\tc@endcnt{+1\tc@ifempty{#2}{\relax}{\count@@toks#2\tc@endcnt}}
\def\tc@ifempty#1{\tc@testxifx{\expandafter\relax\detokenize{#1}\relax}}
\long\def\tc@earg#1#2{\expandafter#1\expandafter{#2}}
\long\def\tctestifnum#1{\tctestifcon{\ifnum#1\relax}}
\long\def\tctestifcon#1{#1\expandafter\tc@exfirst\else\expandafter\tc@exsecond\fi}
\long\def\tc@testxifx{\tc@earg\tctestifx}
\long\def\tctestifx#1{\tctestifcon{\ifx#1}}
\long\def\tc@exfirst#1#2{#1}
\long\def\tc@exsecond#1#2{#2}
\makeatother

\renewcommand{\vec}[1]{\boldsymbol{\mathrm{#1}}}
\usepackage{contour}
\usepackage[normalem]{ulem}
\usepackage{soul}

\contourlength{0.8pt}

\makeatletter
\g@addto@macro\bfseries{\boldmath}
\makeatother

\makeatletter
\DeclareRobustCommand{\cbig}[1]{\mathopen{\setbox0=\hbox{$\m@th\big#1$}\raisebox{-0.55pt}{$\scalebox{1.2}{\copy0}$}}}
\makeatother
\makeatletter
\DeclareRobustCommand{\cBig}[1]{\mathopen{\setbox0=\hbox{$\m@th\Big#1$}\raisebox{-0.4pt}{$\scalebox{1.15}{\copy0}$}}}
\makeatother
\makeatletter
\DeclareRobustCommand{\cbigg}[1]{\mathopen{\setbox0=\hbox{$\m@th\bigg#1$}\raisebox{-0.4pt}{$\scalebox{1.125}{\copy0}$}}}
\makeatother
\makeatletter
\DeclareRobustCommand{\cBigg}[1]{\mathopen{\setbox0=\hbox{$\m@th\Bigg#1$}\raisebox{-0.5pt}{$\scalebox{1.125}{\copy0}$}}}
\makeatother

\newcommand{\mc}[1]{{\color{magenta}{[MC: #1]}}}

\usepackage{setspace}
\definecolor{forestgreen}{rgb}{0.13, 0.55, 0.13}

\usepackage{marginnote}

\begin{document}
	\renewcommand{\headrulewidth}{0pt}
	\pagestyle{fancy}
	\pagenumbering{Roman}
	\fancyhead{}
	\cfoot{$-\;\!$\thepage $\;\!-$}
	
	\begin{flushright}
        {\small
			\textcolor{black}{TUM--HEP--1613/26}\\
			\textcolor{black}{MPP--2026--179}
		}
	\end{flushright}
	\vspace{0.5cm}
	
	\begin{center}
        
        {\LARGE\bf 
        \begin{center}
        On the relaxation dynamics of non-equilibrium quantum systems
        \end{center}}
        
        \vspace{0.5cm}
        
		\textsc{Matthias Carosi}$^{1}$, \textsc{Björn Garbrecht}$^{2}$, \textsc{Silvia Pla}$^{2}$,\\[0.05cm] \textsc{Nils Wagner}$^{2,3}$ and \textsc{Edward Wang}$^{2}$

		\vspace{0.55cm}
        {\it ${}^1$\,International Centre for Theoretical Physics Asia-Pacific (ICTP-AP), University of \\Chinese Academy of Sciences, 100190 Beijing, China\\[0.15cm]}
		{\it ${}^2$\,Department of Physics T70, Technical University of Munich, James--Franck--Stra{\ss}e 1, \\ 85748 Garching, Germany \\[0.15cm]}
		{\it ${}^3$\,Max Planck Institute for Physics (Werner Heisenberg Institute), Boltzmannstraße 8, \\ 85748 Garching, Germany\\}

		\vspace{0.55cm}
		\emph{E--mail:} \href{mailto:matthias.carosi@tum.de}{{\tt matthias.carosi@tum.de}}, \href{mailto:garbrecht@tum.de}{{\tt garbrecht@tum.de}}, \href{mailto:silvia.pla-garcia@tum.de}{{\tt silvia.pla-garcia@tum.de}}, 
		\href{mailto:nils.wagner@tum.de}{{\tt nils.wagner@tum.de}}, \href{mailto:edward.wang@tum.de}{{\tt edward.wang@tum.de}}\hspace{4.05cm}

        \vspace{0.5cm}

		\medskip
		
	\end{center}
	
	\vspace{1cm}
	
	\begin{abstract}
    \noindent 
    We investigate the relaxation of an approximately conserved charge in interacting quantum systems close to local equilibrium. To this end, we provide a pedagogical review of Zubarev’s non-equilibrium statistical operator approach in the minimal setting of a single non-conserved charge and apply it to the problem at hand. We explicitly highlight the physical assumptions that lead to a local relaxation law: weak charge violation, a separation between the short timescale of microscopic correlations and the much longer timescale of charge relaxation, and the resulting loss of microscopic memory. Under these conditions, the leading relaxation rate is determined by an equilibrium correlation function of the charge-violating operator. We show that the same result follows from a simpler local-equilibrium construction based on the system's evolution over an intermediate timescale, providing a direct alternative for practical calculations and making the common physical ingredients of the two approaches explicit. Beyond the decay law itself, we relate the relaxation rate to the equilibrium diffusion of the same charge. We then allow the charge density to vary in space, which leads to a diffusion-relaxation equation. Finally, we illustrate the formalism through electroweak $\mathrm{B+L}$ washout and a perturbative scalar model, where agreement with the linearized Boltzmann equation establishes a direct connection between equilibrium-correlator and kinetic descriptions.
	\end{abstract}
    
    \vfill
    \noindent

\newpage 

\tableofcontents

\newpage

\renewcommand{\headrulewidth}{0pt}
\pagestyle{fancy}
\pagenumbering{arabic}
\fancyhead{}
\cfoot{$-\;\!$\thepage $\;\!-$}

\section{Introduction}
\label{sec:introduction}

\noindent 
 While interacting quantum systems may exhibit extraordinary complexity at the microscopic level, their late-time dynamics can often be described in terms of only a few, slow variables. The presence of conserved charges naturally facilitates such a simplified description. In case a symmetry is weakly broken, the associated charge is no longer exactly conserved, yet it can remain long-lived on timescales over which microscopic correlations have already decayed. This separation of timescales suggests an effective description in which the slow charge obeys a local relaxation law whose parameters are determined by equilibrium properties of the underlying theory. Prominent examples of such approximately conserved charges include the axial (chiral) charge in anomalous relativistic systems~\cite{KhlebnikovShaposhnikov1988,RabyMottola1990,Figueroa:2017hun}, the total momentum in nearly translation-invariant electron fluids~\cite{Lucas:2015sya}, particle number in weakly open quantum gases~\cite{Bouchoule:2020lwr}, quasi-conserved integrals of motion in near-integrable models~\cite{Mallayya:2019uqw,Surace:2023wqq}, and topological or higher-form winding charges in superfluids~\cite{Delacretaz:2019brr,Armas:2023tyx}, all of which play a central role in contemporary studies of non-equilibrium dynamics. \\ 

\noindent 
 The study of slow macroscopic dynamics from microscopic principles has a long history. A foundational step was provided by linear response and fluctuation--dissipation theory, which established a connection between irreversible evolution and equilibrium correlation functions~\cite{Callen:1951,Green:1952,Green:1954ubq,Kubo1957,KuboYokotaNakajima1957,Kubo:1966}. This connection was developed further within projection-operator methods, where a chosen set of slow observables is separated systematically from the remaining microscopic degrees of freedom, whose effects are encoded in memory terms~\cite{Nakajima:1958pnl,ZwanzigProjection1960,Zwanzig:1961zz,Mori:1965oqj,Robertsen1966,Grabert:1975,Vrugt:2019qhr}. These ideas also underlie microscopic approaches to hydrodynamics and local equilibrium~\cite{Kadanoff:1963axw,Forster:1975,vanWeert:1983,Hayata:2015lga,Mabillard2023}, and have been extended to relativistic quantum field theory through real-time and kinetic descriptions~\cite{Hosoya:1983id,Calzetta:1986cq,Jeon:1995zm,Hidaka:2022dmn}. Within this broader framework, approximately conserved quantities provide a particularly simple setting in which a separation between microscopic and macroscopic timescales emerges, and weakly broken symmetries have consequently attracted renewed attention~\cite{Bamler20215,Grozdanov:2018fic,Armas:2021vku,Hongo:2024brb}. In thermal quantum field theory, this separation of scales has been exploited to relate the relaxation of weakly violated quantum numbers directly to equilibrium real-time correlators and susceptibilities~\cite{Bodeker:2012gs,Bodeker:2014hqa,Bodeker:2019ajh}. \\

\noindent 
Another particularly promising framework for describing relaxation dynamics is Zubarev's non-equilibrium statistical operator (NSO) method~\cite{Zub61,ZubarevArticle1970,ZubarevBook1971,Zub80,ZubarevBook1996,ZubarevBook1997,RoepkeBook2013}. In this approach, the macroscopic state is specified by a set of relevant observables and their conjugate thermodynamic parameters, while a retarded construction selects the causal solution of the microscopic dynamics. For an approximately conserved charge, that charge itself may be included among the relevant observables, with a conjugate chemical potential parameterizing its departure from equilibrium. The non-equilibrium statistical operator constructed in this way can then be used to determine the dynamics of the relevant observables and, more generally, to characterize the system's relaxation toward equilibrium. The NSO formalism has been developed extensively in kinetic theory and relativistic hydrodynamics~\cite{Kuz18,BecattiniArticle2019,Harutyunyan:2021rmb}, and has been applied directly to anomalous baryon plus lepton number relaxation by Khlebnikov \& Shaposhnikov~\cite{KhlebnikovShaposhnikov1988} and by Mottola \& Raby~\cite{RabyMottola1990}. \\ 

\noindent 
Despite the systematic nature of the NSO construction, derivations of a local relaxation equation within that framework are often presented in a rather compressed form. This can obscure the role of the hierarchy between microscopic and macroscopic timescales and the suppression of memory effects, as well as the approximations underlying the derivation. We therefore revisit the relaxation of a single weakly violated charge in a minimal setting, making these ingredients explicit and clarifying how they lead to a local description of the late-time dynamics.\\

\noindent 
We complement this analysis with a parallel derivation based directly on local equilibrium. Rather than constructing the full non-equilibrium statistical operator, we use a local-equilibrium state to represent the instantaneous macroscopic state at late times. Since this state does not retain the microscopic correlations of the physical state, we evolve it over a mesoscopic interval $\Delta t$ satisfying $\tau_{\rm micro}\ll\Delta t\ll\tau_{\rm macro}$, long enough for sensitivity to the discarded microscopic correlations to be lost but short enough that the slow, macroscopic variables change only parametrically little. The resulting finite change of the charge over the time interval $\Delta t$ leads directly to the desired relaxation rate. Despite their different treatment of the microscopic history, the NSO and local-equilibrium derivations agree at the order considered.\\

\noindent
The local-equilibrium construction naturally extends beyond the relaxation of a homogeneous charge. Applied to slowly varying densities, it incorporates spatial transport and weak charge violation within a common diffusion--relaxation description, connecting with standard hydrodynamics and more recent Schwinger--Keldysh formulations of charge transport~\cite{Kadanoff:1963axw,Kovtun:2012rj,Liu:2018kfw,Akyuz:2023lsm,Firat:2025upx}. The framework is furthermore applicable to the electroweak baryon plus lepton $(\mathrm{B+L})$ charge, whose non-conservation is tied to the chiral anomaly~\cite{Adler:1969gk,Bell:1969ts,tHooft:1976rip} and whose relaxation rate can be related to the Chern--Simons diffusion rate~\cite{Manton:1983nd,Klinkhamer:1984di,Kuzmin:1985mm,ArnoldMcLerran1987,Arnold:1996dy,Bodeker:1998hm,Moore:1996qs}. In a perturbative scalar model, the agreement with the linearized Boltzmann equation provides an independent check of the correlator-based description in a regime where both approaches apply~\cite{Jeon:1994if,Weinstock:2005jw,Lindner:2005kv}.\\

\noindent 
The paper is organized as follows: We first introduce the general setup, notation, and assumptions in section~\ref{sec:2_SetupSection}. With these conventions in place, we turn to two complementary derivations of the relaxation rate. Sections~\ref{sec:2_Zubarev_Method} and~\ref{sec:Zubarev_Calculation} provide a detailed derivation based on the aforementioned NSO formalism, while section~\ref{sec:3_LEQ_Method} develops the simplified local-equilibrium approach. Section~\ref{sec:details} then establishes the connection between relaxation and diffusion, with section~\ref{sec:applications} applying the results to scalar and electroweak examples. Finally, section~\ref{sec:transport} generalizes the analysis to spatially varying slow variables, before we conclude in section~\ref{sec:conclusions}. Selected technical details are deferred to the appendices.

\section{General setup \& notational conventions}
\label{sec:2_SetupSection}

Let us begin by outlining the class of systems under consideration and introducing the notation used throughout this work. We consider a closed quantum system governed by a time-independent Hamiltonian $H$ and exhibiting an approximately preserved continuous global symmetry generated by the charge operator $Q$. The weak violation of $Q$ is encoded in the decomposition of the Hamiltonian
\begin{equation}
    H = H^{\mathrm{(cons.)}} + \lambda H^{\mathrm{(viol.)}} \,,
    \label{eq:Hamiltonian_Split}
\end{equation}
where $\lambda \ll 1$ is a small dimensionless parameter controlling the strength of the symmetry breaking. Here, $H^{\mathrm{(cons.)}}$ parametrizes the charge-conserving part of the dynamics, while $H^{\mathrm{(viol.)}}$ generates processes that violate $Q$, such that\footnote{The parameter $\lambda$ is chosen such that the commutator between the charge-violating part of the Hamiltonian and $Q$ yields contributions of order $\lambda^0$ in a perturbative expansion in $\lambda$ when evaluated in expectation values or operator insertions.}
\begin{equation}
    \cbig[H^{\mathrm{(cons.)}},Q\cbig] \, =  0 \,, \qquad \cbig[H^{\mathrm{(viol.)}},Q\cbig] \, \neq  0 \,.
    \label{eq:Commutators_Cons_Viol}
\end{equation}
We adopt conventions in which the expectation value of the operator $Q$ is dimensionless. Throughout the general discussion, we restrict ourselves to the minimal case in which the total energy and the approximately conserved charge $Q$ are the only macroscopic conserved or slowly relaxing quantities retained in the statistical description. In particular, we assume that no additional exactly conserved global charges constrain the relaxation of $Q$. Finally, we assume the charge operator $Q$ to be odd under a $\mathcal{C}$ transformation, an assumption maintained throughout this manuscript.\footnote{As it turns out, it would suffice for $Q$ to be odd under the combined $\mathcal{CPT}$ transformation. However, it is more natural to demand $Q$ to break $\mathcal{C}$.}

\subsection{Quantum-mechanical pictures}

While equations~\eqref{eq:Hamiltonian_Split} and~\eqref{eq:Commutators_Cons_Viol} should be understood to be given in the Schrödinger picture, we will henceforth indicate the relevant picture on any operator $O(t)$ explicitly by a subscript
\begin{subequations}
\begin{align}
    &&&&O_{\mathrm{S}}(t)&& &\text{\textbf{S}chrödinger picture}\, , &&&&\\[0.11cm]
    &&&&O_{\mathrm{H}}(t)&\equiv \exp\cBig(\frac{iHt}{\hbar}\cBig) \:\!O_{\mathrm{S}}(t) \:\! \exp\cBig(-\:\!\frac{iHt}{\hbar}\cBig)&  &\text{\textbf{H}eisenberg picture}\, , &&&&\\[0.05cm]
    &&&&O_{\mathrm{I}}(t)&\equiv \exp\cBig(\frac{i}{\hbar}\:\!H^{\mathrm{(cons.)}}_{\!\!\;\mathrm{S}}\:\!t\cBig) \:\!O_{\mathrm{S}}(t) \:\! \exp\cBig(-\:\!\frac{i}{\hbar}\:\!H^{\mathrm{(cons.)}}_{\!\!\;\mathrm{S}}\:\!t\cBig)& &\text{\textbf{I}nteraction picture}\, . &&&& \label{eq:Interaction_Picture}
\end{align}
\end{subequations}
As expressions in the following will involve all three pictures, this convention serves to prevent any ambiguity. It also allows us to clearly distinguish quantum-mechanical operators from ordinary functions, since we refrain from using hats to indicate operators. With the Hamiltonian $H$ being time-independent, we choose to omit its picture label whenever the interaction picture is not being used. We adopt the same convention for the equilibrium density matrix $\rho^{(\mathrm{EQ})}$, which depends solely on $H$.

\subsection{Notation of expectation values} 
Expectation values of operators are marked with a superscript $t$, indicating the time at which the inserted density matrix $\rho$ is evaluated, e.g.
\begin{align}
    \braket{H\:\!}^t \equiv \mathrm{tr}\Big[\rho_\mathrm{S}(t)\:\!H\Big]\, ,\qquad \text{or} \qquad \braket[\cbig]{O_{\mathrm{H}}(t)}^{\!\!\:t'} \equiv \mathrm{tr}\Big[\rho_\mathrm{H}(t')\:\!O_{\mathrm{H}}(t)\Big]\, .
\end{align}
When expectation values are computed in equilibrium, so that the traced density matrix is explicitly time-independent, we omit this time argument. In case a particular density matrix is used to define the desired average, this is indicated by an additional subscript, e.g. 
\begin{align}
    \braket[\cbig]{O_\mathrm{I}(t)}_\mathrm{EQ} \equiv \mathrm{tr} \Big[ \rho^{(\mathrm{EQ})} \:\! O_\mathrm{I}(t) \Big] \,,
    \qquad \text{or} \qquad
    \braket[\cbig]{O_\mathrm{H}(t)}^{\!\!\:t'}_\mathrm{NEQ} \equiv \mathrm{tr}\Big[ \rho^{(\mathrm{NEQ})}_\mathrm{H}(t') \:\! O_\mathrm{H}(t) \Big] \,.
\end{align}
Although we always explicitly indicate the picture we are working in, we remind the reader that expectation values are independent of this choice.

\subsection{Primary objective: Relaxation rate}

The system is initialized at $t=0$ with a nonvanishing expectation value $\braket[\rbig]{Q_{\mathrm{S}}}^{t=0}\neq 0$, thereby placing it out of equilibrium. In particular, the initial density matrix $\rho_{\mathrm{S}}(t\!\!\:=\!\!\:0)$ does not coincide with an equilibrium (EQ) distribution 
\begin{align}
    \rho^{(\text{EQ})}\equiv \frac{1}{Z_\mathrm{EQ}}\exp\cbig(-\beta_\mathrm{EQ}H\cbig)\, ,\qquad Z_\mathrm{EQ}\equiv \mathrm{tr}\Big[\exp\cbig(-\beta_\mathrm{EQ}H\cbig)\Big]\, ,
    \label{eq:EQ_Operator}
\end{align}
for any constant inverse temperature $\smash{\beta_\mathrm{EQ}}$. Our goal is then to investigate the real-time dynamics of the expectation value 
\begin{align}
    \braket[\rbig]{Q_{\mathrm{S}}}^{\!\!\;t} = \mathrm{tr}\Big[\rho_\mathrm{S}(t)\,Q_{\mathrm{S}}\Big] = \mathrm{tr}\Big[e^{-iH t/\hbar}\, \rho_\mathrm{S}(t\!\!\:=\!\!\:0) \:\! e^{iH t/\hbar}Q_{\mathrm{S}}\Big] \,.
\end{align}
As the system evolves and approaches equilibrium, the total charge relaxes to zero\footnote{The equilibrium average of $Q_\mathrm{S}$ vanishes due to the charge operator being odd under a $\mathcal{CPT}$ transformation, whereas the Hamiltonian $H$, and thus the equilibrium density matrix, is even. A more detailed discussion can be found in appendix~\ref{sec:CPT_Arguments}. Note that, if the equilibrium density matrix fails to be $\mathcal{CPT}$-even, e.g. because the equilibrium ensemble is characterized by some non-vanishing charge $\braket{Q_\mathrm{S}'}_\mathrm{EQ}\neq0$ with $Q_\mathrm{S}'$ being a $\mathcal{C}$-odd charge operator that commutes with $H$, the equilibrium average of $Q_\mathrm{S}$ might not vanish.}
\begin{align}
    \braket[\rbig]{Q_{\mathrm{S}}}^{\!\!\:t} \:\!\xrightarrow{\,t\,\to\,\infty\,}\:\! \braket[\rbig]{Q_{\mathrm{S}}}_\mathrm{EQ} = \mathrm{tr}\Big[\rho_\mathrm{S}^{\:(\text{EQ})}\,Q_{\mathrm{S}}\Big] = 0\,.
\end{align}
At sufficiently late times, the relaxation toward equilibrium is expected to be governed by a linear differential equation of the form
\begin{align}
    \label{eq:relaxation_equation}
    \frac{\mathrm{d}}{\mathrm{d}t}\, \braket[\rbig]{Q_{\mathrm{S}}}^{\!\!\:t} = - \:\!\Gamma_{\mathrm{relax}} \, \braket[\rbig]{Q_{\mathrm{S}}}^{\!\!\:t} \:\!\bigg\{1+\mathcal{O}\Big(\!\!\;\braket[\rbig]{Q_{\mathrm{S}}}^{\!\!\:t},\lambda\Big)\!\!\;\!\!\:\bigg\} \,,
\end{align}
where $\Gamma_{\mathrm{relax}}$ is an effective \emph{relaxation rate} that depends on the microscopic details of the theory. As we will see below, corrections to this relation involve both higher powers of $\braket[\rbig]{Q_{\mathrm{S}}}^{\!\!\:t}$, which are subleading in the late-time limit, as well as higher-order contributions in the small parameter $\lambda$. Deriving equation~\eqref{eq:relaxation_equation} and identifying $\Gamma_{\mathrm{relax}}$ is the primary goal of our discussion.

\subsection{Central assumptions}
\label{sec:Central_Assumptions}

As we shall see, deriving the desired relaxation equation~\eqref{eq:relaxation_equation} critically relies on a separation of timescales between equilibration and relaxation, or equivalently, between fast and slow variables (sometimes informally referred to as modes). Close to equilibrium, thermal and quantum effects induce fast fluctuations with a finite correlation time, which we denote as the microscopic \emph{equilibration time} $\tau_{\mathrm{micro}}$. Formulated explicitly, this corresponds to the assumption that any (connected) equilibrium two-point correlator that involves a fast mode decays exponentially over a microscopic timescale $\tau_\mathrm{micro}$, namely
\begin{align}
    \!\!\braket[\cbig]{\!\!\;A^{(\mathrm{fast})}_\mathrm{H}(t)\:\! B_\mathrm{H}(t')\!\!\;}_{\!\!\:\mathrm{EQ}} - \braket[\cbig]{\!\!\;A^{(\mathrm{fast})}_\mathrm{H}\!\!\;(0)\!\!\;}_{\!\!\:\mathrm{EQ}} \braket[\cbig]{\!\!\;B_\mathrm{H}(0)\!\!\;}_{\!\!\:\mathrm{EQ}} \xrightarrow{\,\lvert t-t'\rvert \,\gg \,\tau_\mathrm{micro}\,} \mathrm{const.}\times \exp\!\!\:\bigg(\!\!\!\:-\frac{\lvert t-t'\rvert}{\tau_\mathrm{micro}}\bigg) \, ,
    \label{eq:Behavior_TwoPointFunction}
\end{align}
where the operator $B_{\mathrm{H}}$ can be either a fast or slow variable.\footnote{A trivial example of a slow mode is given by the Hamiltonian $H$, for which the equilibrium two-point correlator is a time-independent non-zero constant. A more interesting example is the Hamiltonian density, as we shall discuss in section~\ref{sec:transport}.} Crucially, this remains true even when $\lambda=0$, i.e. when the charge is exactly conserved. In this case, where the equilibrium inverse temperature sets the only relevant microscopic scale, the equilibration time is naturally of order $\tau_{\mathrm{micro}}\sim \hbar\beta_{\mathrm{EQ}}\sim \hbar (k_\mathrm{B}T)^{-1}$, and acquires perturbative corrections for small nonzero $\lambda$. Not all fluctuation modes, however, relax on this microscopic timescale. In what follows, we will be particularly interested in identifying and isolating those modes that remain parametrically slow. For $\lambda\neq 0$, one expects one-point functions of slow operators $O_\mathrm{S}$ without explicit time dependence to relax toward their equilibrium values on a much longer macroscopic timescale according to
\begin{align}
    \braket[\rbig]{O_{\mathrm{S}}}^t - \braket[\rbig]{O_{\mathrm{S}}}_\mathrm{EQ} \xrightarrow{\,t \,\gg \,\tau_\mathrm{macro}\,} \mathrm{const.} \times \exp\!\!\:\bigg(\!\!\!\:-\!\!\:\frac{t}{\tau_\mathrm{macro}}\bigg)\, .
    \label{eq:Behavior_OnePointFunction}
\end{align}
Here, we have introduced the characteristic \emph{relaxation time} $\tau_{\mathrm{macro}}$, which we will ultimately identify with the inverse relaxation rate $\Gamma_{\mathrm{relax}}^{-1}$. As we will see in due time, the macroscopic timescale $\tau_\mathrm{macro}$ is in fact enhanced by two rather than just one power of the coupling $\lambda$, such that $\tau_\mathrm{macro}\sim \lambda^{-2} \hbar\beta_\mathrm{EQ}$, where again $\hbar\beta_\mathrm{EQ}$ sets the relevant microscopic timescale. The preceding arguments suggest that the system naturally exhibits a hierarchy of timescales
\begin{align}
    \frac{\tau_{\mathrm{micro}}}{\tau_{\mathrm{macro}}} \sim \lambda^2 \ll \: 1  \quad \Longrightarrow \quad \tau_{\mathrm{micro}} \ll \: \tau_{\mathrm{macro}} \,.
\end{align}
When we are only interested in the description of the non-equilibrium dynamics of a few slow modes over time-scales $t\sim\tau_\mathrm{macro}\gg\tau_\mathrm{micro}$, details of the infinitely many fast modes become irrelevant and the analysis simplifies considerably.

\section{Overview of Zubarev's NSO approach}
\label{sec:2_Zubarev_Method}

The non-equilibrium statistical operator (NSO) approach, originally conceptualized by Zubarev~\cite{Zub61,ZubarevArticle1970,ZubarevBook1971}, offers a powerful framework for deriving convenient expressions for relaxation rates in out-of-equilibrium systems. The scenario of interest, namely the relaxation of the expectation value of a weakly non-conserved operator $Q$ toward equilibrium following an initial perturbation, constitutes a minimal and controlled setting for the application of Zubarev’s ansatz.
Closely following the analysis provided by both Khlebnikov \& Shaposhnikov~\cite{KhlebnikovShaposhnikov1988} and Mottola \& Raby~\cite{RabyMottola1990}, we will outline the general idea behind the NSO approach in the generic setting introduced in section~\ref{sec:2_SetupSection}. While we will eventually reproduce the results reported in the aforementioned works, we mainly aim to offer a clearer and more explicit derivation, filling in some intermediate steps omitted in earlier presentations while making the underlying assumptions more transparent. For alternative perspectives and generalizations beyond the scope of this discussion, we refer to the comprehensive textbook treatments~\cite{ZubarevBook1996,ZubarevBook1997,RoepkeBook2013}. However, we believe that Zubarev’s method leaves certain important questions unresolved, rendering the framework somewhat ambiguous when computing next-to-leading order (NLO) corrections in the parameter $\lambda$. Therefore, even readers already familiar with the procedure may find it worthwhile to consult section~\ref{sec:ZubarevDiscussion}.

\subsection{Conceptual foundations of the NSO approach}

Before constructing a suitable density matrix to describe non-equilibrium systems, let us briefly outline the guiding principles of the framework. If one wishes to study the real-time evolution of the expectation value $\braket[\rbig]{Q_\mathrm{S}}^{\!\!\;t}$ of a (weakly) non-conserved operator $Q_\mathrm{S}$, it would formally suffice to specify the density matrix at some initial reference time $t_\mathrm{ref}$, granting the exact relation 
\begin{align}
    \braket[\rbig]{Q_{\mathrm{S}}}^{\!\!\;t} &=  \mathrm{tr}\Bigg\{\exp\!\bigg[\!-\!\!\:\frac{iH}{\hbar}\:\!\big(t-t_\mathrm{ref}\big)\bigg] \:\! \rho_\mathrm{S}\big(t\!\:\!=\!\:\!t_\mathrm{ref}\big) \:\! \exp\!\bigg[\frac{iH}{\hbar}\:\!\big(t-t_\mathrm{ref}\big)\bigg]Q_{\mathrm{S}}\Bigg\} \, .
\end{align}
While formally exact, this expression immediately raises two fundamental difficulties. First, which initial density matrix $\rho_\mathrm{S}\big(t\!\:\!=\!\:\!t_\mathrm{ref}\big)$ appropriately characterizes the non-equilibrium situation of interest? Second, even if such a state were known, the explicit evaluation of the time-evolution operators is generally intractable in interacting systems.\\

\noindent 
Zubarev’s construction circumvents these issues by introducing an appropriately structured density matrix whose time dependence is encoded entirely in a set of coefficient functions. These functions are fixed self-consistently through a simple system of constraint equations, whose solutions can be reliably approximated. Crucially, as will become clear below, the construction avoids the need to specify the initial conditions explicitly. Instead, the method directly targets the universal relaxation behavior of the system under small perturbations around equilibrium, thereby exploiting the natural separation between microscopic and macroscopic timescales. \\ 

\noindent 
The underlying philosophy bears some resemblance to the treatment of resonant states in quantum mechanics. When analyzing the decay of a metastable system, one is typically not concerned with resolving the full microscopic time evolution. Rather, one focuses on the intermediate temporal regime in which the decay becomes effectively exponential and can be characterized by stationary states with purely outgoing (Gamow--Siegert) boundary conditions~\cite{GamowAlphaDecay,SiegertRadiativeStates}. These resonant states provide an idealized, yet highly efficient description of the decaying system~\cite{IntroductionGamovVectors,Hatano:2026tdm}. In a similar spirit, Zubarev’s approach leverages scale separation in a relaxing system to introduce a physically motivated ansatz that captures its universal, coarse-grained dynamics without requiring explicit control over the full microscopic evolution.

\subsection{Construction of the non-equilibrium statistical operator}
\label{sec:Construction_NSO}

\noindent
The central assumption underlying the NSO method is the existence of a hierarchy of timescales, which enables the macroscopic evolution of the system to be effectively described in terms of expectation values of a set of so-called \emph{relevant} dynamical operators. These operators are associated with the slow variables introduced in section~\ref{sec:Central_Assumptions}, which in our case are the system's total energy $\braket{H\:\!}\raisebox{5pt}{\scalebox{0.75}{$t$}}$ and charge $\braket[\rbig]{Q_{\mathrm{S}}}\raisebox{5pt}{\scalebox{0.75}{$t$}}$.\footnote{Note that, in general, one works with densities rather than their integrated quantities; however, since we assume a homogeneous system, it is sufficient to use the total quantities. The discussion can be easily extended to inhomogeneous systems, see section~\ref{sec:transport}.} If the system of interest exhibits additional slow observables, they should be included in the effective description following the same reasoning outlined below.\\ 

\noindent 
To motivate the subsequent construction of the NEQ operator $\smash{\rho_\mathrm{S}^{(\mathrm{NEQ})}}$, let us begin by considering the following scenario: Suppose a benevolent deity granted us complete knowledge of the physical density matrix $\smash{\rho_\mathrm{S}^{(\mathrm{phys})}\!\!\:(t)}$ governing the system’s evolution during the desired uniform relaxation phase. How might we obtain a reasonably accurate approximation to it? As we shall see, once the desired approximation is formulated self-consistently and the relevant assumptions are satisfied, explicit reference to $\smash{\rho_\mathrm{S}^{(\mathrm{phys})}\!\!\:(t)}$ is no longer required. This is essential, since in practice $\smash{\rho_\mathrm{S}^{(\mathrm{phys})}\!\!\:(t)}$ is, of course, unknown. The NSO approach is motivated by the idea that, in the absence of any additional information about the initial conditions, the NEQ operator furnishes an effective representation of $\rho_{\mathrm S}^{(\mathrm{phys})}(t)$ for the description of the relevant observables. \\

\noindent 
Having (hypothetical) access to the full density matrix $\smash{\rho_\mathrm{S}^{(\mathrm{phys})}\!\!\:(t)}$ grants us access to the relevant expectation values 
\begin{align}
    \braket{H\:\!}^{t}_{\mathrm{phys}} = \mathrm{tr}\Big[\rho_\mathrm{S}^{(\mathrm{phys})}(t)\:\!H\Big] \qquad \text{and}\qquad \braket[\rbig]{Q_{\mathrm{S}}}^{\!\!\;t}_{\mathrm{phys}} = \mathrm{tr}\Big[\rho_\mathrm{S}^{(\mathrm{phys})}(t)\,Q_{\mathrm{S}}\Big]\, ,
    \label{eq:RelevantExpectationValues}
\end{align}
which constitute the only information about $\smash{\rho_\mathrm{S}^{(\mathrm{phys})}\!\!\:(t)}$ that we will use in the subsequent construction, reflecting our assumption that the macroscopic state is adequately characterized by these slow modes alone. Therefore, we look for a non-equilibrium statistical operator $\smash{\rho_\mathrm{S}^{(\mathrm{NEQ})}}$ that fulfills the following requirements:
\begin{enumerate}[label=(\roman*)]
    \item The to-be-determined approximation should be formulated as a functional of the two averages $\braket{H\:\!}^{t}_{\mathrm{phys}}$ and $\braket[\rbig]{Q_\mathrm{S}}^{\!\!\;t}_{\mathrm{phys}}$. Moreover, causality requires that the resulting density matrix at time $t$ depend only on the averages at times $t'\leq t$, and never on future values.
    \item Since the approximation is intended to encode the generic relaxation behavior of the system, it should exhibit time-translation invariance in the sense that the functional form of the density matrix is preserved over time.
    \item The approximation should satisfy the von Neumann equation, thereby possessing the correct time-evolution under the full Hamiltonian $H$.\footnote{Note that this condition will be relaxed in the alternative construction presented in section~\ref{sec:3_LEQ_Method}, as discussed in more detail below.}
\end{enumerate}
\noindent 
The construction of a suitable idealized density matrix describing the relaxation process proceeds in two steps. One first introduces an auxiliary density matrix $\smash{\rho_\mathrm{S}^{(\mathrm{LEQ})}\!\!\:(t)}$, referred to as the \emph{local equilibrium (LEQ) statistical operator}, which is then used to construct the desired \emph{non-equilibrium (NEQ) statistical operator} $\smash{\rho_\mathrm{S}^{(\mathrm{NEQ})}\!\!\:(t)}$.\footnote{Beware that the LEQ operator $\smash{\rho_\mathrm{S}^{(\mathrm{LEQ})}\!\!\:(t)}$ is also sometimes referred to as the \emph{quasi-equilibrium density matrix}~\cite{ZubarevArticle1970} or as the \emph{relevant distribution}~\cite{RoepkeBook2013}.} As previous expositions of the method have not always been entirely transparent to us, we take the opportunity to restate its underlying logic in a manner that we hope is more natural and clarifies its essential ideas. Before proceeding with the construction of the LEQ and NEQ operators, let us first briefly recall how the situation would appear if the operator $Q_\mathrm{S}$ were fully conserved.

\subsubsection{Equilibrium statistical operator}

\noindent 
We first deal with the situation in which the operator $Q_\mathrm{S}$ commutes with the full Hamiltonian $H$, amounting to $\lambda=0$ in the previously introduced split~\eqref{eq:Hamiltonian_Split}. In this case, both the system's total energy $E_\mathrm{tot}$ as well as the total charge $Q_\mathrm{tot}$ are exactly conserved, allowing the system to attain thermal and chemical equilibrium. Since the time evolution is governed by the von Neumann equation
\begin{align}
    \frac{\mathrm{d}\rho_{\:\!\mathrm{S}}(t)}{\mathrm{d}{t}}+\frac{i}{\hbar}\:\!\cbig[H,\rho_{\:\!\mathrm{S}}(t)\cbig]=0\, ,
    \label{eq:VonNeumannEquation}
\end{align}
a stationary equilibrium (EQ) density matrix $\smash{\rho_\mathrm{S}^{(\mathrm{EQ})}}$ must commute with the Hamiltonian $H$. Avoiding pathological cases, this requirement confines the EQ operator $\smash{\rho_\mathrm{S}^{(\mathrm{EQ})}}$ to be an arbitrary function of both the Hamiltonian $H$ as well as the conserved operator $Q_\mathrm{S}$. The conserved quantities, fixed by the initial state, together with an overall normalization condition, impose the additional constraints
\begin{align}
    \mathrm{tr}\cbig(\rho_\mathrm{S}^{(\mathrm{EQ})}H\cbig)\,\overset{!}{=} E_{\scalebox{0.7}{\text{tot}}}  \, , \qquad \mathrm{tr}\cbig(\rho_\mathrm{S}^{(\mathrm{EQ})}Q_{\mathrm{S}}\cbig)\,\overset{!}{=} Q_\mathrm{tot} \, , \qquad \mathrm{tr}\cbig(\rho_\mathrm{S}^{(\mathrm{EQ})}\cbig)\,=1\, .
    \label{eq:Constraints_EQ_Operator}
\end{align}
However, these conditions alone do not uniquely determine the equilibrium state. Instead, an additional physical principle is required to unambiguously specify a desirable candidate. For a system with sufficiently many microscopic degrees of freedom, one therefore invokes the maximum-entropy principle~\cite{Jaynes:1957,Jaynes:1957II} and selects the density matrix that maximizes the information entropy functional
\begin{align}
    S[\rho_\mathrm{S}]\equiv -k_{\mathrm{B}} \,\mathrm{tr}\Big[\rho_\mathrm{S} \log(\rho_\mathrm{S})\Big]\, ,
    \label{eq:EntropyFunctional}
\end{align}
subject to the previously stated constraints~\eqref{eq:Constraints_EQ_Operator}. The corresponding variational problem results in the well-known Gibbs ensemble
\begin{align}
    \rho_\mathrm{S}^{(\mathrm{EQ})} = \exp\!\bigg\{\!-\Phi-\beta_\mathrm{EQ}\Big[H-\mu_\mathrm{EQ} \:\!Q_\mathrm{S}\Big]\!\!\:\bigg\}\, ,
    \label{eq:EQ_Density_Matrix_Mu}
\end{align}
with the Lagrange multipliers $\Phi$, $\beta_\mathrm{EQ}$ and $\mu_\mathrm{EQ}$ uniquely fixed by the constraints~\eqref{eq:Constraints_EQ_Operator}. While the so-called Massieu--Planck-potential $\Phi$ takes care of the normalization, $\beta_\mathrm{EQ}$ is interpreted as the inverse (equilibrium) temperature $T$ of the system via the usual relation $\smash{\beta_\mathrm{EQ}=(k_{\!\!\;B} T)^{-1}}$. Similarly, the remaining thermodynamic parameter $\mu_\mathrm{EQ}$ describes the rate of change of the system’s energy with respect to the externally set conserved quantity $Q_\mathrm{tot}$ and is therefore identified as the corresponding chemical potential in equilibrium.

\subsubsection{Local equilibrium statistical operator}

\begin{figure}[H]
    \centering
    \includegraphics[width=0.95\textwidth]{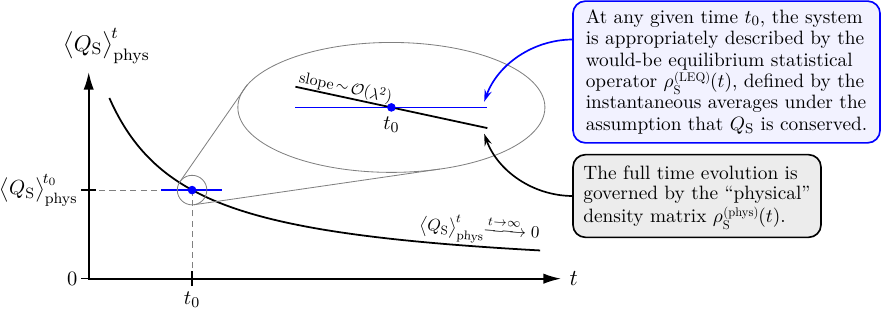}
    \caption{Schematic depiction of the uniform relaxation of the charge expectation value $\braket[\rbig]{Q_\mathrm{S}}^{\!\!\:t}_{\mathrm{phys}}$. Although the physical time evolution is governed by the density operator $\rho_\mathrm{S}^{(\mathrm{phys})}\!\!\:(t)$, at any instant $t_0$ the system may be viewed as occupying an effective equilibrium state with charge $\braket[\rbig]{Q_\mathrm{S}}^{\!\!\:t\:=\:t_0}_{\mathrm{phys}}$, corresponding to the configuration that would arise if the operator $Q_\mathrm{S}$ were conserved.}
    \label{fig:LEQ operator-SlopePlot}
\end{figure}

The system we aim to describe is slightly more complicated, as the operator $Q_{\mathrm{S}}$ is non-conserved. Accordingly, thermal equilibrium is characterized not by relation~\eqref{eq:EQ_Density_Matrix_Mu}, but rather by the thermal distribution~\eqref{eq:EQ_Operator}. In case the system is initially brought out of equilibrium, i.e. $\braket[\rbig]{Q_\mathrm{S}}^{\!\!\:t\:\!=\:\!t_0}_{\mathrm{phys}}\neq 0$, non-conservation of the system's total charge will lead it to relax slowly back to equilibrium, as illustrated in figure~\ref{fig:LEQ operator-SlopePlot}. 

\noindent 
Owing to the assumption that the non-conservation of $Q_\mathrm{S}$ is very weak, the system at any fixed time $t_0$ can be regarded as quasi-static. On sufficiently short timescales, it behaves approximately as if $Q_\mathrm{S}$ were conserved, as depicted in figure~\ref{fig:LEQ operator-SlopePlot}. It is therefore natural to adopt, as a first approximation, the equilibrium counterpart~\eqref{eq:EQ_Density_Matrix_Mu} derived previously, namely 
\begin{align}
    \rho_\mathrm{S}^{(\mathrm{phys})}\!\!\:(t_0) \approx \rho_\mathrm{S}^{(\mathrm{LEQ})}\!\!\:(t_0)=\exp\!\!\:\bigg\{\!-\Phi_0-\!\!\:\beta_0\Big[H-\mu_0 Q_{\mathrm{S}}\Big]\!\!\:\bigg\}\, .
\end{align}
As before, the Massieu--Planck-potential $\Phi_0$ fixes the normalization, whereas the instantaneous inverse temperature $\beta_0$ and chemical potential $\mu_0$ are uniquely determined by demanding the correct averages at time $t_0$ to be reproduced, i.e. demanding $\smash{\rho_\mathrm{S}^{(\mathrm{LEQ})}\!\!\:(t_0)}$ to satisfy
\begin{subequations}
\begin{align}
    \braket{H\:\!}^{t_0}_{\text{LEQ}} &= \mathrm{tr}\Big[\rho_\mathrm{S}^{(\mathrm{LEQ})}(t_0)\:\!H\Big]\:\,\overset{!}{=}\braket{H\:\!}^{t_0}_{\text{phys}}\, , \qquad \text{and} \\[0.15cm] 
    \braket[\rbig]{Q_{\mathrm{S}}}^{\!\!\;t_0}_{\text{LEQ}} &= \mathrm{tr}\Big[\rho_\mathrm{S}^{(\mathrm{LEQ})}(t_0)\,Q_{\mathrm{S}}\Big]\overset{!}{=}\braket[\rbig]{Q_{\mathrm{S}}}^{\!\!\;t_0}_{\text{phys}} \, .
\end{align}
\label{eq:LEQ_Matching_t0}%
\end{subequations}
The underlying idea is that, if we evolve $\smash{\rho_\mathrm{S}^{(\mathrm{LEQ})}\!\!\:(t_0)}$ forward or backward in time over a sufficiently short interval, the resulting state should remain close to the true physical density matrix. Significant deviations are expected to emerge only on longer timescales. Taking the above ansatz to its logical conclusion, one may impose the matching procedure~\eqref{eq:LEQ_Matching_t0} separately at each point in time, thereby promoting the Lagrange multipliers $\Phi(t)$, $\beta(t)$ and $\mu(t)$ to time-dependent quantities that are fixed independently at every instant. The arising density matrix
\begin{align}
    \rho_\mathrm{S}^{(\mathrm{LEQ})}\!\!\:(t)=\exp\!\!\:\bigg\{\!-\Phi(t)-\!\!\:\beta(t)\Big[H-\mu(t)Q_{\mathrm{S}}\Big]\!\!\:\bigg\}\, ,
    \label{eq:LEQ_Operator}
\end{align}
is dubbed the local equilibrium (LEQ) statistical operator, mimicking the desired evolution of the relevant averages by imposing an equilibrium distribution at each point in time individually, thereby maximizing the entropy~\eqref{eq:EntropyFunctional} locally in time. By construction, the so-defined density matrix satisfies    
\begin{align}
    \braket{H\:\!}^{t}_{\text{LEQ}} &\overset{!}{=}\braket{H\:\!}^{t}_{\text{phys}}\, , \qquad \braket[\rbig]{Q_{\mathrm{S}}}^{\!\!\;t}_{\text{LEQ}} \overset{!}{=}\braket[\rbig]{Q_{\mathrm{S}}}^{\!\!\;t}_{\text{phys}} \, ,
    \label{eq:SelfConsistency_LEQ}
\end{align}
for all $t$. Unfortunately, the so-defined LEQ operator $\smash{\rho_\mathrm{S}^{(\mathrm{LEQ})}\!\!\:(t)}$ does not satisfy the von Neumann equation~\eqref{eq:VonNeumannEquation}, thus does not appropriately reflect the time-evolution of the system. An even more dire issue is the fact that the functions $\beta(t)$ and $\mu(t)$, entering the ansatz~\eqref{eq:LEQ_Operator}, stand in a one-to-one correspondence to the physical averages $\braket{H\:\!}^{t}_{\text{phys}}$ and $\braket[\rbig]{Q_{\mathrm{S}}}^{\!\!\;t}_{\text{phys}}$ through the self-consistency relations~\eqref{eq:SelfConsistency_LEQ}. However, access to these quantities was granted only within our hypothetical setup, for which $\smash{\rho_\mathrm{S}^{(\mathrm{phys})}\!\!\:(t)}$ was assumed to be god-given. This is precisely the issue, as in any physically relevant scenario, these expectation values are the quantities we seek to determine rather than prescribed inputs that can be exploited from the outset. \\ 

\noindent 
Zubarev’s proposal for resolving both issues is conceptually straightforward: construct a second improved approximation to $\smash{\rho_\mathrm{S}^{(\mathrm{phys})}\!\!\:(t)}$ based on $\smash{\rho_\mathrm{S}^{(\mathrm{LEQ})}\!\!\:(t)}$, then require that the relevant expectation values coincide in both descriptions. The arising matching conditions then self-consistently determine the functions $\beta(t)$ and $\mu(t)$.

\subsubsection{Non-equilibrium statistical operator}

\noindent
A natural candidate for the NEQ operator is obtained by preparing the system in local equilibrium at some past reference time $t_\mathrm{ref}$ and subsequently evolving it unitarily as
\begin{align}
    \rho_\mathrm{S}^{(\text{NEQ-cand.})}\!\!\:(t) &= \exp\!\bigg[\!-\!\!\:\frac{iH}{\hbar}\:\!\big(t-t_\mathrm{ref}\big)\bigg] \:\!\rho_\mathrm{S}^{(\mathrm{LEQ})}\!\!\:(t_\mathrm{ref}) \:\! \exp\!\bigg[\frac{iH}{\hbar}\:\!\big(t-t_\mathrm{ref}\big)\bigg] \, ,
    \label{eq:NaiveNEQ operator}
\end{align}
where we fix the initial conditions $\beta(t_\mathrm{ref})$ and $\mu(t_\mathrm{ref})$ from the system's energy and charge content at the reference time $t_\mathrm{ref}$. Initializing the system with the correct macroscopic correlations should reproduce the expectation values of generic physical observables at late times $t\gg t_\mathrm{ref}$, as the missing microscopic correlations are generated dynamically. In particular, one expects excellent agreement when inferring the limit $t_\mathrm{ref}\to -\infty$, which would also eliminate any residual dependence on the arbitrary reference time $t_\mathrm{ref}$.\footnote{In fact, the ansatz~\eqref{eq:NaiveNEQ operator} in the respective limit $t_\mathrm{ref}\to -\infty$ is sometimes chosen as the definition of the NEQ operator, see e.g. reference~\cite{BecattiniArticle2019}.} However, the prescribed limit may be ill-defined, as there generally exists no stationary non-equilibrium state that the system could asymptotically approach in the distant past, and hence no well-defined limiting values for $\beta(t_\mathrm{ref})$ and $\mu(t_\mathrm{ref})$. To avoid these issues, the limit is regularized by averaging over the system's possible preparation times.\footnote{In the relevant literature, this principle is often dubbed ``weakening of initial correlations''~\cite{Bogoliubov:1946,RoepkeBook2013}.} Rather than selecting a single reference time $t_\mathrm{ref}$, one averages over a continuum of past reference times $t'$, setting
\begin{align}
    \rho_\mathrm{S}^{(\text{NEQ})}\!\!\:(t) &= \frac{1}{t-t_\mathrm{ref}}\mathlarger{\mathlarger{\int}}_{t_\mathrm{ref}}^t \mathrm{d}t'\: \exp\!\bigg[\!-\!\!\:\frac{iH}{\hbar}\:\!\big(t-t'\big)\bigg] \:\!\rho_\mathrm{S}^{(\mathrm{LEQ})}\!\!\:(t') \:\! \exp\!\bigg[\frac{iH}{\hbar}\:\!\big(t-t'\big)\bigg]\, .
    \label{eq:Definition_NEQ operator_simple}
\end{align}
Again, we refer to the starting point of the integral as $t_\mathrm{ref}$, which subsequently should be taken to $-\infty$ in order to avoid dependence on this arbitrary choice, while making sure long-lived correlations are properly erased. In the present context, the limit $t_\mathrm{ref}\to-\infty$ is also required to recover the von Neumann equation~\eqref{eq:VonNeumannEquation}, since for finite $t_\mathrm{ref}$ the NEQ operator~\eqref{eq:Definition_NEQ operator_simple} obeys
\begin{align}
	\frac{\mathrm{d}\rho_\mathrm{S}^{(\mathrm{NEQ})}(t)}{\mathrm{d}t} +\frac{i}{\hbar}\:\!\Big[H,\rho_\mathrm{S}^{(\mathrm{NEQ})}(t)\Big] &=  \frac{\rho_\mathrm{S}^{(\mathrm{LEQ})}(t)-\rho_\mathrm{S}^{(\mathrm{NEQ})}(t)}{t-t_\mathrm{ref}} \,\xrightarrow{\,\scalebox{0.8}{$t_\mathrm{ref}\!\!\:\to\!\!\: -\infty$}\,} \, 0 \, .
\end{align}
It proves convenient to reformulate the limit $t_\mathrm{ref}\to -\infty$ in terms of a regulator $\varepsilon\to 0^+$ through the relation 
\begin{align}
	\lim_{T\to\infty} \frac{1}{T}\int_{-T}^0\mathrm{d}t'\: f(t') = \lim_{\varepsilon\to 0^+} \varepsilon \int_{-\infty}^0 \mathrm{d}t'\:e^{\varepsilon t'}f(t')\, ,
\end{align}
which allows us to represent the sought-after NEQ operator in its final form 
\begin{align}
    \rho_\mathrm{S}^{(\text{NEQ})}\!\!\:(t) &= \lim_{\varepsilon \to 0^+} \!\left\{\varepsilon \mathlarger{\mathlarger{\int}}_{-\infty}^t \mathrm{d}t'\: e^{\varepsilon (t'-t)} \exp\!\bigg[\!-\!\!\:\frac{iH}{\hbar}\:\!\big(t-t'\big)\bigg] \:\!\rho_\mathrm{S}^{(\mathrm{LEQ})}\!\!\:(t') \:\! \exp\!\bigg[\frac{iH}{\hbar}\:\!\big(t-t'\big)\bigg]\right\} \, .
    \label{eq:Definition_NEQ operator_full}
\end{align}
Note that the integral will be of order $\varepsilon^{-1}$, as it receives $\mathcal{O}(1)$ contributions from timescales $t-t'\lesssim \varepsilon^{-1}$. The small positive $\varepsilon$ acts as a regulator, suppressing contributions from large time separations $t-t'$, thereby allowing us to avoid the aforementioned issues. By construction of the NEQ operator~\eqref{eq:Definition_NEQ operator_full}, it is natural to require that it also reproduces the relevant physical averages $\braket{H\:\!}^{t}_{\mathrm{phys}}$ and $\braket[\rbig]{Q_\mathrm{S}}^{\!\!\;t}_{\mathrm{phys}}$. This requirement gives rise to the two self-consistency relations
\begin{subequations}
\begin{align}
	\braket{H\:\!}_\mathrm{LEQ}^t=\mathrm{tr}\Big[\rho_{\mathrm{S}}^{\,(\mathrm{LEQ})}(t)\:\!H\Big]&\overset{!}{=} \braket{H\:\!}^{t}_{\mathrm{phys}} \overset{!}{=}\mathrm{tr}\Big[\rho_{\mathrm{S}}^{\,(\mathrm{NEQ})}(t)\:\!H\Big]=\braket{H\:\!}_\mathrm{NEQ}^t\, , \label{eq:HConstraint_Schrödinger}\\[0.1cm] 
	\braket[\rbig]{Q_\mathrm{S}}_\mathrm{LEQ}^t=\mathrm{tr}\Big[\rho_{\mathrm{S}}^{\,(\mathrm{LEQ})}(t)\,Q_{\mathrm{S}}\Big]&\overset{!}{=} \braket[\rbig]{Q_\mathrm{S}}^{\!\!\;t}_{\mathrm{phys}} \overset{!}{=} \mathrm{tr}\Big[\rho_{\mathrm{S}}^{\,(\mathrm{NEQ})}(t)\,Q_{\mathrm{S}}\Big]=\braket[\rbig]{Q_\mathrm{S}}_\mathrm{NEQ}^t\, , \label{eq:NConstraint_Schrödinger}
\end{align} 
\label{eq:Constraints_Schrödinger}%
\end{subequations}
which must hold at all times and thereby self-consistently determine the a priori unknown functions $\beta(t)$ and $\mu(t)$ entering the LEQ operator.\footnote{The resulting solutions for $\beta(t)$ and $\mu(t)$ are unique up to time translations, reflecting the formal limit in which the reference time is sent to $-\infty$. Note that, for our purposes, no initial condition needs to be specified, since we are only interested in the late-time behavior of $\beta(t)$ and $\mu(t)$, which are related to expectation values in our theory. If one instead wishes to access the full time evolution for the NEQ-operator, $\beta(t)$ and $\mu(t)$ can be fixed at a chosen reference time, thereby selecting a unique solution from the one-parameter family allowed by the constraint equations~\eqref{eq:Constraints_Schrödinger}.} While seemingly innocuous, imposing the self-consistency conditions~\eqref{eq:Constraints_Schrödinger} is a crucial and generally nontrivial step, as it eliminates the need for explicit knowledge of the unknown physical averages $\braket{H\:\!}^{t}_{\text{phys}}$ and $\braket[\rbig]{Q_{\mathrm{S}}}^{\!\!\;t}_{\text{phys}}$. The assumptions under which this identification is justified are, however, far from obvious, which we will briefly discuss in section~\ref{sec:ZubarevDiscussion}.\\

\noindent 
The key feature of Zubarev's NSO construction therefore lies in the fact that two functionally distinct approximations $\smash{\rho_\mathrm{S}^{(\mathrm{LEQ})}\!\!\:(t)}$ and $\smash{\rho_\mathrm{S}^{(\mathrm{NEQ})}\!\!\:(t)}$ to the physical density matrix $\smash{\rho_\mathrm{S}^{(\mathrm{phys})}\!\!\:(t)}$ depend on the same set of functions $\beta(t)$ and $\mu(t)$, which are determined by requiring both approximations to reproduce the relevant averages. Once $\beta(t)$ and $\mu(t)$ have been determined by the self-consistency relations~\eqref{eq:Constraints_Schrödinger}, all sought-after correlators should be evaluated using the NEQ operator, which is purported to provide the optimal approximation given the chosen set of relevant operators. \\

\noindent 
This concludes our extensive interlude on the construction of Zubarev's NSO operator $\smash{\rho_\mathrm{S}^{(\mathrm{NEQ})}\!\!\:(t)}$. Note that the so-defined NEQ operator checks all demands initially put forth in section~\ref{sec:Construction_NSO}, as the given density matrix~\eqref{eq:Definition_NEQ operator_full} is a causal functional of the relevant averages $\braket{H\:\!}^{t}$ and $\braket[\rbig]{Q_\mathrm{S}}^{\!\!\;t}$, formally satisfies the von Neumann equation 
\begin{align}
	\frac{\mathrm{d}\rho_\mathrm{S}^{(\mathrm{NEQ})}(t)}{\mathrm{d}t} +\frac{i}{\hbar}\:\!\Big[H,\rho_\mathrm{S}^{(\mathrm{NEQ})}(t)\Big] &=\varepsilon \Big[\rho_{\mathrm{S}}^{\,(\mathrm{LEQ})}(t)- \rho_{\mathrm{S}}^{\,(\mathrm{NEQ})}(t)\Big] \,\xrightarrow{\,\scalebox{0.8}{$\varepsilon\!\!\:\to\!\!\: 0^+$}\,} \, 0 \, ,
    \label{eq:Modified_von_Neumann_Equation}
\end{align}
and retains its functional form under the full time-evolution of the system. Moreover, through its relation to the LEQ operator, $\beta(t)$ and $\mu(t)$ retain their useful interpretation as the instantaneous temperature and chemical potential of the system. Note that the infinitesimal source term on the right-hand side of the modified von Neumann equation~\eqref{eq:Modified_von_Neumann_Equation} breaks the time-reversal symmetry of the usual von Neumann equation~\eqref{eq:VonNeumannEquation}, thereby formally selecting the retarded solution appropriate for the non-equilibrium system at hand. For a more in-depth discussion, we refer the reader to the abundant literature on the subject, see e.g. the exhaustive textbook treatments~\cite{ZubarevBook1996,ZubarevBook1997,RoepkeBook2013}.

\subsection{Open questions surrounding Zubarev's method}
\label{sec:ZubarevDiscussion}

Let us conclude our introduction to the NSO method by outlining the limitations of the approach as perceived by the authors. To this end, note that the construction of the NEQ operator $\rho_\mathrm{S}^{(\mathrm{NEQ})}\!\!\:(t)$ is, in fact, not uniquely tied to the particular choice of the associated LEQ operator $\rho_\mathrm{S}^{(\mathrm{LEQ})}\!\!\:(t)$. Owing to the averaging procedure underlying the formalism, one could equally well define the LEQ operator as
\begin{align}
    \rho_\mathrm{S}^{(\mathrm{LEQ,\,altered})}\!\!\:(t)=\exp\!\!\:\bigg\{\!-\Phi(t)-\!\!\:\beta(t)\Big[H_\mathrm{S}^{(\mathrm{cons.})}-\mu(t)Q_{\mathrm{S}}\Big]\!\!\:\bigg\} = \rho_\mathrm{S}^{(\mathrm{LEQ})}\!\!\:(t)\Big[1+\mathcal{O}(\lambda)\Big]\, ,
    \label{eq:LEQ_Operator_altered}
\end{align}
which can be argued to satisfy the identical requirements as the original LEQ operator. Similar alternative definitions can also be formulated using the interaction picture~\eqref{eq:Interaction_Picture}, yielding further viable candidates for the LEQ operator, all granting slightly different versions of the resulting NEQ operator. The crucial observation is that these different constructions agree only up to corrections of order $\lambda$. Consequently, while the NSO formalism provides a self-consistent approximation scheme for the full, physical non-equilibrium operator $\smash{\rho_\mathrm{S}^{(\mathrm{phys})}\!\!\:(t)}$, it inherits an ambiguity at next-to-leading order (NLO), since distinct but seemingly equally justified choices of the LEQ operator generally lead to different NLO contributions. Since our analysis is restricted to leading-order effects, this ambiguity does not affect the subsequent results. Nevertheless, it should be kept in mind when interpreting the predictive scope of the method beyond leading order.\\

\noindent 
A more fundamental question concerns the role of the self-consistency relations~\eqref{eq:Constraints_Schrödinger} in selecting the uniform relaxation regime that the NSO method aims to describe. Starting from an arbitrary physical density matrix $\smash{\rho_\mathrm{S}^{(\mathrm{phys})}\!\!\:(t)}$, one can always construct the associated LEQ operator~\eqref{eq:LEQ_Operator} that reproduces its relevant expectation values and, from this, the corresponding NEQ operator~\eqref{eq:Definition_NEQ operator_full}. In general, however, there is no reason why the relevant averages computed with the NEQ operator should agree with those of the LEQ operator. Imposing $\braket[\rbig]{O_\mathrm{S}}_{\mathrm{NEQ}}^{\!\!\: t}=\braket[\rbig]{O_\mathrm{S}}_{\mathrm{LEQ}}^{\!\!\: t}$ for all relevant operators $O_\mathrm{S}$ therefore places a nontrivial restriction on the dynamics. The apparent philosophy of the NSO method is that, after microscopic transients and detailed memory of the initial state have decayed, the system enters a universal relaxation regime, and that this regime is precisely characterized by the above self-consistency conditions~\eqref{eq:Constraints_Schrödinger}. The self-consistency relations thus do more than determine the time-dependent Lagrange multipliers, as they effectively select this uniform relaxation regime from the possible microscopic evolutions. However, it is unclear to us why equality of the relevant LEQ and NEQ averages provides the sought-after physical criterion for this selection. The usual arguments based on scale separation, weakening of initial correlations, and restriction to slow variables do not appear to establish this identification. We therefore regard the self-consistency relations as an additional closure assumption whose precise physical justification and domain of validity are not transparent from the standard construction alone.

\section{Relaxation rate from Zubarev's NSO approach}
\label{sec:Zubarev_Calculation}

\noindent 
With the construction of the NEQ operator in place, we can proceed to study the desired time-evolution of the average $\braket[\rbig]{Q_\mathrm{S}}^{\!\!\;t}$. For convenience, we switch to the Heisenberg picture, which avoids the need to keep track of multiple time-evolution operators and keeps the following calculation concise. The starting point of all subsequent considerations are the previously introduced statistical operators
\begin{subequations}
\begin{align}
    \rho_{\mathrm{H}}^{\,(\mathrm{LEQ})}(t)&=\frac{1}{Z^{(\mathrm{LEQ})}(t)}\,\exp\!\!\:\bigg\{\!-\beta(t)\Big[H -\mu(t)\:\!Q_{\mathrm{H}}(t)\Big]\!\!\:\bigg\} \, , \\[0.25cm] 
    \rho_{\mathrm{H}}^{\,(\mathrm{NEQ})}(t)
    &= \lim_{\varepsilon \to 0^+} \cBigg\{\varepsilon \mathlarger{\mathlarger{\int}}_{-\infty}^t \mathrm{d}t'\: e^{\varepsilon (t'-t)} \:\!\rho_\mathrm{H}^{(\mathrm{LEQ})}\!\!\:(t') \cBigg\} \notag \\ 
    &= \rho_{\mathrm{H}}^{\,(\mathrm{LEQ})}(t)-\lim_{\varepsilon\to 0^+}\mathlarger{\mathlarger{\int}}_{-\infty}^t\mathrm{d}t'\: e^{\varepsilon (t'-t)} \,\frac{\mathrm{d} \rho_{\mathrm{H}}^{\,(\mathrm{LEQ})}(t')}{\mathrm{d}t'}\, , \label{eq:NEQ operator_Heisenberg}
\end{align}%
\end{subequations}
with the functions $\beta(t)$ and $\mu(t)$ fixed by the two self-consistency relations 
\begin{align}
	\braket{H\:\!}_\mathrm{LEQ}^t &\overset{!}{=} \braket{H\:\!}_\mathrm{NEQ}^t\, , & \mathrm{and} && \braket[\rbig]{Q_\mathrm{H}(t)}_\mathrm{LEQ}^t &\overset{!}{=} \braket[\rbig]{Q_\mathrm{H}(t)}_\mathrm{NEQ}^t \, . 
    \label{eq:Constraints_Heisenberg}
\end{align}
Note that in the last line of equation~\eqref{eq:NEQ operator_Heisenberg}, we have integrated by parts to resolve the $\varepsilon \times \varepsilon^{-1}$ structure, thereby expressing the result entirely in terms of quantities of order unity. Note that when prematurely taking the limit $\varepsilon\to 0^+$ in the integral expression~\eqref{eq:NEQ operator_Heisenberg}, one ends up with the na\"{\i}ve NEQ operator~\eqref{eq:NaiveNEQ operator} arising from the possibly ill-defined limit $t_\mathrm{ref}\to -\infty$. Utilizing the modified von Neumann equation~\eqref{eq:Modified_von_Neumann_Equation} satisfied by the NEQ operator
\begin{equation}
    \frac{\mathrm{d} \rho_{\mathrm{H}}^{\,(\mathrm{NEQ})}(t)}{\mathrm{d}t} = \varepsilon \Big[\rho_{\mathrm{H}}^{\,(\mathrm{LEQ})}(t)- \rho_{\mathrm{H}}^{\,(\mathrm{NEQ})}(t)\Big] \,\xrightarrow{\,\scalebox{0.8}{$\varepsilon\!\!\:\to\!\!\: 0^+$}\,} \, 0\, ,
\end{equation}
it becomes evident that, due to the Hamiltonian $H$ being time-independent, the system’s total energy $E_\mathrm{tot}$ is conserved, as follows from
\begin{align}
    \frac{\mathrm{d}}{\mathrm{d}t}\, \braket{H\:\!}_\mathrm{NEQ}^t &= \mathrm{tr}\Bigg\{\frac{\mathrm{d}\:\! \rho_{\mathrm{H}}^{\,(\mathrm{NEQ})}(t)}{\mathrm{d}t}\:\!H\Bigg\} \notag \\ 
    &= \varepsilon \: \mathrm{tr}\bigg\{\!\!\:\Big[\rho_{\mathrm{H}}^{\,(\mathrm{LEQ})}(t)-\rho_{\mathrm{H}}^{\,(\mathrm{NEQ})}(t)\Big]H\bigg\} = \varepsilon \Big[\braket{H\:\!}_\mathrm{LEQ}^t-\braket{H\:\!}_\mathrm{NEQ}^t\Big] = 0\, ,
\end{align}
with the last equality following from the self-consistency relation~\eqref{eq:Constraints_Heisenberg}. For late times, the system slowly approaches thermal equilibrium, which is characterized by
\begin{align}
    \!\!\beta(t)\xrightarrow{\scalebox{0.8}{$t \!\!\:\to \!\!\:\infty$}} \beta_\mathrm{EQ}\, , \;\;\!\!\: \mu(t)\xrightarrow{\scalebox{0.8}{$t \!\!\:\to\!\!\: \infty$}} 0 \quad\!\!\: \Longleftrightarrow \quad\!\!\!\: \rho_{\mathrm{H}}^{(\mathrm{LEQ})}(t)\xrightarrow{\scalebox{0.8}{$t \!\!\:\to\!\!\: \infty$}} \rho^{(\mathrm{EQ})} = \frac{1}{Z^{(\mathrm{EQ})}}\:\!\exp\!\!\:\Big(\!\!\!\;-\!\!\:\beta_\mathrm{EQ}H\Big)\, .
\end{align}
The equilibrium temperature $\beta_\mathrm{EQ}$, defined in the previous equation, is therefore related to the total energy $E_\mathrm{tot}$ of the system via the implicit relation
\begin{align}
    E_\mathrm{tot} = \braket{H\:\!}_\mathrm{NEQ}^t = \braket{H\:\!}_\mathrm{LEQ}^t = \braket{H\:\!}_\mathrm{EQ} = \mathrm{tr}\Big[\rho^{(\mathrm{EQ})}H\Big] \, .
    \label{eq:Energy_Conservation_Constraint}
\end{align}

\subsection{Linearized near-equilibrium dynamics}

To study the system’s relaxation behavior, we restrict ourselves to small deviations from thermal equilibrium, assuming $\big\lvert\beta(t)-\beta_\mathrm{EQ}\big\rvert\ll \beta_\mathrm{EQ}$ and $\big\lvert\beta_\mathrm{EQ}\:\!\mu(t)\big\rvert\ll 1$. This corresponds to working in the late-time limit, where the system has relaxed sufficiently close to thermal equilibrium so that the deviation is small. This allows all observables to be expanded in the small parameters $\Delta \beta(t)\equiv \beta(t)-\beta_\mathrm{EQ}$ and $\mu(t)$, where we will only keep the leading-order terms.\footnote{Rather than working with the small dimensionless parameters $\Delta \beta(t)/\beta_\mathrm{EQ}$ and $\beta_\mathrm{EQ}\:\!\mu(t)$, we will instead expand directly in $\Delta \beta(t)$ and $\mu(t)$. The perturbative series may, of course, be rearranged so as to be expressed in terms of the specified dimensionless parameters. The same convention will be understood in the error estimates below.} Expanding the LEQ operator around its equilibrium value yields a Dyson-series-type expansion of the exponential operator, which is discussed in appendix~\ref{sec:Dyson_Expansion}. A straightforward computation using the general expression~\eqref{eq:Combined_Expansion_Dyson} grants the representation 
\begin{align}
    \rho_{\mathrm{H}}^{\,(\mathrm{LEQ})}(t)= \rho^{(\mathrm{EQ})} \Bigg\{1&-\mathlarger{\int}_0^1  \mathrm{d}s\; e^{\scalebox{0.75}{$s\beta_\mathrm{EQ}H$}}\!\!\;\cBig[\Delta\beta(t) H-\beta_\mathrm{EQ}\:\!\mu(t) Q_{\mathrm{H}}(t)\cBig] e^{\scalebox{0.75}{$-s\beta_\mathrm{EQ}H$}} \notag \\[-0.05cm]
    &+\braket[\cBig]{\Delta\beta(t) H-\beta_\mathrm{EQ}\:\!\mu(t) Q_{\mathrm{H}}(t)}_{\!\!\!\:\mathrm{EQ}} +\mathcal{O}\Big[\mu(t)^2,\Delta\beta(t)^2,\mu(t)\Delta\beta(t)\Big]\!\!\:\Bigg\} \, .
    \label{eq:Expanded_LEQ_Operator}
\end{align}
One can simplify this expression further by noting that the equilibrium average $\smash{\braket[\rbig]{Q_\mathrm{H}(t)}_\mathrm{EQ}}$ vanishes, which one can formally show using $\mathcal{CPT}$-arguments, as illustrated in appendix~\ref{sec:CPT_Arguments}. Inserting the above expansion into the energy conservation constraint~\eqref{eq:Energy_Conservation_Constraint} and subtracting the leading-order contribution common to both sides yields the relation\footnote{Beware that, when rewriting the trace as an equilibrium average, in light of equation~\eqref{eq:Expanded_LEQ_Operator}, careful attention must be paid to operator ordering. Although this subtlety is immaterial for the present case of an $H$-insertion, it becomes essential for an insertion of $Q_\mathrm{H}(t)$, where the ordering does affect the result.}
\begin{align}
    0 &= \braket{H\:\!}_\mathrm{EQ} - \braket{H\:\!}_\mathrm{LEQ}^t \notag \\[0.1cm] 
    &= \mathlarger{\int}_0^1  \mathrm{d}s\: \braket[\rbigg]{e^{\scalebox{0.75}{$s\beta_\mathrm{EQ}H$}} \!\!\:\cBig[\Delta\beta(t) H-\beta_\mathrm{EQ}\:\!\mu(t) Q_{\mathrm{H}}(t)\cBig] e^{\scalebox{0.75}{$-s\beta_\mathrm{EQ}H$}}H}_{\!\!\mathrm{EQ}} \notag \\[-0.1cm] 
    &\qquad\qquad\qquad\quad\!\!\! - \Delta\beta(t)\cbig[\braket{H\:\!}_{\mathrm{EQ}}\cbig]^{\!\!\:2} \,+ \: \mathcal{O}\Big[\mu(t)^2,\Delta\beta(t)^2,\mu(t)\Delta\beta(t)\Big] \notag \\[0.15cm] 
    &= \Delta\beta(t) \:\!\bigg\{\braket[\cbig]{H^{2}}_{\mathrm{EQ}} -\cbig[\braket{H\:\!}_{\mathrm{EQ}}\cbig]^{\!\!\:2}\cBig\} \,+ \:\mathcal{O}\Big[\mu(t)^2,\Delta\beta(t)^2,\mu(t)\Delta\beta(t)\Big] \, .
    \label{eq:Variance_H}
\end{align}
In going from the second to the last line, we again used the cyclicity of the trace to show that the integrand is independent of $s$ by commuting the $s$-dependent exponential pieces through $H$. The remaining contribution $\braket[\rbig]{HQ_\mathrm{H}(t)}_\mathrm{EQ}$ proportional to $\mu(t)$ vanishes similarly to $\braket[\rbig]{Q_\mathrm{H}(t)}_\mathrm{EQ}$, see appendix~\ref{sec:CPT_Arguments}. Since the variance of the Hamiltonian $H$ computed in equilibrium generally does not vanish, the only valid conclusion from equation~\eqref{eq:Variance_H} is that $\Delta\beta(t)$ is of order $\mu(t)^2$, i.e. a change in temperature is energetically much more costly for the system than a change in the chemical potential. As a consequence, when working to leading order in $\mu(t)$, we can neglect deviations of the temperature from its equilibrium value. This allows us to greatly simplify our previous expansion~\eqref{eq:Expanded_LEQ_Operator} of the LEQ operator, which now reads 
\begin{align}
    \rho_{\mathrm{H}}^{\,(\mathrm{LEQ})}(t)= \rho^{(\mathrm{EQ})} \Bigg\{1+\beta_\mathrm{EQ}\:\!\mu(t)\!\!\:\mathlarger{\int}_0^1 \mathrm{d}s\: e^{\scalebox{0.75}{$s\beta_\mathrm{EQ}H$}} Q_{\mathrm{H}}(t) \:\! e^{\scalebox{0.75}{$-s\beta_\mathrm{EQ}H$}} \,+ \: \mathcal{O}\cbig[\mu(t)^2\cbig]\!\!\:\Bigg\} \, .
    \label{eq:Expanded_LEQ_Operator_Simplified}
\end{align}
Computing the LEQ-average $\braket[\rbig]{Q_\mathrm{H}(t)}_\mathrm{LEQ}^t$ then grants the important relation 
\begin{align}
    \braket[\cbig]{Q_\mathrm{H}(t)\!\!\;}_\mathrm{LEQ}^t = \beta_\mathrm{EQ}\:\! \mu(t) \mathlarger{\int}_0^1 \mathrm{d}s\: \braket[\cBig]{e^{\scalebox{0.75}{$s\beta_\mathrm{EQ}H$}} Q_{\mathrm{H}}(t) \:\!e^{\scalebox{0.75}{$-s\beta_\mathrm{EQ}H$}} Q_{\mathrm{H}}(t)}_{\!\!\mathrm{EQ}} \!+\mathcal{O}\cbig[\mu(t)^2\cbig]\, , \;\,
    \label{eq:QLEQ_Propto_mu}
\end{align}
where, once more, we dropped the vanishing contribution $\braket[\rbig]{Q_\mathrm{H}(t)}_\mathrm{EQ}$. We note that the correlator in the integrand is independent of $t$, as can be seen by switching to the Schrödinger picture, where all time-evolution operators cancel due to the cyclicity of the trace and the fact that the Hamiltonian is time-independent, therefore commuting with itself. Phrased differently, the equilibrium ensemble is time-translation invariant, such that any two-point function $\braket[\rbig]{A_\mathrm{H}(t_1)B_\mathrm{H}(t_2)}_{\mathrm{EQ}}$ only depends on the time difference $t_1-t_2$. Using the self-consistency relation~\eqref{eq:Constraints_Heisenberg} allows us to replace the LEQ average with its NEQ counterpart, thereby concluding
\begin{align}
    \;\,\braket[\cbig]{Q_\mathrm{H}(t)\!\!\;}_\mathrm{NEQ}^t = \beta_\mathrm{EQ}\:\! \mu(t) \mathlarger{\int}_0^1 \mathrm{d}s\: \braket[\cBig]{e^{\scalebox{0.75}{$s\beta_\mathrm{EQ}H$}} Q_{\mathrm{H}}(0) \:\! e^{\scalebox{0.75}{$-s\beta_\mathrm{EQ}H$}} Q_{\mathrm{H}}(0)\!\!\;}_{\!\!\mathrm{EQ}} +\mathcal{O}\cbig[\mu(t)^2\cbig]\, ,
    \label{eq:QNEQ_Propto_mu}
\end{align}
where the only time-dependence resides inside $\mu(t)$. Thus, to leading order, the total charge is proportional to the instantaneous chemical potential $\mu(t)$. For future convenience, we rewrite this relation and its time derivative as
\begin{subequations}
\begin{align}
    \mu(t) &= \braket[\cbig]{Q_\mathrm{H}(t)}_\mathrm{NEQ}^t \,\Bigg\{\beta_\mathrm{EQ}\mathlarger{\int}_0^1 \mathrm{d}s\: \braket[\cBig]{e^{\scalebox{0.75}{$s\beta_\mathrm{EQ}H$}} Q_{\mathrm{H}}(0) \:\!e^{\scalebox{0.75}{$-s\beta_\mathrm{EQ}H$}} Q_{\mathrm{H}}(0)}_{\!\!\mathrm{EQ}} \Bigg\}^{\!\!\!\:-1}\!\!\:\cBig\{1\!\!\:+\!\!\:\mathcal{O}\big[\mu(t)\big]\!\!\:\cBig\}\, , 
    \label{eq:NProptoMu} \\
    \dot{\mu}(t) &= \frac{\mathrm{d}\:\!\braket[\rbig]{Q_\mathrm{H}(t)}_\mathrm{NEQ}^t}{\mathrm{d}t} \,\Bigg\{\beta_\mathrm{EQ}\mathlarger{\int}_0^1 \mathrm{d}s\: \braket[\cBig]{e^{\scalebox{0.75}{$s\beta_\mathrm{EQ}H$}}Q_{\mathrm{H}}(0) \:\!e^{\scalebox{0.75}{$-s\beta_\mathrm{EQ}H$}} Q_{\mathrm{H}}(0)}_{\!\!\mathrm{EQ}} \Bigg\}^{\!\!\!\:-1}\!\!\:\cBig\{1\!\!\:+\!\!\:\mathcal{O}\big[\mu(t)\big]\!\!\:\cBig\} \, . 
    \label{eq:DerivativeNProptoMudot}
\end{align}%
\end{subequations}
In what follows, we will take the time derivative of the non-equilibrium expectation value $\smash{\braket[\rbig]{Q_{\mathrm{H}}(t)}_{\mathrm{NEQ}}^t}$, finding that, to leading order in the deviation from equilibrium parametrized by $\mu(t)$, the derivative is proportional to the expectation value itself. Comparison with equation~\eqref{eq:relaxation_equation} will then allow us to identify the proportionality constant as the relaxation rate $\Gamma_\mathrm{relax}$. Starting from the equality
\begin{align}
    \frac{\mathrm{d}}{\mathrm{d}t}\, \braket[\cbig]{Q_\mathrm{H}(t)}_\mathrm{NEQ}^t &= \mathrm{tr}\Bigg\{\frac{\mathrm{d}\rho_{\mathrm{H}}^{\,(\mathrm{NEQ})}(t)}{\mathrm{d}t}\:\!Q_\mathrm{H}(t)\Bigg\} +  \mathrm{tr}\Bigg\{\rho_{\mathrm{H}}^{\,(\mathrm{NEQ})}(t)\:\!\frac{\mathrm{d}Q_\mathrm{H}(t)}{\mathrm{d}t}\Bigg\} \notag \\ 
    &= \!\underbrace{\varepsilon \: \mathrm{tr}\bigg\{\!\!\:\Big[\rho_{\mathrm{H}}^{\,(\mathrm{LEQ})}(t)-\rho_{\mathrm{H}}^{\,(\mathrm{NEQ})}(t)\Big]Q_\mathrm{H}(t)\bigg\}}_{\displaystyle{=0\; [\text{due to constraint equation~\eqref{eq:Constraints_Heisenberg}}]}} \!+  \:\braket[\rbigg]{\frac{\mathrm{d}Q_\mathrm{H}(t)}{\mathrm{d}t}}_{\!\!\!\:\mathrm{NEQ}}^{\!\!t} \, ,
    \label{eq:DerivativeN_FirstStep}
\end{align} 
we can insert the NEQ operator~\eqref{eq:NEQ operator_Heisenberg} to arrive at
\begin{align}
    \frac{\mathrm{d}}{\mathrm{d}t}\, \braket[\cbig]{Q_\mathrm{H}(t)}_\mathrm{NEQ}^t &= \braket[\rbigg]{\frac{\mathrm{d}Q_\mathrm{H}(t)}{\mathrm{d}t}}_{\!\!\!\:\mathrm{LEQ}}^{\!\!t}  -\lim_{\varepsilon\to 0^+}\mathlarger{\mathlarger{\int}}_{-\infty}^t\mathrm{d}t'\: e^{\varepsilon (t'-t)} \,\mathrm{tr}\cbigg\{\frac{\mathrm{d} \rho_{\mathrm{H}}^{\,(\mathrm{LEQ})}(t')}{\mathrm{d}t'} \frac{\mathrm{d}Q_\mathrm{H}(t)}{\mathrm{d}t}\cbigg\}\, .
    \label{eq:DerivativeN_SecondStep}
\end{align} 
At this point, we may employ our previously obtained expansion~\eqref{eq:Expanded_LEQ_Operator_Simplified} for the LEQ operator $\smash{\rho_{\mathrm{H}}^{\,(\mathrm{LEQ})}(t)}$. A straightforward insertion grants 
\begin{align}
    \braket[\rbigg]{\frac{\mathrm{d}Q_\mathrm{H}(t)}{\mathrm{d}t}}_{\!\!\!\:\mathrm{LEQ}}^{\!\!t} = \braket[\rbigg]{\frac{\mathrm{d}Q_\mathrm{H}(t)}{\mathrm{d}t}}_{\!\!\!\:\mathrm{EQ}} + \:\beta_\mathrm{EQ}\:\! \mu(t) \!\!\:\mathlarger{\mathlarger{\int}}_0^1 \!\!\:\mathrm{d}s\: \braket[\rbigg]{e^{\scalebox{0.75}{$s\beta_\mathrm{EQ}H$}} Q_{\mathrm{H}}(t) \:\! e^{\scalebox{0.75}{$-s\beta_\mathrm{EQ}H$}} \frac{\mathrm{d}Q_{\mathrm{H}}(t)}{\mathrm{d}t}}_{\!\!\mathrm{EQ}} +\mathcal{O}\cbig[\mu(t)^2\cbig] \, , \label{eq:LEQ_Average_DerivativeN_Expanded}
\end{align}
where both matrix elements on the right-hand side of equation~\eqref{eq:LEQ_Average_DerivativeN_Expanded} vanish identically. Hence, the LEQ expectation value on the left-hand side vanishes to the desired order, and therefore the first term on the right-hand side of equation~\eqref{eq:DerivativeN_SecondStep} does not contribute. Here, we have again used that $H$ is even under a joint $\mathcal{CPT}$ transformation, whereas the charge operator $Q_\mathrm{H}(t)$ is odd. These transformation properties suffice to establish the vanishing of the given equilibrium correlators, see appendix~\ref{sec:CPT_Arguments} for details. Differentiating relation~\eqref{eq:Expanded_LEQ_Operator_Simplified}, the remaining contribution in equation~\eqref{eq:DerivativeN_SecondStep} takes the form
\begin{align}
    \!\!\frac{\mathrm{d}}{\mathrm{d}t}\, \braket[\cbig]{Q_\mathrm{H}(t)\!\!\;}_\mathrm{NEQ}^t &= -\lim_{\varepsilon\to 0^+} \beta_\mathrm{EQ} \mathlarger{\mathlarger{\int}}_{-\infty}^t\mathrm{d}t'\: e^{\varepsilon (t'-t)}\mathlarger{\mathlarger{\int}}_{0}^1\mathrm{d}s \label{eq:ConstraintMassaged2}  \\[0.1cm]
    &\qquad\qquad\qquad \times \braket[\cbigg]{e^{\scalebox{0.75}{$s\beta_\mathrm{EQ}H$}} \frac{\mathrm{d}}{\mathrm{d}t'}\Big[\mu(t')Q_{\mathrm{H}}(t')\Big] e^{\scalebox{0.75}{$-s\beta_\mathrm{EQ}H$}}\frac{\mathrm{d}Q_{\mathrm{H}}(t)}{\mathrm{d}t}}_{\!\!\mathrm{EQ}} \cBig\{1+\mathcal{O}\big[\mu(t)\big]\!\!\:\cBig\} \,. \notag 
\end{align}

\subsection{Applying the Markovian approximation}

Thus far, the separation of timescales has been used only implicitly in the construction of the NEQ operator, see section~\ref{sec:2_Zubarev_Method}. To make further progress, we now explicitly invoke the scale hierarchy between the internal (microscopic) equilibration time $\tau_\mathrm{micro}$ and the (macroscopic) relaxation time $\tau_\mathrm{macro}$. As we shall see, the rapid loss of memory on microscopic timescales renders the effective evolution on macroscopic timescales Markovian. Since the hierarchy of timescales is controlled by the previously introduced parameter $\lambda$, see the defining relation~\eqref{eq:Hamiltonian_Split}, we can formally expand the right-hand side of equation~\eqref{eq:ConstraintMassaged2} in powers of $\lambda \ll 1$. To this end, we assume that (connected) equilibrium two-time correlation functions involving a fast operator decay exponentially on the microscopic timescale $\tau_\mathrm{micro}$ and are therefore strongly suppressed for $\lvert t-t'\rvert \gg \tau_\mathrm{micro}$, as stated in equation~\eqref{eq:Behavior_TwoPointFunction}. Consequently, correlations with the distant past are rapidly erased compared with the macroscopic evolution of the system, providing the basis for the emergence of Markovian dynamics. In particular, we have\footnote{Note that $\mathrm{d}Q_\mathrm{H}(t)/\mathrm{d}t$ is a fast variable as it does not commute with $H^{(\mathrm{cons.})}$.} 
\begin{align}
    \!\!\mathcal{C}_1\big(s;t,t'\big)&\equiv \braket[\rbigg]{e^{\scalebox{0.75}{$s\beta_\mathrm{EQ}H$}} Q_{\mathrm{H}}(t') \:\! e^{\scalebox{0.75}{$-s\beta_\mathrm{EQ}H$}} \frac{\mathrm{d}Q_{\mathrm{H}}(t)}{\mathrm{d}t}}_{\!\!\mathrm{EQ}} \xrightarrow{\,\lvert t-t'\rvert \:\gg \:\tau_\mathrm{micro}\,} \: \mathrm{const.}\times\exp \!\bigg(\!\!\!\:-\frac{|t-t'|}{\tau_{\mathrm{micro}}} \bigg) \, . \label{eq:Definition_C1}
\end{align}
Differentiating in the case $t>t'$ then yields the important relation 
\begin{align}
    \frac{\mathrm{d}\:\! \mathcal{C}_1\big(s;t,t'\big)}{\mathrm{d}t'} &= \braket[\rbigg]{e^{\scalebox{0.75}{$s\beta_\mathrm{EQ}H$}} \frac{\mathrm{d}Q_{\mathrm{H}}(t')}{\mathrm{d}t'} \:\! e^{\scalebox{0.75}{$-s\beta_\mathrm{EQ}H$}}\frac{\mathrm{d}Q_{\mathrm{H}}(t)}{\mathrm{d}t}}_{\!\!\mathrm{EQ}} \xrightarrow{\,\lvert t-t'\rvert \:\gg \:\tau_\mathrm{micro}\,} \,\frac{\mathcal{C}_1(s;t,t')}{\tau_\mathrm{micro}}\, ,
\end{align}
showing that $\mathcal{C}_1$ and $\mathrm{d}\mathcal{C}_1/\mathrm{d}t'$, for large time separation, are related by a factor $\tau_\mathrm{micro}$. While this hierarchy is expected to hold in the stated limiting case $\lvert t-t'\rvert \gg \tau_\mathrm{micro}$, the ratio of the two terms is assumed to remain of the same parametric order throughout the pertinent domain, motivating the definition
\begin{align}
    \mathcal{C}_2\big(s;t,t'\big)&\equiv \tau_{\mathrm{micro}}\braket[\rbigg]{e^{\scalebox{0.75}{$s\beta_\mathrm{EQ}H$}} \frac{\mathrm{d}Q_{\mathrm{H}}(t')}{\mathrm{d}t'} \:\! e^{\scalebox{0.75}{$-s\beta_\mathrm{EQ}H$}} \frac{\mathrm{d}Q_{\mathrm{H}}(t)}{\mathrm{d}t}}_{\!\!\mathrm{EQ}} 
    \, ,
    \label{eq:Definition_C2}
\end{align}
with $\mathcal{C}_1/\mathcal{C}_2=\mathcal{O}(1)$ within the range under consideration.\footnote{Although $\mathcal{C}_1$ contains only one explicit factor of $\mathrm{d}Q_\mathrm{H}(t)/\mathrm{d}t = \mathcal{O}(\lambda)$, its $\mathcal{O}(\lambda)$ contribution vanishes. For $\lambda=0$, $Q_\mathrm{H}(t)$ commutes with $H$, causing both $\mathcal{C}_1$ and $\mathcal{C}_2$ to vanish identically. We therefore conclude that their leading contributions are of order $\mathcal{O}(\lambda^2)$. Together with $\mathrm{d} \mathcal{C}_1(s;t,t')/\mathrm{d}t'=\mathcal{C}_2(s;t,t')/\tau_\mathrm{micro}$ and the assumption that $\mathcal{C}_1$ varies on the scale $\tau_\mathrm{micro}$, this implies $\mathcal{C}_1/\mathcal{C}_2=\mathcal{O}(1)$.} Expressing the rate of change of the total charge in terms of $\mathcal{C}_1$ and $\mathcal{C}_2$ yields 
\begin{align}
    \!\!\frac{\mathrm{d}}{\mathrm{d}t}\, \braket[\cbig]{Q_\mathrm{H}(t)}_\mathrm{NEQ}^t &= -\lim_{\varepsilon\to 0^+} \beta_\mathrm{EQ} \mathlarger{\int}_{-\infty}^t\mathrm{d}t'\: e^{\varepsilon (t'-t)}\mathlarger{\int}_{0}^1\mathrm{d}s \label{eq:ConstraintMassaged3}\\ 
    &\qquad\qquad\qquad\qquad\; \times \Big[\dot{\mu}(t')\,\mathcal{C}_1\big(s;t,t'\big)+\tau_{\mathrm{micro}}^{-1}\,\mu(t')\,\mathcal{C}_2\big(s;t,t'\big)\Big] \cBig\{1+\mathcal{O}\big[\mu(t)\big]\!\!\:\cBig\} \,. \notag
\end{align}
To proceed, we turn to the analysis of the scaling of the mesoscopic quantity $\mu(t)$. Let us recall that, at the order considered, $\mu(t)$ is proportional to the total charge $\braket[\rbig]{Q_\mathrm{H}(t)}_\mathrm{NEQ}^t$, see equation~\eqref{eq:QNEQ_Propto_mu}. At late times, consistent with the assumption $\mu(t) \ll 1$ employed in the preceding analysis, we anticipate that $\mu(t)$, similarly to $\braket[\rbig]{Q_\mathrm{H}(t)}_\mathrm{NEQ}^t$, follows an exponential decay law of the form considered in equation~\eqref{eq:Behavior_OnePointFunction}. The above assumption is encapsulated in the limiting relation
\begin{align}
    \mu(t) \xrightarrow{\,t\:\gg\:\tau_\mathrm{macro}\,} \mathrm{const.}\times \exp\!\bigg(\!\!-\!\!\:\frac{t}{\tau_\mathrm{macro}}\bigg)\, ,
\end{align}
once more granting the relation that at late times, we have $\big\lvert\dot{\mu}(t)\big\rvert \sim \big\lvert\tau_\mathrm{macro}^{-1}\,\mu(t)\big\rvert$. As before, let us make the scaling explicit by factoring out the characteristic timescale, defining
\begin{align}
    \mu_0(t)&\equiv\mu(t)\, ,& \mu_1(t)&\equiv \tau_\mathrm{macro}\,\dot{\mu}(t)\, , & \mu_2(t)&\equiv \tau_\mathrm{macro}^2\,\ddot{\mu}(t)\, , &\ldots& &\mu_n(t) &\equiv \tau_\mathrm{macro}^n \:\! \frac{\mathrm{d}^n \mu(t)}{\mathrm{d}t^n}\, .
\end{align}
By assumption, all $\mu_n(t)$ scale similarly in the given late-time limit, for which the deviation from equilibrium is assumed to be small. Since the integral~\eqref{eq:ConstraintMassaged3} is dominated by contributions from the parametric region $t-t'\sim \tau_\mathrm{micro}\ll \tau_\mathrm{macro}$, owing to the peaked nature of the correlators $\mathcal{C}_1$ and $\mathcal{C}_2$, we may expand $\mu(t')$ and $\dot{\mu}(t')$ about $t$ as  
\begin{align}
    \mu(t') &= \mathlarger{\sum}_{n=0}^\infty \; \frac{\mu_n(t)}{n!} \bigg(\frac{t'-t}{\tau_\mathrm{macro}}\bigg)^{\!\!n}\, ,
    &\tau_\mathrm{macro}\,\dot{\mu}(t') &= \mathlarger{\sum}_{n=0}^\infty \; \frac{\mu_{n+1}(t)}{n!} \bigg(\frac{t'-t}{\tau_\mathrm{macro}}\bigg)^{\!\!n}\, .
\end{align}
Inserting the above identities into equation~\eqref{eq:ConstraintMassaged3}, we arrive at the intermediate expression
\begin{align}
    \!\!\frac{\mathrm{d}}{\mathrm{d}t}\, \braket[\cbig]{Q_\mathrm{H}(t)}_\mathrm{NEQ}^t &= -\!\lim_{\varepsilon\to 0^+} \beta_\mathrm{EQ}\:\! \tau_\mathrm{macro}^{-1} \:\mathlarger{\sum}_{n=0}^\infty\:\frac{1}{n!}\mathlarger{\mathlarger{\int}}_{-\infty}^t\mathrm{d}t'\: e^{\varepsilon (t'-t)}\bigg(\frac{t'-t}{\tau_\mathrm{macro}}\bigg)^{\!\!n}\mathlarger{\mathlarger{\int}}_{0}^1\mathrm{d}s \label{eq:ConstraintMassaged4}\\ 
    &\qquad\qquad\qquad\times \bigg[\mu_{n+1}(t)\,\mathcal{C}_1\big(s;t,t'\big)+\frac{\tau_\mathrm{macro}}{\tau_\mathrm{micro}}\:\!\mu_{n}(t)\,\mathcal{C}_2\big(s;t,t'\big)\bigg] \cBig\{1+\mathcal{O}\big[\mu(t)\big]\!\!\:\cBig\} \,, \notag
\end{align}
where we readily switched summation and integration. With sizable contributions to the integral only arising from the region $t-t'\sim \tau_\mathrm{micro}$, we observe that the leading-order expression arises from the sole $\mu_0(t)$-contribution, which entails no additional suppression factors of $\tau_\mathrm{micro}/\tau_\mathrm{macro} \ll 1$. Therefore, we obtain the leading-order relation 
\begin{align}
    \!\!\frac{\mathrm{d}}{\mathrm{d}t}\, \braket[\cbig]{Q_\mathrm{H}(t)}_\mathrm{NEQ}^t = -\!\lim_{\varepsilon\to 0^+} \beta_\mathrm{EQ} \:\!\tau_{\mathrm{micro}}^{-1}\,\mu(t)\!\mathlarger{\mathlarger{\int}}_{-\infty}^t\mathrm{d}t'\: e^{\varepsilon (t'-t)}\!\mathlarger{\mathlarger{\int}}_{0}^1\mathrm{d}s \: \mathcal{C}_2\big(s;t,t'\big) \Bigg\{1+\mathcal{O}\bigg[\mu(t),\frac{\tau_\mathrm{micro}}{\tau_\mathrm{macro}}\bigg]\!\!\:\Bigg\} \,, 
    \label{eq:ConstraintMassaged5}
\end{align}
which essentially amounts to replacing $\mu(t')\mapsto \mu(t)$ and $\dot{\mu}(t)\mapsto 0$ in the original relation~\eqref{eq:ConstraintMassaged3}~\cite{RabyMottola1990}. We may also undo our previously introduced definition~\eqref{eq:Definition_C2} for $\mathcal{C}_2$ to express the relation in terms of the original correlator again. As a last step, we utilize the fact that $\mu(t)$ is proportional to the charge average through relation~\eqref{eq:NProptoMu}, which finally grants the desired linear relaxation equation of the form\footnote{Note that when higher-order terms in the small parameter $\tau_\mathrm{micro}/\tau_\mathrm{macro}$ are retained, the relaxation equation remains linear but may involve higher-order time derivatives. At next-to-leading order, the contributions from $\mu_1(t)$ introduce first-derivative terms that can be absorbed into the left-hand side, effectively preserving the structure of the relaxation equation~\eqref{eq:Relaxation_Gamma_Full} while yielding an improved $\Gamma_\mathrm{relax}$. However, this simplification no longer holds once $\mu_2(t)$ contributions are included, as they generate genuinely higher-order derivative terms that cannot be eliminated in the same manner. Despite this, due to linearity, an exponential ansatz for the total charge $\braket[\rbig]{Q_\mathrm{H}(t)}_\mathrm{NEQ}^t$ still solves the ODE, it is simply that the relaxation rate $\Gamma_\mathrm{relax}$ has to be computed from a polynomial equation of higher degree. As a word of caution, however, it remains unclear whether the NSO approach is valid beyond leading order in $\lambda$, as we have touched upon in section~\ref{sec:ZubarevDiscussion}. Since the construction of the NEQ operator appears to be ambiguous at NLO, it is unclear whether this ambiguity also affects higher-order predictions.}
\begin{equation}
    \frac{\mathrm{d}}{\mathrm{d}t}\,\braket[\cbig]{Q_\mathrm{H}(t)}_\mathrm{NEQ}^t 
    = - \,\Gamma_{\!\!\:\mathrm{relax}} \,\braket[\cbig]{Q_\mathrm{H}(t)}_\mathrm{NEQ}^t \Bigg\{1+\mathcal{O}\bigg[\braket[\cbig]{Q_\mathrm{H}(t)}_\mathrm{NEQ}^t\:\!,\frac{\tau_\mathrm{micro}}{\tau_\mathrm{macro}}\bigg]\!\!\:\Bigg\}\,,
    \label{eq:RelaxEquation}
\end{equation}
for which we identify the relaxation rate 
\begin{align}
    \scalebox{0.975}{$\displaystyle{\Gamma_{\!\!\:\mathrm{relax}} = \!\!\:\lim_{\varepsilon\to 0^+}\!\!\:\frac{\displaystyle{\mathlarger{\mathlarger{\int}}_{-\infty}^t\mathrm{d}t'\:e^{\varepsilon (t'-t)} \mathlarger{\mathlarger{\int}}_{0}^1\mathrm{d}s \:
    \braket[\rbigg]{e^{\scalebox{0.75}{$s\beta_\mathrm{EQ}H$}} \:\!\frac{\mathrm{d}Q_{\mathrm{H}}(t')}{\mathrm{d}t'} \:\!e^{\scalebox{0.75}{$-s\beta_\mathrm{EQ}H$}} \:\!\frac{\mathrm{d}Q_{\mathrm{H}}(t)}{\mathrm{d}t}}_{\!\!\mathrm{EQ}}}}{\displaystyle{\mathlarger{\mathlarger{\int}}_0^1 \mathrm{d}s\: \braket[\rbigg]{e^{\scalebox{0.75}{$s\beta_\mathrm{EQ}H$}} Q_{\mathrm{H}}(0)\:\! e^{\scalebox{0.75}{$-s\beta_\mathrm{EQ}H$}} Q_{\mathrm{H}}(0)}_{\!\!\mathrm{EQ}}}}\:\!\Big[1+\mathcal{O}(\lambda)\Big] \:\! .}$}
    \label{eq:Relaxation_Gamma_Full}
\end{align}
At this stage, the limit $\varepsilon\to 0^+$ can be safely taken, as the integral is regular. This follows from the strongly peaked nature of the correlator, which suppresses potentially large contributions from early times that would otherwise require regulation. The given expression matches the one derived in previous works, see e.g. Khlebnikov \& Shaposhnikov~\cite{KhlebnikovShaposhnikov1988} or Mottola \& Raby~\cite{RabyMottola1990}. It is worth noting that, just as the denominator, the numerator arising in equation~\eqref{eq:Relaxation_Gamma_Full} is independent of $t$. We can make this manifest by employing the operator equation
\begin{equation}
    \frac{\mathrm{d}Q_{\mathrm{H}}(t)}{\mathrm{d}t} = \frac{i}{\hbar} \Big[H,Q_{\mathrm{H}}(t) \Big]\, ,
\end{equation}
after which we can commute the time evolution operators entering the Heisenberg operator $Q_\mathrm{H}(t)$ through the $H$-dependent exponentials. Thus, the integrand depends solely on the relative time difference $t_\mathrm{rel}=t'-t$, such that after a change of variables, any residual $t$-dependence drops out. This leaves us with the manifestly $t$-independent expression 
\begin{align}
    \Gamma_{\mathrm{relax}} &= -\,  \frac{\displaystyle{\mathlarger{\mathlarger{\int}}_{-\infty}^0\mathrm{d}t_\mathrm{rel} \mathlarger{\mathlarger{\int}}_{0}^1\mathrm{d}s \:
    \braket[\rbigg]{e^{\scalebox{0.75}{$s\beta_\mathrm{EQ}H$}} \!\!\;\Big[H,Q_\mathrm{H}(t_\mathrm{rel})\Big] e^{\scalebox{0.75}{$-s\beta_\mathrm{EQ}H$}} \!\!\;\Big[H,Q_\mathrm{H}(0)\Big] }_{\!\!\mathrm{EQ}}}}{\displaystyle{\hbar^2\mathlarger{\mathlarger{\int}}_0^1 \mathrm{d}s\: \braket[\rbigg]{e^{\scalebox{0.75}{$s\beta_\mathrm{EQ}H$}} Q_{\mathrm{H}}(0) \:\! e^{\scalebox{0.75}{$-s\beta_\mathrm{EQ}H$}} Q_{\mathrm{H}}(0)}_{\!\!\mathrm{EQ}}}}\:\!\Big[1+\mathcal{O}(\lambda)\Big]\:\! .
    \label{eq:Relaxation_Gamma_Full2}
\end{align}
One can further simplify the relaxation rate~\eqref{eq:Relaxation_Gamma_Full2} by expanding the denominator in $\lambda$. Note that, since $\tau_\mathrm{micro}/\tau_\mathrm{macro}$ is related to $\lambda$, the preceding expansion~\eqref{eq:Relaxation_Gamma_Full} already constituted an implicit expansion in $\lambda$. Thus, we are not adding a further expansion layer, but instead proceed in accordance with the previous line of reasoning. For the denominator, the leading-order result can be obtained by simply dropping the $s$-integration, which is easily spotted by utilizing the Campbell identity in guise of
\begin{align}
    \mathlarger{\int}_{0}^1\mathrm{d}s\: e^{\scalebox{0.75}{$s\beta_\mathrm{EQ}H$}} Q_\mathrm{H}(0) e^{\scalebox{0.75}{$-s\beta_\mathrm{EQ}H$}} &= \mathlarger{\int}_{0}^1\mathrm{d}s\: \mathlarger{\sum}_{n=0}^\infty \:\frac{1}{n!} \cbig[s\beta_\mathrm{EQ}H,Q_\mathrm{H}(0)\cbig]_n = \mathlarger{\sum}_{n=0}^\infty\:\mathlarger{\int}_{0}^1\mathrm{d}s \:\frac{\big(s\beta_\mathrm{EQ}\big)^n}{n!} \cbig[H,Q_\mathrm{H}(0)\cbig]_n \notag \\ 
    &= Q_\mathrm{H}(0)+\underbrace{\mathlarger{\sum}_{n=1}^\infty\:\frac{\big(\beta_\mathrm{EQ}\big)^n}{(n+1)!} \cbig[H,Q_\mathrm{H}(0)\cbig]_n}_{\displaystyle{=\mathcal{O}(\lambda)}}\, ,
\end{align}
where $\big[H,Q_\mathrm{H}(0)\big]_n$ denotes the iterated commutator 
\begin{align}
   \cbig[H,Q_\mathrm{S}\cbig]_n \equiv \Big[\underbrace{\!\!\:H,\dotsb\cbig[H,\big[H}_{\displaystyle{n \text{ times}}},Q_\mathrm{H}(0)\big]\!\!\;\cbig]\dotsb\Big]\, .
\end{align}
However, in the numerator, one cannot discard the $s$-dependent contributions by the same argument, since the charge operator $Q_\mathrm{S}$ occurs only in combination with commutators, which are already $\mathcal{O}(\lambda)$. As there is no basis for assuming that a double commutator is suppressed compared to a single one, we are required to retain the full structure. Therefore, one arrives at the final result 
\begin{align}
    \Gamma_{\mathrm{relax}} &= -\,  \frac{\displaystyle{\mathlarger{\mathlarger{\int}}_{-\infty}^0\mathrm{d}t_\mathrm{rel} \mathlarger{\mathlarger{\int}}_{0}^1\mathrm{d}s \:
    \braket[\rbigg]{e^{\scalebox{0.75}{$s\beta_\mathrm{EQ}H$}} \!\!\:\Big[H,Q_\mathrm{H}(t_\mathrm{rel})\Big] e^{\scalebox{0.75}{$-s\beta_\mathrm{EQ}H$}} \!\!\:\Big[H,Q_\mathrm{H}(0)\Big] }_{\!\!\mathrm{EQ}}}}{\displaystyle{\hbar^2\braket[\cbig]{Q_{\mathrm{H}}(0)^2}_{\!\mathrm{EQ}}}}\:\!\Big[1+\mathcal{O}(\lambda)\Big] \, .
    \label{eq:Final_Result_Zubarev}
\end{align}

\subsection{Equivalent representations of the relaxation rate}
\label{sec:Representations_RelaxRate}

Given the extensive literature on non-equilibrium statistical physics and field theory, relation~\eqref{eq:Final_Result_Zubarev} appears in a variety of equivalent forms, involving, for example, commutators with the Hamiltonian~\cite{KhlebnikovShaposhnikov1988}, imaginary-time-shifted correlators~\cite{KuboYokotaNakajima1957}, or the Bogoliubov inner product~\cite{RoepkeArticle2018,Forster:1975}. In this section, we present a non-exhaustive list of representations obtained through straightforward manipulations of equation~\eqref{eq:Final_Result_Zubarev}. \\ 

\noindent 
Simplifying our notation, let us introduce the charge-violation operator $J_\mathrm{S}$ by defining
\begin{align}
    J_\mathrm{S} \equiv  \frac{i\lambda}{\hbar} \Big[H_\mathrm{S}^{(\mathrm{viol.})},Q_{\mathrm{S}}\Big] = \frac{i}{\hbar} \cbig[H,Q_{\mathrm{S}}\cbig]\, ,
\end{align}
whose Heisenberg- and interaction-picture representations are given by
\begin{subequations}
    \begin{align}
        J_\mathrm{H}(t) &= \frac{i}{\hbar} \Big[H,Q_{\mathrm{H}}(t) \Big] = \frac{\mathrm{d}Q_{\mathrm{H}}(t)}{\mathrm{d}t}\, , \label{eq:ChargeViolationOperator_J} \\[0.05cm]
        J_\mathrm{I}(t) & = \frac{i\lambda}{\hbar} \Big[H_\mathrm{I}^{(\mathrm{viol.})}(t),Q_{\mathrm{I}} \Big]\, .
        \label{eq:ChargeViolationOperator_J_Interaction}
    \end{align}
\end{subequations}
Beware that this newly introduced operator does not commute with $H$. Utilizing $J_\mathrm{H}(t)$, we arrive at the two equivalent representations 
\begin{align}
    \Gamma_{\mathrm{relax}} &= \Big[\braket[\cbig]{Q_{\mathrm{H}}(0)^2}_{\!\mathrm{EQ}}\Big]^{-1}\mathlarger{\int}_{-\infty}^0\mathrm{d}t \mathlarger{\int}_{0}^1\mathrm{d}s \:
    \braket[\cBig]{e^{\scalebox{0.75}{$s\beta_\mathrm{EQ}H$}} J_\mathrm{H}(t) \:\! e^{\scalebox{0.75}{$-s\beta_\mathrm{EQ}H$}} J_\mathrm{H}(0)}_{\!\!\mathrm{EQ}}\:\!\Big[1+\mathcal{O}(\lambda)\Big] \notag \\
    &=  \Big[\braket[\cbig]{Q_{\mathrm{H}}(0)^2}_{\!\mathrm{EQ}}\Big]^{-1} \mathlarger{\int}^{\infty}_0\mathrm{d}t \mathlarger{\int}_{0}^1\mathrm{d}s \:
    \braket[\cBig]{e^{\scalebox{0.75}{$s\beta_\mathrm{EQ}H$}}J_\mathrm{H}(0) \:\! e^{\scalebox{0.75}{$-s\beta_\mathrm{EQ}H$}}J_\mathrm{H}(t)}_{\!\!\mathrm{EQ}}\:\!\Big[1+\mathcal{O}(\lambda)\Big]\, ,
    \label{eq:Relaxation_Rate_Intermediate}
\end{align}
where in the last line we changed the integration variable $t\mapsto -t$ and shifted the time-dependence inside $J_\mathrm{H}(-t)$ to the second operator. Using the Bogoliubov inner product
\begin{align}
    \!\braket[\rbig]{X,Y}^{\!\!\:\text{(Bogoliubov)}}_{\mathrm{EQ}}&\equiv  \mathlarger{\int}_0^1 \mathrm{d}s \; \braket[\rBig]{e^{\scalebox{0.775}{$s\beta_\mathrm{EQ} H$}} X^\dagger e^{\scalebox{0.775}{$-s\beta_\mathrm{EQ} H$}} Y}_{\!\!\:\mathrm{EQ}} \, ,
\end{align} 
one can express the relaxation rate as  
\begin{align}
    \Gamma_{\mathrm{relax}} &=\Big[\braket[\cbig]{Q_{\mathrm{H}}(0)^2}_{\!\mathrm{EQ}}\Big]^{-1} \mathlarger{\int}^{\infty}_0\mathrm{d}t \mathlarger{\int}_{0}^1\mathrm{d}s \:
    \braket[\rBig]{J_\mathrm{H}\big(\!-\!\!\:i\beta_\mathrm{EQ}\hbar s\big) \:\!J_\mathrm{H}(t)}_{\!\mathrm{EQ}}\:\!\Big[1+\mathcal{O}(\lambda)\Big] \notag \\[0.1cm]
    &=\Big[\braket[\cbig]{Q_{\mathrm{H}}(0)^2}_{\!\mathrm{EQ}}\Big]^{-1} \mathlarger{\int}^{\infty}_0\mathrm{d}t \:
    \braket[\rBig]{J_\mathrm{H}(0), J_\mathrm{H}(t)}_{\!\mathrm{EQ}}^{\!\text{(Bogoliubov)}}\:\!\Big[1+\mathcal{O}(\lambda)\Big]\, .
    \label{eq:Reformulation_Zubarev}
\end{align}
Note that in all above expressions, we may use a coordinate transformation $s\mapsto 1-s$ to switch the role of $J_\mathrm{H}(0)$ and $J_\mathrm{H}(t)$, showing that the expression is indeed symmetric under this exchange. Since the Bogoliubov inner product satisfies $\smash{\braket[\rbig]{X,Y}=\braket[\rbig]{Y,X}^{\!\!\:*}}$, the aforementioned property guarantees that $\Gamma_\mathrm{relax}$ is real. The most commonly encountered representation of the decay rate is given by the Green--Kubo relation
\begin{align}
    \tcboxmath{\Gamma_{\mathrm{relax}} = \frac{1}{\braket[\rbig]{Q_\mathrm{H}(0)^2}_\mathrm{EQ}}\:\mathlarger{\int}^{\infty}_0\mathrm{d}t \:
    \braket[\rBig]{J_\mathrm{H}(t), J_\mathrm{H}(0)}_{\!\mathrm{EQ}}^{\!\text{(Bogoliubov)}}\:\!\Big[1+\mathcal{O}(\lambda)\Big]\, .}
    \label{eq:Final_Relaxation_Rate}
\end{align}
As shown later in section~\ref{sec:Frequency_Space_Relaxation}, under the assumptions relevant here, the Bogoliubov inner product may be replaced by the corresponding ordinary equilibrium correlation function. In frequency space, the imaginary-time smearing that defines the Bogoliubov product introduces a thermal prefactor that tends to unity in the zero-frequency limit, see equation~\eqref{eq:Relaxation_PowerSpectrum}. Since the dissipative part of the Green--Kubo integral is determined by this limit as illustrated in equation~\eqref{eqn:power_spectrum_definition}, the two expressions yield the same relaxation rate. We have retained the Bogoliubov form at this stage because it arises naturally in the derivation and may provide better spectral regularity in cases where the unsmeared correlation function is not sufficiently well behaved.

\section{A simplified local equilibrium approach}
\label{sec:3_LEQ_Method}

Having reviewed the conventional Zubarev derivation in the preceding sections, we now propose a simpler alternative approach that yields the desired relaxation rate. To clarify the key differences between this approach and Zubarev’s, we first briefly discuss the shift in underlying philosophy before turning to the actual computation.

\subsection{Conceptual framework}
\label{sec:LEQ_Concept}
As before, our goal is to isolate the slow relaxation of the approximately conserved charge $Q$ from the remaining fast microscopic dynamics. In Zubarev's approach, this is accomplished by constructing the full non-equilibrium statistical operator $\smash{\rho^{(\mathrm{NEQ})}\!\!\:(t)}$, which encodes the relevant history of the system and in principle allows one to evaluate arbitrary observables. For our more restricted purpose of solely determining the late-time evolution of the slow average $\braket{Q}^t$, this construction retains substantially more information than necessary. We can instead take a simpler route that exploits the strong separation of timescales directly.\\ 

\noindent 
As seen before, the late-time evolution of the slow variable $Q$ does not require the full physical density matrix $\smash{\rho^{(\mathrm{phys})}\!\!\:(t)}$. Once the dynamics is effectively Markovian, the sought-after rate of change $\mathrm{d}\braket{Q}^t/\mathrm{d}t$ is determined solely by the macroscopic state at that time. To determine the relaxation law, it is therefore sufficient to answer two questions at each time $t$: what are the instantaneous values of the slow variables $H$ and $Q$ that specify the macroscopic state, and what are their instantaneous rates of change? The first question is answered by the local-equilibrium ansatz: by construction, $\smash{\rho^{(\mathrm{LEQ})}\!\!\:(t)}$ reproduces the macroscopic averages of the physical state while dispensing with its detailed microscopic information. As for the second question, the total energy is exactly conserved, so that $\mathrm{d}\braket{H}^t/\mathrm{d}t=0$, leaving only the slow drift of $\braket{Q}^t$ to be determined.\\ 

\noindent 
Evaluating this rate of change, however, requires some care. Although the LEQ and physical states agree on the retained macroscopic variables at time $t$, they generally differ in their microscopic correlations. As a result, the relevant averages need not evolve in the same way immediately after $t$. We therefore cannot infer the physical relaxation rate simply from an infinitesimal evolution of the LEQ state. Instead, we evolve it over a mesoscopic interval
\begin{equation}
    \tau_{\mathrm{micro}} \ll \Delta t \ll \tau_{\mathrm{macro}} \, ,
    \label{eq:TimeScales}
\end{equation}
long enough for the system to lose sensitivity to the microscopic correlations discarded at the initial time and to establish the short-lived correlations governing the evolution of the slow charge, but short enough that $\braket{Q}^t$ and the other macroscopic variables change only parametrically little. The resulting finite change $\braket{Q}^{t+\Delta t}-\braket{Q}^t$ therefore isolates the slow drift associated with the state at time $t$, and may be identified with its coarse-grained instantaneous relaxation rate $\mathrm{d}\braket{Q}^t/\mathrm{d}t$. Repeating this construction at arbitrary times in the late-time regime yields the desired local relaxation equation.\\

\noindent 
It is particularly convenient to formulate this procedure in the interaction picture with respect to $H^{(\mathrm{cons})}$, as defined in equation~\eqref{eq:Interaction_Picture}. In this picture, the charge $Q_\mathrm{I}$ is time independent, while the explicit evolution of the density matrix is generated solely by the weak charge-violating interaction $\lambda H_\mathrm{I}^{(\mathrm{viol})}(t)$. This formulation makes both the perturbative expansion in $\lambda$ and the separation of fast equilibration from slow relaxation manifest.

\subsection{Obtaining a relaxation equation}

\noindent 
We now turn to the explicit determination of the relaxation rate within the framework outlined above. In the interaction picture, we consider the evolution of the charge expectation value
\begin{equation}
    \braket[\rbig]{Q_\mathrm{I}}^t=\mathrm{tr}\cbig[\rho_\mathrm{I}(t)\:\!Q_{\mathrm{I}}\cbig]\,.
\end{equation}
Since $Q_\mathrm{I}$ is time independent, its evolution arises entirely from that of the density matrix, which obeys the von Neumann equation
\begin{equation} \label{eq:drhodt0}
    \frac{\mathrm{d}\rho_\mathrm{I}(t)}{\mathrm{d}t} + \frac{i \lambda }{\hbar }\Big[H^{\mathrm{(viol.)}}_\mathrm{I}(t),\rho_\mathrm{I}(t)\Big] = 0\, .
\end{equation}
The evolution of the density matrix over the mesoscopic interval $\Delta t$ is then given by the formal solution
\begin{equation}
    \rho_{\mathrm{I}}\big(t+\Delta t\big)= U_\mathrm{I}\big(t+\Delta t, t\big)\:\! \rho_{\mathrm{I}} (t)\,U_\mathrm{I}^{\dagger}\:\!\!\big(t+\Delta t, t\big)\, ,
\end{equation}
where the corresponding interaction-picture time-evolution operator reads
\begin{equation}
    U_\mathrm{I}\big(t+\Delta t, t\big)= \mathcal{T}\!\!\:\exp\cbigg\{\!\!\:- \,\frac{i \lambda}{\hbar }\!\!\:\int_{t}^{t+\Delta t} \mathrm{d}t'\,H^{\mathrm{(viol.)}}_\mathrm{I}(t') \cbigg\} \, .
\end{equation}
One might be tempted to expand the time-evolution operator to second order in $\lambda$. However, such an expansion is not obviously justified at the operator level. Indeed, the dimensionless parameter controlling a naïve estimate of the Dyson series is of order $\lambda \Delta t/ \tau_\mathrm{micro}$. For the mesoscopic interval $ \Delta t\sim \lambda^{-1}\tau_\mathrm{micro}$ adopted below, this quantity is of order unity. Consequently, higher-order terms in the Dyson expansion are not parametrically suppressed at the operator level and cannot be discarded a priori. Instead, one should investigate the nested commutator terms arising in the expansion of the desired expectation value difference
\begin{align}
    \braket[\rbig]{Q_{\mathrm I}}^{t+\Delta t} - \braket[\rbig]{Q_{\mathrm I}}^{t} &= \mathrm{tr}\bigg\{\!\!\:\Big[\rho_{\mathrm I}\big(t+\Delta t\big)-\rho_{\mathrm I}(t)
    \Big] Q_{\mathrm I} \bigg\} \notag \\ 
    &= \mathrm{tr}\cbigg\{\rho_{\mathrm I}(t) \Big[U_\mathrm{I}^{\dagger}\:\!\!\big(t+\Delta t, t\big) \:\!Q_{\mathrm I} \:\! U_\mathrm{I}\:\!\!\big(t+\Delta t, t\big)-Q_\mathrm{I}\Big]\!\!\:\cbigg\} \notag \\ 
    &= \mathlarger{\sum}_{n\:\!=\:\!1}^\infty \,\cBig(\frac{i\lambda}{\hbar}\cBig)^{\!\!\!\:n}\mathlarger{\int}_{t}^{t+\Delta t}\mathrm{d}t_1\mathlarger{\int}_{t}^{t_1}\mathrm{d}t_2\ldots \mathlarger{\int}_{t}^{t_{n-1}}\mathrm{d}t_n \label{eq:Expansion_QDifference_allOrders} \\[-0.1cm]
    &\qquad\qquad\qquad\!\!\!\! \times \mathrm{tr}\Bigg\{\rho_{\mathrm I}(t)\bigg[H^{\mathrm{(viol.)}}_\mathrm{I}(t_n),\!\!\:\cBig[H^{\mathrm{(viol.)}}_\mathrm{I}(t_{n-1}),\:\!\ldots\!\!\: \Big[H^{\mathrm{(viol.)}}_\mathrm{I}(t_1),Q_\mathrm{I} \Big]\!\!\:\ldots\!\!\:\cBig]\!\!\:\bigg] \!\!\!\;\Bigg\}\, . \notag
\end{align}
The naïve counting described above changes once the expanded operator is evaluated inside an expectation value, where microscopic clustering constrains the relevant time integrations. In analogy with the two-time correlators discussed around equation~\eqref{eq:Behavior_TwoPointFunction}, we assume that the connected correlation functions of the fast interaction-picture operators decay once their relative time separations exceed the microscopic equilibration time $\tau_\mathrm{micro}$. Consequently, only one of the $n$ time integrations remains extensive in $\Delta t$, whereas the remaining $n-1$ integrations are effectively restricted to relative time separations of order $\tau_{\mathrm{micro}}$. The corresponding integrals therefore scale parametrically as $\Delta t\,\tau_{\mathrm{micro}}^{\,n-1}$, rather than as $\Delta t^n$. Since each additional insertion of $\smash{H_{\mathrm I}^{(\mathrm{viol.})}}$ is accompanied by a factor $\lambda/\hbar$, successive terms in the expansion are indeed suppressed by additional powers of $\lambda$, allowing for a consistent truncation. Keeping terms up to second order, we obtain 
\begin{align}
    \!\!\braket[\rbig]{Q_{\mathrm I}}^{t+\Delta t} - \braket[\rbig]{Q_{\mathrm I}}^{t} &= \phantom{+}\frac{i\lambda}{\hbar} \mathlarger{\int}_t^{t+\Delta t}\mathrm{d}t_1\,\mathrm{tr}\bigg\{\rho_{\mathrm I}(t)\Big[H^{(\mathrm{viol.})}_{\mathrm I}(t_1),Q_{\mathrm I}\Big]\!\!\:\bigg\}  \label{eq:Expansion_QDifference} \\ 
    &\phantom{=} -\frac{\lambda^2}{\hbar^2} \mathlarger{\int}_t^{t+\Delta t}\mathrm{d}t_1 \mathlarger{\int}_t^{t_1}\mathrm{d}t_2\, \mathrm{tr}\cbigg\{\rho_{\mathrm I}(t)\cBig[H^{(\mathrm{viol.})}_{\mathrm I}(t_2),\Big[H^{(\mathrm{viol.})}_{\mathrm I}(t_1) , Q_{\mathrm I} \Big]\!\!\:\cBig]\!\cbigg\} \:+\: \mathcal{O}\big(\lambda^3\Delta t\big) \, .\notag
\end{align}
With these points clarified, we now implement the local-equilibrium prescription motivated in section~\ref{sec:LEQ_Concept}. At an arbitrary late time $t$, taken as the beginning of the mesoscopic evolution interval, we represent the instantaneous macroscopic state by
\begin{equation} \label{eq:rhoIrhoLEQ}
    \rho_\mathrm{I}(t) \longrightarrow \rho^{(\mathrm{LEQ})}_{\mathrm{I}}(t)= \frac{1}{Z_\mathrm{I}^{(\mathrm{LEQ})}\!\!\:(t)}\,\exp\!\bigg\{\!\!\:- \!\!\:\beta(t) \Big[H^{\mathrm{(cons.)}}_\mathrm{I} - \mu(t) \:\! Q_\mathrm{I}\Big]\!\!\:\bigg\}\, .
\end{equation}
As in Zubarev's approach, the parameters $\beta(t)$ and $\mu(t)$ are fixed by matching to the instantaneous expectation values of $H_\mathrm{I}^{(\mathrm{cons.})}$ and $Q_\mathrm{I}$  
\begin{subequations}
    \begin{align}
        \braket[\cbig]{H_\mathrm{I}^{(\mathrm{cons.})}}_{\!\!\:\mathrm{LEQ}}^{\!\!\:t} = \mathrm{tr}\Big[\rho^{(\mathrm{LEQ})}_{\mathrm{I}}\!\!\;(t)\:\!H_\mathrm{I}^{(\mathrm{cons.})}\Big] &\overset{!}{=} \mathrm{tr}\Big[\rho^{(\mathrm{phys.})}_{\mathrm{I}}\!\!\;(t)\:\!H_\mathrm{I}^{(\mathrm{cons.})}\Big] = \braket[\cbig]{H_\mathrm{I}^{(\mathrm{cons.})}}_{\!\!\:\mathrm{phys.}}^{\!\!\:t}\, , \label{eq:Constraint_InteractionPic1} \\[0.05cm] 
        \braket[\rbig]{Q_\mathrm{I}}_{\mathrm{LEQ}}^{t} = \mathrm{tr}\Big[\rho^{(\mathrm{LEQ})}_{\mathrm{I}}\!\!\;(t)\:\!Q_\mathrm{I}\Big] &\overset{!}{=} \mathrm{tr}\Big[\rho^{(\mathrm{phys.})}_{\mathrm{I}}\!\!\;(t)\:\!Q_\mathrm{I}\Big] = \braket[\rbig]{Q_\mathrm{I}}_{\!\!\:\mathrm{phys.}}^{\!\!\:t}\, .
    \label{eq:Constraint_InteractionPic0}
    \end{align}
\end{subequations}
For convenience, we depart from the previous definition of the LEQ operator~\eqref{eq:LEQ_Operator} in terms of the full Hamiltonian $H_\mathrm{I}$ and instead construct~\eqref{eq:rhoIrhoLEQ} using only its charge-conserving part $\smash{H^{\mathrm{(cons.)}}_\mathrm{I}}$. This choice is not essential, as the alternative LEQ ansatz constructed using $H_\mathrm{I}$, when formulated consistently in the interaction picture,\footnote{Here, a consistent formulation requires choosing the reference time at which the Schrödinger and interaction pictures coincide to be the time $t$ at which the LEQ state is constructed. Simply transforming the previously defined Schrödinger-picture LEQ operator~\eqref{eq:LEQ_Operator} from a fixed reference time $t_\mathrm{ref}=0$ would retain the coherent microscopic evolution accumulated between 0 and $t$, contrary to the local-equilibrium prescription.} leads to the same leading-order relaxation rate, with the two constructions differing only at higher orders in $\lambda$. Note also that, since we only require $\smash{\rho^{(\mathrm{LEQ})}_{\mathrm{I}}(t)}$ to serve as the initial condition for the evolution over the mesoscopic interval, we do not require it to satisfy the von Neumann equation~\eqref{eq:drhodt0}, in contrast to the NEQ operator. \\

\noindent 
The provided ansatz~\eqref{eq:rhoIrhoLEQ} greatly simplifies the evolution equation~\eqref{eq:Expansion_QDifference}. To this end, note that $Q_{\mathrm I}$ commutes with both $H^{(\mathrm{cons.)}}_\mathrm{I}$ and with itself, implying
    \begin{equation}
    \Big[\rho_{\mathrm I}^{(\mathrm{LEQ})}(t),Q_{\mathrm I}\Big] =0 \, .
\end{equation}
As a consequence, the first-order term in equation~\eqref{eq:Expansion_QDifference} vanishes identically, as the first-order contribution describes coherent, reversible dynamics and therefore does not contribute to the relaxation of the charge. The first irreversible change arises only at second order in $\lambda$, which in turn determines the leading-order relaxation rate. With the second-order term being invariant under a simultaneous shift of $t_1$ and $t_2$, it depends solely on the relative time coordinate $t_\mathrm{rel}=t_1-t_2$. This allows the time-ordered double integral to be rewritten as a single integral over the relative time difference, according to
\begin{equation}
    \int_t^{t+\Delta t}\mathrm{d}t_1\int_t^{t_1}\mathrm{d}t_2\:f\big(t_1-t_2\big) = \int_0^{\Delta t}\mathrm{d}t_{\mathrm{rel}}\:\big(\Delta t-t_{\mathrm{rel}}\big)\:\! f(t_{\mathrm{rel}}) \, .
\end{equation}
Instating these simplifications and dividing both sides by $\Delta t$ grants the finite-difference quotient 
\begin{align}
    \scalebox{0.953}{$\displaystyle{\frac{\braket[\rbig]{Q_{\mathrm I}}^{t+\Delta t} - \braket[\rbig]{Q_{\mathrm I}}^{t}}{\Delta t} = - \frac{\lambda^2}{\hbar^2} \mathlarger{\int}_0^{\Delta t}\mathrm{d}t_{\mathrm{rel}}\,   \cBig(1-\frac{t_{\mathrm{rel}}}{\Delta t}\cBig)\,  \mathrm{tr}\cbigg\{\rho^{(\mathrm{LEQ})}_{\mathrm{I}}(t)\cBig[H^{(\mathrm{viol.})}_{\mathrm I}(t_\mathrm{rel}),\!\!\:\Big[H^{(\mathrm{viol.})}_{\mathrm I}(0) , Q_{\mathrm I} \Big]\!\!\:\cBig]\!\cbigg\} \:+\: \mathcal{O}\big(\lambda^3\big) \, .}$}
    \label{eq:Finite_Difference_Quotient}
\end{align}
We now utilize the scale separation $\tau_{\mathrm{micro}} \ll \Delta t$. Since the correlator is assumed to be strongly peaked around $t_{\mathrm{rel}}=0$, decaying on the microscopic scale $\tau_{\mathrm{micro}}$, the integral receives sizable contributions only from the parametric region $t_{\mathrm{rel}}\sim \tau_{\mathrm{micro}}\ll \Delta t$. Within this effective support, we may drop the small corrections arising through the term $t_\mathrm{rel}/\Delta t\ll 1$. We can also extend the upper integration limit to infinity, as the contributions from $t_{\mathrm{rel}}\gg \tau_{{\mathrm{micro}}}$ are exponentially suppressed. This lets us arrive at the coarse-grained evolution equation
\begin{align}
    \frac{\langle Q_{\mathrm I}\rangle^{t+\Delta t}-\langle Q_{\mathrm I}\rangle^t}{\Delta t} = -\frac{\lambda^2}{\hbar^2} \mathlarger{\int}_{0}^{\infty}\mathrm{d}t_{\mathrm{rel}}\;\mathrm{tr}\cbigg\{\rho^{(\mathrm{LEQ})}_{\mathrm{I}}(t)\cBig[H^{(\mathrm{viol.})}_{\mathrm I}(t_\mathrm{rel}),\!\!\:\Big[H^{(\mathrm{viol.})}_{\mathrm I}(0) , Q_{\mathrm I} \Big]\!\!\:\cBig]\!\cbigg\} \phantom{\, .} \notag  \\[0.1cm]
     \times \bigg\{1+\mathcal{O}\cBig(\lambda\:\!,\frac{\tau_{\mathrm{micro}}}{\Delta t}\cBig)\!\!\:\bigg\}\, .
\end{align}
The remaining separation of scales $\Delta t \ll \tau_{\mathrm{macro}}$ ensures that the expectation value $\langle Q_{\mathrm I}\rangle^t$ changes only negligibly over the mesoscopic interval $\Delta t$. The left-hand side may therefore be identified with the time derivative, up to corrections of order $\Delta t/\tau_{\mathrm{macro}}$. We therefore finally arrive at 
\begin{align}
   \frac{\mathrm{d}}{\mathrm{d}t} \braket[\rbig]{Q_{\mathrm I}}^{\!\!\:t} = -\frac{\lambda^2}{\hbar^2} \mathlarger{\int}_0^\infty \mathrm{d}t_{\mathrm{rel}}\;\mathrm{tr}\bigg\{\!\!\:\Big[\rho^{(\mathrm{LEQ})}_{\mathrm{I}}(t),H^{(\mathrm{viol.})}_{\mathrm I}(0)\Big]\!\!\:\Big[H^{(\mathrm{viol.})}_{\mathrm I}(t_\mathrm{rel}) , Q_{\mathrm I} \Big]\!\!\:\bigg\} \phantom{\, .}\notag \\ 
   \times \cbigg\{1+\mathcal{O}\cBig(\lambda\:\!,\frac{\tau_{\mathrm{micro}}}{\Delta t},\frac{\Delta t}{\tau_{\mathrm{macro}}}\cBig)\!\!\:\cbigg\}\, ,
    \label{eq:EDOQ-leq00}
\end{align}
where we have rearranged the commutators into a more convenient form. At this stage, the evolution has become local in time. The microscopic correlator entering the time integral decays on the short timescale $\tau_{\mathrm{micro}}$, long before the macroscopic variables exhibit any appreciable change. As a result, the system effectively loses memory of its detailed past evolution, and the rate at time $t$ depends only on the instantaneous state, here encoded in the density matrix $\rho^{(\mathrm{LEQ})}_\mathrm{I}(t)$. This is the origin of the Markovian approximation within this local equilibrium framework. Choosing $\Delta t\sim \lambda^{-1}\hbar \beta_\mathrm{EQ}$, in accordance with equation~\eqref{eq:TimeScales}, ensures that the corrections in $\tau_\mathrm{micro}/\Delta t$ and $\Delta t/\tau_\mathrm{macro}$ are of order $\lambda$, greatly simplifying the error estimate. \\

\noindent 
Before expanding expression~\eqref{eq:EDOQ-leq00} in small deviations from equilibrium, we can further massage the commutator between the LEQ operator and $\smash{H^{(\mathrm{viol.})}_{\mathrm I}(0)}$. Using the operator equation 
\begin{align}
    \cbig[e^{A},B\cbig] = \int_0^1 \mathrm{d}s \: e^{(1-s)A}\big[A,B\big] e^{sA}\, ,
\end{align}
we can recast the desired commutator into the form
\begin{align}
\!\Big[\rho_{\mathrm{I}}^{(\mathrm{LEQ})}(t),H_{\mathrm{I}}^{(\mathrm{viol.})}(0)\Big] &= \frac{1}{Z_\mathrm{I}^{(\mathrm{LEQ})}\!\!\:(t)}\,\Bigg[\exp\!\bigg\{\!\!\:- \!\!\:\beta(t) \Big[H^{\mathrm{(cons.)}}_\mathrm{I} - \mu(t) \:\! Q_\mathrm{I}\Big]\!\!\:\bigg\}\:\!,H_{\mathrm{I}}^{(\mathrm{viol.})}(0)\Bigg] \\ 
&=-\beta(t) \mathlarger{\int}_0^1 \mathrm{d}s\,\Big(\rho_{\mathrm{I}}^{(\mathrm{LEQ})}\!\!\;(t)\!\!\:\Big)^{\!1-s} \Big[H^{\mathrm{(cons.)}}_\mathrm{I} - \mu(t) \:\! Q_\mathrm{I},H_{\mathrm{I}}^{(\mathrm{viol.})}(0)\Big] \Big(\rho_{\mathrm{I}}^{(\mathrm{LEQ})}\!\!\;(t)\!\!\:\Big)^{\!s}\,. \notag
\end{align}
One may notice that the contribution arising from the commutator term involving $\smash{H^{\mathrm{(cons.)}}_\mathrm{I}}$ ultimately vanishes. To see this explicitly, we rearrange the corresponding term as follows
\begin{align}
   &\:\mathlarger{\int}_0^1 \mathrm{d}s\mathlarger{\int}_0^\infty \mathrm{d}t_{\mathrm{rel}}\;\mathrm{tr}\bigg\{\!\!\:\Big(\rho_{\mathrm{I}}^{(\mathrm{LEQ})}\!\!\;(t)\!\!\:\Big)^{\!1-s} \Big[H^{\mathrm{(cons.)}}_\mathrm{I},H_{\mathrm{I}}^{(\mathrm{viol.})}(0)\Big] \Big(\rho_{\mathrm{I}}^{(\mathrm{LEQ})}\!\!\;(t)\!\!\:\Big)^{\!s}\Big[H^{(\mathrm{viol.})}_{\mathrm I}(t_\mathrm{rel}) , Q_{\mathrm I} \Big]\!\!\:\bigg\} \notag \\ 
   =&\: \mathlarger{\int}_0^1 \mathrm{d}s\mathlarger{\int}_0^\infty \mathrm{d}t_{\mathrm{rel}}\;\mathrm{tr}\bigg\{\!\!\:\Big(\rho_{\mathrm{I}}^{(\mathrm{LEQ})}\!\!\;(t)\!\!\:\Big)^{\!1-s} \Big[H^{\mathrm{(cons.)}}_\mathrm{I},H_{\mathrm{I}}^{(\mathrm{viol.})}(-t_\mathrm{rel})\Big] \Big(\rho_{\mathrm{I}}^{(\mathrm{LEQ})}\!\!\;(t)\!\!\:\Big)^{\!s}\Big[H^{(\mathrm{viol.})}_{\mathrm I}(0) , Q_{\mathrm I} \Big]\!\!\:\bigg\}  \notag \\ 
   =&\:\mathlarger{\int}_0^1 \mathrm{d}s\mathlarger{\int}_0^\infty \mathrm{d}t_{\mathrm{rel}}\;\mathrm{tr}\bigg\{\!\!\:\Big(\rho_{\mathrm{I}}^{(\mathrm{LEQ})}\!\!\;(t)\!\!\:\Big)^{\!1-s} (-i\hbar) \, \frac{\mathrm{d}H^{(\mathrm{viol.})}_{\mathrm I}(-t_\mathrm{rel})}{\mathrm{d}(-t_\mathrm{rel})} \Big(\rho_{\mathrm{I}}^{(\mathrm{LEQ})}\!\!\;(t)\!\!\:\Big)^{\!s}\Big[H^{(\mathrm{viol.})}_{\mathrm I}(0) , Q_{\mathrm I} \Big]\!\!\:\bigg\} \notag \\ 
   =&\: i\hbar \mathlarger{\int}_0^1 \mathrm{d}s\mathlarger{\int}_0^\infty \mathrm{d}t_{\mathrm{rel}}\;\frac{\mathrm{d}}{\mathrm{d}t_\mathrm{rel}}\:\!\mathrm{tr}\bigg\{\!\!\:\Big(\rho_{\mathrm{I}}^{(\mathrm{LEQ})}\!\!\;(t)\!\!\:\Big)^{\!1-s} H^{(\mathrm{viol.})}_{\mathrm I}(-t_\mathrm{rel}) \Big(\rho_{\mathrm{I}}^{(\mathrm{LEQ})}\!\!\;(t)\!\!\:\Big)^{\!s}\Big[H^{(\mathrm{viol.})}_{\mathrm I}(0) , Q_{\mathrm I} \Big]\!\!\:\bigg\} \notag \\ 
   =&\:\! -\:\!\!i\hbar \mathlarger{\int}_0^1 \mathrm{d}s\; \mathrm{tr}\bigg\{\!\!\:\Big(\rho_{\mathrm{I}}^{(\mathrm{LEQ})}\!\!\;(t)\!\!\:\Big)^{\!1-s} H^{(\mathrm{viol.})}_{\mathrm I}(0) \Big(\rho_{\mathrm{I}}^{(\mathrm{LEQ})}\!\!\;(t)\!\!\:\Big)^{\!s}\Big[H^{(\mathrm{viol.})}_{\mathrm I}(0) , Q_{\mathrm I} \Big]\!\!\:\bigg\} \, .
   \label{eqn:getting_rid_of_H_cons_contribution}
\end{align}
In the last line, we have utilized that the contribution from the upper boundary at large relative time vanishes due to the microscopic decay of the correlator. Meanwhile, the equal-time term vanishes by cyclicity of the trace together with the change of variables $s\to 1-s$, again using that $Q_\mathrm{I}$ commutes with $\smash{\rho_{\mathrm{I}}^{(\mathrm{LEQ})}\!\!\;(t)}$. Thus, we end up with the intermediate relation 
\begin{align}
   \frac{\mathrm{d}}{\mathrm{d}t} \braket[\rbig]{Q_{\mathrm I}}^{\!\!\:t} = -\frac{\beta(t)\mu(t)\lambda^2}{\hbar^2} \mathlarger{\int}_0^1 \mathrm{d}s\mathlarger{\int}_0^\infty \mathrm{d}t_{\mathrm{rel}}\;\mathrm{tr}\bigg\{\!\!\:&\Big(\rho_{\mathrm{I}}^{(\mathrm{LEQ})}\!\!\;(t)\!\!\:\Big)^{\!1-s} \Big[Q_\mathrm{I},H_{\mathrm{I}}^{(\mathrm{viol.})}(0)\Big] \notag \\ 
   \times &\Big(\rho_{\mathrm{I}}^{(\mathrm{LEQ})}\!\!\;(t)\!\!\:\Big)^{\!s}\Big[H^{(\mathrm{viol.})}_{\mathrm I}(t_\mathrm{rel}) , Q_{\mathrm I} \Big]\!\!\:\bigg\} \Big[1+\mathcal{O}(\lambda)\Big]\, ,
   \label{eq:CoarseGrained_EvolutionEquation}
\end{align}
which remains valid for arbitrary $\beta(t)$ and $\mu(t)$, since no expansion around equilibrium has been invoked.\\

\noindent 
As a final remark, the result derived above may be obtained through several equivalent routes. Instead of starting from the finite-difference expression $\langle Q_{\mathrm I}\rangle^{t+\Delta t}-\langle Q_{\mathrm I}\rangle^{t}$, one may begin directly with the von Neumann equation~\eqref{eq:drhodt0} for the interaction-picture density matrix $\rho_{\mathrm I}(t)$, multiply by $Q_{\mathrm I}$, and subsequently take the trace. Because $Q_{\mathrm I}$ is time-independent in the interaction picture, all nontrivial time dependence is again encoded in the weakly charge-violating interaction Hamiltonian $\smash{H_\mathrm{I}^{(\mathrm{viol.})}(t)}$. Proceeding in this way, and again exploiting the hierarchy $\tau_{\mathrm{micro}}\ll\Delta t\ll\tau_{\mathrm{macro}}$, one finds that the resulting charge evolution is governed by the same microscopic correlators and becomes local in time for the same reason as in the derivation above: short-lived correlations decay within the interval $\Delta t$, whereas the macroscopic charge changes only weakly over it. This alternative route therefore recovers the same coarse-grained late-time differential equation~\eqref{eq:CoarseGrained_EvolutionEquation}.

\subsection{Linear relaxation regime}

As before, we finally restrict our analysis to the linear relaxation regime, characterized by $\big\lvert \beta_\mathrm{EQ}\mu(t)\big\rvert\ll 1$ and $\big\lvert\beta(t)-\beta_\mathrm{EQ}\big\rvert\ll \beta_\mathrm{EQ}$. Inspecting equation~\eqref{eq:CoarseGrained_EvolutionEquation} shows that the leading-order behavior is trivially attained by replacing the LEQ operator $\smash{\rho^{(\mathrm{LEQ})}_\mathrm{I}(t)}$ with its equilibrium counterpart\footnote{Note that, analogously to the LEQ operator~\eqref{eq:rhoIrhoLEQ}, the equilibrium operator $\smash{\rho^{(\mathrm{EQ})}_{\mathrm{I}}}$ is constructed from the charge-conserving Hamiltonian $\smash{H^{\mathrm{(cons.)}}_\mathrm{I}}$. In this way, the equilibrium density matrix is time independent and does not evolve in the interaction picture.}
\begin{equation} \label{eq:rhoIrhoEQ}
    \rho^{(\mathrm{EQ})}_{\mathrm{I}}= \frac{1}{Z_\mathrm{I}^{(\mathrm{EQ})}}\,\exp\!\bigg\{\!\!\:-\!\!\: \beta_\mathrm{EQ} H^{\mathrm{(cons.)}}_\mathrm{I}\!\!\:\bigg\}\, ,
\end{equation}
while simultaneously setting $\beta(t)\to\beta_\mathrm{EQ}$ in the overall prefactor, granting the relation 
\begin{align}
   \frac{\mathrm{d}}{\mathrm{d}t} \braket[\rbig]{Q_{\mathrm I}}^{\!\!\:t} = \frac{\beta_\mathrm{EQ}\mu(t)\lambda^2}{\hbar^2} \mathlarger{\int}_0^1 \mathrm{d}s\mathlarger{\int}_0^\infty \mathrm{d}t_{\mathrm{rel}}\;\mathrm{tr}\bigg\{\!\!\:&\Big(\rho^{(\mathrm{EQ})}_{\mathrm{I}}\Big)^{\!1-s} \Big[H_{\mathrm{I}}^{(\mathrm{viol.})}(0),Q_\mathrm{I}\Big] \label{eq:CoarseGrained_EvolutionEquation_LO}\\ 
   \times \!&\:\Big(\rho^{(\mathrm{EQ})}_{\mathrm{I}}\Big)^{\!s}\Big[H^{(\mathrm{viol.})}_{\mathrm I}(t_\mathrm{rel}) , Q_{\mathrm I} \Big]\!\!\:\bigg\} \cBig\{1+\mathcal{O}\cbig[\lambda,\mu(t),\Delta\beta(t)\cbig]\!\cBig\}\, . \notag 
\end{align} 
As before, we have abbreviated the (inverse) temperature difference $\beta(t)-\beta_{\mathrm{EQ}}$ by $\Delta \beta(t)$. At last, we derive the equivalent of equation~\eqref{eq:QNEQ_Propto_mu}, relating $\mu(t)$ to the expectation value $\braket[\rbig]{Q_{\mathrm I}}^{\!\!\:t}$. Since the derivation is identical to that in Zubarev's approach, let us only sketch the main steps. As in the NSO derivation, requiring the ansatz~\eqref{eq:rhoIrhoLEQ} to reproduce the conserved total energy of the system implies that the temperature difference $\beta(t)-\beta_\mathrm{EQ}$ is of order $\mu(t)^2$. At the order considered here, this may be implemented using $H^{(\mathrm{cons.})}$ as 
\begin{align}
    \braket[\cbig]{H_\mathrm{I}^{(\mathrm{cons.})}}_{\!\!\:\mathrm{LEQ}}^{\!\!\:t} = \braket[\cbig]{H_\mathrm{I}^{(\mathrm{cons.})}}_{\!\!\:\mathrm{EQ}} + \mathcal{O}(\lambda)\, ,
    \label{eq:Constraint_InteractionPic}
\end{align}
where we used the defining relation~\eqref{eq:Constraint_InteractionPic1} and the fact that the expectation value of the charge-conserving part of the Hamiltonian $\smash{H^{\mathrm{(cons.)}}_\mathrm{I}}$ agrees with the total energy $E_\mathrm{tot}$ up to corrections of order $\lambda$. Linearizing the LEQ operator as previously seen in equation~\eqref{eq:Expanded_LEQ_Operator} gives
\begin{align}
    \scalebox{0.985}{$\displaystyle{\rho_{\mathrm{I}}^{\,(\mathrm{LEQ})}(t)\!\!\;=\!\!\; \rho^{(\mathrm{EQ})}_\mathrm{I} \Bigg\{1}$}\:&\scalebox{0.985}{$\displaystyle{-\:\!\!\;\cBig[\Delta\beta(t) H^{(\mathrm{cons.})}_\mathrm{I}\!\!\;-\!\!\;\beta_\mathrm{EQ}\:\!\mu(t) Q_{\mathrm{I}}\cBig]}$} \label{eq:Expanded_LEQ_InteractionPic} \\[-0.15cm]
    &\scalebox{0.985}{$\displaystyle{+\:\!\!\;\braket[\cBig]{\Delta\beta(t) H^{(\mathrm{cons.})}_\mathrm{I}\!\!\;-\!\!\;\beta_\mathrm{EQ}\:\!\mu(t) Q_{\mathrm{I}}}_{\!\!\!\:\mathrm{EQ}} \!+\mathcal{O}\Big[\mu(t)^2,\Delta\beta(t)^2,\mu(t)\Delta\beta(t)\Big]\!\!\:\Bigg\} \:\! , }$}\notag
\end{align}
where we readily dropped the $s$-integral due to $Q_\mathrm{I}$ commuting with $H_\mathrm{I}^{(\mathrm{cons.})}$. By the same $\mathcal{CPT}$ arguments employed previously, one finds
\begin{equation}
    \braket[\cbig]{H^{(\mathrm{cons.})}_{\mathrm{I}}Q_{\mathrm{I}}}_{\mathrm{EQ}} = 0\, ,
\end{equation}
as discussed in appendix~\ref{sec:CPT_Arguments}. Inserting relation~\eqref{eq:Expanded_LEQ_InteractionPic} into the equality~\eqref{eq:Constraint_InteractionPic}, one thus arrives at the identity
\begin{align}
    \mathcal{O}(\lambda) &= \braket[\cbig]{H_\mathrm{I}^{(\mathrm{cons.})}}_{\!\!\:\mathrm{EQ}} - \braket[\cbig]{H_\mathrm{I}^{(\mathrm{cons.})}}_{\!\!\:\mathrm{LEQ}}^{\!\!\:t\:\!} \\[0.1cm] 
    &= \Delta\beta(t) \:\!\cbigg\{\braket[\cBig]{\Big[H_\mathrm{I}^{(\mathrm{cons.})}\Big]^{\!\!\:2}}_{\!\!\:\mathrm{EQ}} -\cBig[\braket[\cbig]{H_\mathrm{I}^{(\mathrm{cons.})}}_{\!\!\:\mathrm{EQ}} \cBig]^2\cbigg\} \: +\: \mathcal{O}\Big[\mu(t)^2,\Delta\beta(t)^2,\mu(t)\Delta\beta(t)\Big] \, .\notag
\end{align}
Since the expansion in $\mu(t)$ characterizes the late-time limit and is independent of the expansion in $\lambda$, the above relation implies that, at fixed order in $\lambda$, the temperature deviation $\Delta \beta(t)$ enters only at higher order in $\mu(t)$. This subsequently grants the simplified expansion
\begin{align}
    \rho_{\mathrm{I}}^{\,(\mathrm{LEQ})}(t)\!\!\;=\!\!\; \rho^{(\mathrm{EQ})}_\mathrm{I} \cBig\{1&+\beta_\mathrm{EQ}\:\!\mu(t) Q_{\mathrm{I}} +\mathcal{O}\cbig[\mu(t)^2\cbig]\!\cBig\} \, , 
\end{align}
which, upon employing equation~\eqref{eq:Constraint_InteractionPic0}, gives
\begin{align}
    \braket[\rbig]{Q_\mathrm{I}}^{\!\!\:t} = \braket[\rbig]{Q_\mathrm{I}}_{\mathrm{LEQ}}^{t} = \mathrm{tr}\Big[\rho^{(\mathrm{LEQ})}_{\mathrm{I}}\!\!\;(t)\:\!Q_\mathrm{I}\Big] = \beta_{\mathrm{EQ}}\mu(t)\, \braket[\rbig]{Q_{\mathrm I}^2}_\mathrm{EQ} + \mathcal{O}\cbig[\mu(t)^2\cbig] \, .
    \label{eq:Relation_mu_LEQ_approach}
\end{align}
In the last step, we utilized that the equilibrium expectation value of $Q_\mathrm{I}$ vanishes, which can again be established using $\mathcal{CPT}$ arguments. The given relation is the analogue of the previous equality~\eqref{eq:QNEQ_Propto_mu}, for which the $s$-integral could ultimately also be dropped to leading order in $\lambda$. Solving for $\mu(t)$ and inserting the arising relation into equation~\eqref{eq:CoarseGrained_EvolutionEquation_LO} finally yields the desired identity
\begin{align}
   \frac{\mathrm{d}}{\mathrm{d}t} \braket[\rbig]{Q_{\mathrm I}}^{\!\!\:t} = \frac{\lambda^2\braket[\rbig]{Q_{\mathrm I}}^{\!\!\:t}}{\hbar^2\braket[\rbig]{Q_{\mathrm I}^2}_\mathrm{EQ}} \mathlarger{\mathlarger{\int}}_0^1 \mathrm{d}s\mathlarger{\mathlarger{\int}}_0^\infty \mathrm{d}t_{\mathrm{rel}}\;\mathrm{tr}\cbigg\{\!\!\:&\Big(\rho^{(\mathrm{EQ})}_{\mathrm{I}}\Big)^{\!1-s} \Big[H_{\mathrm{I}}^{(\mathrm{viol.})}(0),Q_\mathrm{I}\Big] \label{eq:Relax_Equation_LEQ_Method}\\[-0.15cm] 
   \times \!&\:\Big(\rho^{(\mathrm{EQ})}_{\mathrm{I}}\Big)^{\!s}\Big[H^{(\mathrm{viol.})}_{\mathrm I}(t_\mathrm{rel}) , Q_{\mathrm I} \Big]\!\!\:\cbigg\} \Big[1+\mathcal{O}\cbig(\lambda,\braket[\rbig]{Q_{\mathrm I}}^{\!\!\:t}\:\!\cbig)\Big]\, . \notag 
\end{align}
Again, the leading-order relaxation equation takes the form previously observed in equation~\eqref{eq:RelaxEquation}, with $\Gamma_\mathrm{relax}$ given by
\begin{align}
   \Gamma_{\!\!\:\mathrm{relax}} = -\frac{\lambda^2}{\hbar^2\braket[\rbig]{Q_{\mathrm I}^2}_\mathrm{EQ}} \mathlarger{\mathlarger{\int}}_0^1 \mathrm{d}s\mathlarger{\mathlarger{\int}}_0^\infty \mathrm{d}t\;\mathrm{tr}\cbigg\{\:&\!e^{\scalebox{0.775}{$(s\!\!\:-\!\!\:1)\beta_\mathrm{EQ} H_\mathrm{I}^{(\mathrm{cons.})}$}} \!\!\;\Big[H_{\mathrm{I}}^{(\mathrm{viol.})}\!\!\:(0),Q_\mathrm{I}\Big] \label{eq:Relax_Rate_Alternative}\\[-0.15cm] 
   \times\, & e^{\scalebox{0.775}{$-s\beta_\mathrm{EQ} H_\mathrm{I}^{(\mathrm{cons.})}$}} \Big[H^{(\mathrm{viol.})}_{\mathrm I}\!\!\:(t) , Q_{\mathrm I} \Big] \!\!\:\cbigg\} \Big[1+\mathcal{O}(\lambda)\Big]\, .  \notag
\end{align}
With our convention of defining both LEQ and EQ operators using $\smash{H^{\mathrm{(cons.)}}_\mathrm{I}}$ rather than the full Hamiltonian $H_\mathrm{I}$, the resulting expression~\eqref{eq:Relax_Rate_Alternative} is manifestly of fixed order in $\lambda$. Nonetheless, we may freely replace every occurrence of $H_\mathrm{I}^{(\mathrm{cons.})}$ with $H_\mathrm{I}(0)$, only resulting in subleading corrections in $\lambda$. Together with the previously introduced charge-violation operator $J_\mathrm{I}(t)$, see equation~\eqref{eq:ChargeViolationOperator_J_Interaction}, one therefore arrives at the final result 
\begin{align}
   \Gamma_{\!\!\:\mathrm{relax}} = \mathrm{tr}\Big(Q_\mathrm{I}^2 e^{\scalebox{0.775}{$-\beta_\mathrm{EQ} H_\mathrm{I}(0)$}}\Big)^{\!-1}\mathlarger{\mathlarger{\int}}_0^1 \mathrm{d}s\mathlarger{\mathlarger{\int}}_0^\infty \mathrm{d}t\;\:\!\mathrm{tr}\bigg\{e^{\scalebox{0.775}{$(s\!\!\:-\!\!\:1)\beta_\mathrm{EQ} H_\mathrm{I}(0)$}} \!\!\;J_\mathrm{I}(0) \:\! e^{\scalebox{0.775}{$-s\beta_\mathrm{EQ} H_\mathrm{I}(0)$}}J_\mathrm{I}(t) \!\!\:\bigg\} \Big[1+\mathcal{O}(\lambda)\Big]\, .  \label{eq:Relaxation_Rate_InteractionPic}
\end{align}
With $Q_\mathrm{I}=Q_\mathrm{H}(0)$, $H_\mathrm{I}(0)=H$ and $J_\mathrm{I}(0)=J_\mathrm{H}(0)$, we may note that this expression is identical to equation~\eqref{eq:Relaxation_Rate_Intermediate}, with the sole difference being the appearance of $J_\mathrm{I}(t)$ instead of $J_\mathrm{H}(t)$. Due to the peaked nature of the given equilibrium correlator, the relative time $t$ is of order $\tau_\mathrm{micro}\sim \hbar \beta_\mathrm{EQ}$, such that we can infer the relation 
\begin{align}
    J_\mathrm{I}(t)=\frac{i\lambda}{\hbar} \Big[H^{(\mathrm{viol.})}_\mathrm{I}\!\!\;(t),Q_\mathrm{I}\Big] &= \exp\cBig(\frac{i}{\hbar}\:\!H^{\mathrm{(cons.)}}_{\!\!\;\mathrm{S}}\:\!t\cBig) \frac{i}{\hbar} \Big[\lambda H^{(\mathrm{viol.})}_\mathrm{S},Q_\mathrm{S}\Big] \exp\cBig(-\:\!\frac{i}{\hbar}\:\!H^{\mathrm{(cons.)}}_{\!\!\;\mathrm{S}}\:\!t\cBig) \notag \\ 
    &= \exp\cBig(\frac{i}{\hbar}\:\!H^{\mathrm{(cons.)}}_{\!\!\;\mathrm{S}}\:\!t\cBig) \frac{i}{\hbar} \Big[H,Q_\mathrm{S}\Big] \exp\cBig(-\:\!\frac{i}{\hbar}\:\!H^{\mathrm{(cons.)}}_{\!\!\;\mathrm{S}}\:\!t\cBig) \notag \\ 
    &=\exp\cBig(\frac{iHt}{\hbar}\cBig) \frac{i}{\hbar}\Big[H,Q_\mathrm{S}\Big] \exp\cBig(-\:\!\frac{iHt}{\hbar}\cBig) \Big[1+\mathcal{O}(\lambda t)\Big] \notag \\ 
    &= J_\mathrm{H}(t) \Big[1+\mathcal{O}\big(\lambda \tau_\mathrm{micro}\big)\Big]= J_\mathrm{H}(t) \Big[1+\mathcal{O}(\lambda)\Big]\, .
    \label{eq:Rewriting_Commutator}
\end{align}
Inserted into relation~\eqref{eq:Relaxation_Rate_InteractionPic}, we find complete accordance with the previous result~\eqref{eq:Relaxation_Rate_Intermediate}, and therefore equation~\eqref{eq:Final_Relaxation_Rate}.

\section{From relaxation to diffusion}
\label{sec:details}

Before turning to the applications of the relaxation rate~\eqref{eq:Final_Relaxation_Rate} illustrated in section~\ref{sec:applications}, it is instructive to establish its connection to the equilibrium diffusion rate $\gamma_\mathrm{diff}$. The latter characterizes the linear growth of the equilibrium mean-square displacement on timescales that are long compared with the microscopic correlation time $\tau_\mathrm{micro}$, but short compared with the macroscopic relaxation time $\tau_\mathrm{macro}$, being formally defined as
\begin{equation}
    \label{eqn:diffusion_rate_definition}
    \gamma_\mathrm{diff} = \lim_{\substack{t/\tau_\mathrm{micro}\to\infty\\ t/\tau_\mathrm{macro}\to 0\;\,}} \frac{1}{2Vt} \:\!\braket[\rBig]{\!\!\;\cbig[Q_\mathrm{H}(t)-Q_\mathrm{H}(0)\cbig]^{\!\!\:2}\:\!}_{\mathrm{EQ}} \, .
\end{equation}
Owing to the homogeneity of the system, the factor $V$ accounts for the extensive scaling of the correlation function with the spatial volume. In order to relate the diffusion rate $\gamma_\mathrm{diff}$ to the relaxation rate $\Gamma_\mathrm{relax}$, we first express the latter in terms of the power spectrum $S(\omega)$ of the two-time correlator $\braket[\rbig]{J_\mathrm{H}(t)J_\mathrm{H}(0)}_\mathrm{EQ}$. For completeness, and to make contact with formulations commonly encountered in the literature (see e.g. references~\cite{RabyMottola1990,Bodeker:2012gs,Hartnoll:2012rj,Bodeker:2014hqa}), we also provide equivalent expressions for the relaxation rate in terms of the retarded Green's function and the spectral function associated with the correlator $\braket[\rbig]{J_\mathrm{H}(t)J_\mathrm{H}(0)}_\mathrm{EQ}$.

\subsection{Frequency-space representations of the relaxation rate}
\label{sec:Frequency_Space_Relaxation}

Following the notation of section~\ref{sec:Representations_RelaxRate}, we express the relaxation rate~\eqref{eq:Reformulation_Zubarev} in terms of the charge-violating operator $J_\mathrm{H}(t)$, defined in equation~\eqref{eq:ChargeViolationOperator_J}. As we have briefly commented on below equation~\eqref{eq:Reformulation_Zubarev}, the given Bogoliubov inner product is symmetric under time reversal, as one can utilize the substitution $s\mapsto 1-s$ to switch the roles of $J_\mathrm{H}(t)$ and $J_\mathrm{H}(0)$. This allows us to conveniently represent the relaxation rate as a time integral over the full real line, granting the representation
\begin{align}
    \Gamma_{\mathrm{relax}} = \frac{1}{2\,\braket[\rbig]{Q_\mathrm{H}(0)^2}_\mathrm{EQ}}\:\mathlarger{\int}^{\infty}_{-\infty}\mathrm{d}t \mathlarger{\int}_{0}^1\mathrm{d}s \:
    \braket[\rBig]{J_\mathrm{H}\big(t-i\beta_\mathrm{EQ}\hbar s\big) J_\mathrm{H}(0)}_{\!\mathrm{EQ}}\, ,
    \label{eq:Final_Relaxation_Rate_symmetric}
\end{align}
where here and in the following, we omit the error estimate. Introducing the power spectrum $S(\omega)$ as the Fourier transform of the two-time equilibrium correlator
\begin{equation}
    \label{eqn:power_spectrum_definition}
    S(\omega) = \mathlarger{\int}_{-\infty}^\infty \mathrm{d}t \: e^{i\omega t} \braket[\cbig]{J_\mathrm{H}(t) J_\mathrm{H}(0)}_{\!\!\:\mathrm{EQ}} \,,
\end{equation}
we can represent the relaxation rate as the zero-frequency limit of the power spectrum\footnote{This relation establishes the previously stated equivalence between the expressions for the relaxation rate~\eqref{eq:Final_Relaxation_Rate} in terms of the Bogoliubov and the ordinary equilibrium inner products.}
\begin{align}
    \Gamma_{\mathrm{relax}} &= \frac{1}{2\,\braket[\rbig]{Q_\mathrm{H}(0)^2}_\mathrm{EQ}}\:\mathlarger{\int}^{\infty}_{-\infty}\mathrm{d}t \mathlarger{\int}_{0}^1\mathrm{d}s \mathlarger{\int}^{\infty}_{-\infty}\frac{\mathrm{d}\omega}{2\pi}\: e^{-i\omega t} e^{-\beta_\mathrm{EQ}\hbar \omega s} S(\omega) \notag \\ 
    &= \frac{1}{2\,\braket[\rbig]{Q_\mathrm{H}(0)^2}_\mathrm{EQ}}\: \mathlarger{\int}^{\infty}_{-\infty}\mathrm{d}\omega\: \frac{1-e^{-\beta_\mathrm{EQ}\hbar \omega}}{\beta_\mathrm{EQ}\hbar \omega} S(\omega) \left(\mathlarger{\int}^{\infty}_{-\infty}\frac{\mathrm{d}t}{2\pi}\: e^{-i\omega t} \right) \notag \\ 
    &= \frac{1}{2\,\braket[\rbig]{Q_\mathrm{H}(0)^2}_\mathrm{EQ}}\: \lim_{\omega \to 0}\cbigg\{ \frac{1-e^{-\beta_\mathrm{EQ}\hbar \omega}}{\beta_\mathrm{EQ}\hbar \omega} S(\omega)\cbigg\} \: = \frac{1}{2\,\braket[\rbig]{Q_\mathrm{H}(0)^2}_\mathrm{EQ}}\,\lim_{\omega \to 0} S(\omega)\, .
    \label{eq:Relaxation_PowerSpectrum}
\end{align}
Before demonstrating in section~\ref{sec:Relation_Relaxation_Diffusion} that the diffusion rate $\gamma_\mathrm{diff}$ attains a similar form, let us express the relaxation rate in terms of the retarded Green's function $G^{\mathrm{R}}(t)$ as well as the spectral density $\rho(\omega)$, which may be useful to some readers. We start with the former and define the retarded Green's function as
\begin{equation}
    G^{\mathrm{R}}(t) = - i\:\! \theta(t)  \braket[\rBig]{\cbig[ J_\mathrm{H}(t), J_\mathrm{H}(0) \cbig]} _{\!\!\:\mathrm{EQ}}\, , 
\end{equation}
where $\theta(t)$ denotes a Heaviside function. Note that we have already used time-translation invariance of the equilibrium ensemble to write the correlator as a function of a single time variable. Its Fourier transform satisfies 
\begin{align}
    \widetilde{G}^{\mathrm{R}}(\omega) =  \mathlarger{\int}_{-\infty}^\infty \mathrm{d}t \: e^{i\omega t} G^{\mathrm{R}}(t) &=  -i \mathlarger{\int}_0^\infty \mathrm{d}t \: e^{i\omega t} \cBig( \!\!\;\braket[\cbig]{J_\mathrm{H}(t) J_\mathrm{H}(0)}_{\!\!\:\mathrm{EQ}} - \braket[\cbig]{J_\mathrm{H}(0) J_\mathrm{H}(t)}_{\!\!\:\mathrm{EQ}}\cBig) \notag \\
    &= -i \mathlarger{\int}_0^\infty \mathrm{d}t \:  e^{i\omega t} \int_{-\infty}^\infty \frac{\mathrm{d}\omega'}{2\pi} \Big( e^{-i\omega' t} - e^{i\omega' t} \Big) S(\omega') \notag \\
    &= - \frac{i}{2} \Big[ S(\omega) - S(-\omega) \Big] + \mathcal{P}\mathlarger{\int}_{-\infty  }^\infty \frac{\mathrm{d}\omega'}{2\pi} \bigg[ \frac{S(\omega')}{\omega-\omega'} - \frac{S(\omega')}{\omega+\omega'} \bigg] \,,
\end{align}
where in the last line we utilized the Sokhotski--Plemelj formula in guise of 
\begin{align}
    \int_0^\infty \mathrm{d}t \: e^{i(\omega-\omega')t} = \pi\:\! \delta(\omega-\omega')+ i\:\!\mathcal{P}\cBig(\frac{1}{\omega-\omega'}\cBig)\, ,
\end{align}
with $\mathcal{P}$ denoting the Cauchy principal value. Because $J_\mathrm{H}(t)$ is Hermitian, the power spectrum $S(\omega)$ is real, such that taking the imaginary part of the Fourier-transformed retarded Green's function $\widetilde{G}^{\mathrm{R}}(\omega)$ selects only the first term. This grants the useful relation 
\begin{equation}
    \mathrm{Im} \cbig[\widetilde{G}^{\mathrm{R}}(\omega)\cbig] = \frac{S(-\omega) - S(\omega)}{2}  = \frac{e^{-\beta_{\mathrm{EQ}}\hbar \omega}-1}{2} S(\omega) \,,
    \label{eq:RetardedGreen_ImaginaryPart}
\end{equation}
where we employed the relation $S(-\omega)=e^{-\beta_{\mathrm{EQ}}\hbar\omega}S(\omega)$ to express the result entirely in terms of $S(\omega)$. This equality directly follows from the Kubo--Martin--Schwinger (KMS) relation for the equilibrium correlator $\braket[\rbig]{J_\mathrm{H}(t)J_\mathrm{H}(0)}_\mathrm{EQ}$, which in its spectral representation takes the form 
\begin{align}
    \braket[\cbig]{J_\mathrm{H}(t) J_\mathrm{H}(0)}_{\!\!\:\mathrm{EQ}} = \frac{1}{Z_\mathrm{EQ}} \: \mathlarger{\sum}_{n,m}\; e^{-\beta_\mathrm{EQ}E_m} e^{i(E_m-E_n)t/\hbar} \:\! \cbig\lvert \braket[\rbig]{m\!\!\:|J_\mathrm{H}(0)\!\!\:|n}\cbig\rvert^2\, ,
\end{align}
where $\ket{n}$ and $\ket{m}$ constitute complete sets of eigenstates of the Hamiltonian. Given this representation, the power spectrum attains the simple form 
\begin{align}
    S(\omega) &= \frac{2\pi}{Z_\mathrm{EQ}}\:  \mathlarger{\sum}_{n,m}\; e^{-\beta_{\mathrm{EQ}}E_m} \delta\cBig(\omega-\frac{E_n-E_m}{\hbar}\cBig)\cbig\lvert \braket[\rbig]{m\!\!\:|J_\mathrm{H}(0)\!\!\:|n}\cbig\rvert^2 \,.
\end{align}
Inverting the sign of the argument and flipping the role of $n$ and $m$, one arrives at the desired relation
\begin{align}
    S(-\omega) &= \frac{2\pi}{Z_\mathrm{EQ}}\:  \mathlarger{\sum}_{n,m}\; e^{-\beta_{\mathrm{EQ}}E_m} \delta\cBig(-\,\omega-\frac{E_n-E_m}{\hbar}\cBig)\cbig\lvert \braket[\rbig]{m\!\!\:|J_\mathrm{H}(0)\!\!\:|n}\cbig\rvert^2 \notag \\ 
    &= \frac{2\pi}{Z_\mathrm{EQ}}\:  \mathlarger{\sum}_{n,m}\; e^{-\beta_{\mathrm{EQ}}(E_n+\hbar \omega)} \delta\cBig(\omega-\frac{E_m-E_n}{\hbar}\cBig)\cbig\lvert \braket[\rbig]{m\!\!\:|J_\mathrm{H}(0)\!\!\:|n}\cbig\rvert^2 = e^{-\beta_{\mathrm{EQ}}\hbar \omega}S(\omega)\, .
\end{align}
Utilizing equality~\eqref{eq:RetardedGreen_ImaginaryPart} together with equation~\eqref{eq:Relaxation_PowerSpectrum}, one directly finds
\begin{equation}
\label{eq:Relaxation_RetardedGreen}
    \Gamma_\mathrm{relax} = \frac{1}{2\,\braket[\rbig]{Q_\mathrm{H}(0)^2}_\mathrm{EQ}}\,\lim_{\omega \to 0} S(\omega) =  -\frac{1}{\beta_\mathrm{EQ}\hbar\,\braket[\rbig]{Q_\mathrm{H}(0)^2}_\mathrm{EQ}}\,\lim_{\omega\to 0} \!\!\;\bigg\{\frac{1}{\omega} \,\mathrm{Im} \cbig[\widetilde{G}^{\mathrm{R}}(\omega)\cbig]\!\!\:\bigg\}\,.
\end{equation}
Lastly, let us turn to the spectral density $\rho(\omega)$, which is formally given by 
\begin{equation}
    \rho(\omega) = \mathlarger{\int}_{-\infty}^\infty \mathrm{d}t \: e^{i\omega t} \braket[\rBig]{\cbig[J_\mathrm{H}(t), J_\mathrm{H}(0)\cbig]}_{\!\mathrm{EQ}} = \cbig(1-e^{-\beta_{\mathrm{EQ}}\hbar \omega}\cbig) S(\omega) \,,
\end{equation}
where the relation to the power spectrum is derived in the same way as for the retarded Green's function, the only simplification being that the time integral readily extends over the entire real line. This lets us arrive at the desired relation
\begin{equation}
\label{eq:Relaxation_SpectralDensity}
    \Gamma_\mathrm{relax} = \frac{1}{2\beta_\mathrm{EQ}\hbar\,\braket[\rbig]{Q_\mathrm{H}(0)^2}_\mathrm{EQ}}\,\lim_{\omega \to 0} \frac{\rho(\omega)}{\omega}  \,.
\end{equation}

\subsection{The diffusion rate}
\label{sec:Relation_Relaxation_Diffusion}

We now turn to the diffusion rate $\gamma_\mathrm{diff}$, which, as we shall see, can be expressed in terms of the zero-frequency limit of the power spectrum, thereby establishing a relation between the diffusion and relaxation rates. Representing the charge difference entering definition~\eqref{eqn:diffusion_rate_definition} in terms of the operator $J_\mathrm{H}(t)$ as
\begin{equation}
    Q_\mathrm{H}(t) - Q_\mathrm{H}(0)  = \mathlarger{\int}_0^t \mathrm{d}t' \, \frac{\mathrm{d}Q_\mathrm{H}(t')}{\mathrm{d}t'} = \mathlarger{\int}_0^t \mathrm{d}t' \, J_H(t') \, ,
\end{equation}
we can express the charge variance in equilibrium as
\begin{align}
    \braket[\rBig]{\cbig[Q_\mathrm{H}(t)-Q_\mathrm{H}(0)\cbig]^{\!\!\:2}\:\!}_{\mathrm{EQ}} &=  \mathlarger{\int}_0^t \mathrm{d}t_1 \mathlarger{\int}_0^t \mathrm{d}t_2 \: \braket[\cbig]{J_\mathrm{H}(t_2) J_\mathrm{H}(t_1)}_{\!\!\:\mathrm{EQ}} \notag \\
    &=  \mathlarger{\int}_{-t}^t \mathrm{d}t_\mathrm{rel} \mathlarger{\int}_{\lvert t_\mathrm{rel}\rvert/2}^{t-\lvert t_\mathrm{rel}\rvert/2}\mathrm{d}t_\mathrm{avg} \: \braket[\rbigg]{\!\!\:J_\mathrm{H}\cBig(t_\mathrm{avg}+\frac{t_\mathrm{rel}}{2}\cBig) J_\mathrm{H}\cBig(t_\mathrm{avg}-\frac{t_\mathrm{rel}}{2}\cBig)\!\!\:}_{\!\mathrm{EQ}} \notag \\
    &=  \mathlarger{\int}_{-t}^{t} \mathrm{d}t_\mathrm{rel} \: \big(t-\lvert t_\mathrm{rel}\rvert \big) \braket[\cbig]{J_\mathrm{H}(t_\mathrm{rel}) J_\mathrm{H}(0)}_{\!\!\:\mathrm{EQ}}  \, ,
\end{align}
where we have transformed from the variables $(t_1,t_2)$ to the relative and average time coordinates $(t_\mathrm{rel},t_\mathrm{avg})$, defined as $t_\mathrm{rel}=t_2-t_1$ and $t_\mathrm{avg}=(t_1+t_2)/2$. The integration over $t_\mathrm{avg}$ becomes trivial by exploiting the time-translation invariance of the equilibrium correlator $\braket[\rbig]{J_\mathrm{H}(t_2) J_\mathrm{H}(t_1)}_{\mathrm{EQ}}$. In the limit $t/\tau_{\mathrm{micro}} \to\infty$, the factor $t-\lvert t_\mathrm{rel}\rvert$ may be replaced by $t$, since the dominant contribution to the integral arises from relative times $t_{\mathrm{rel}}\sim \tau_{\mathrm{micro}}$ due to the peaked nature of the two-time correlator, see equation~\eqref{eq:Behavior_TwoPointFunction}. Furthermore, the same argument justifies extending the integration bounds to $\pm\infty$, such that we arrive at\footnote{Strictly speaking, the limit considered here is an intermediate-time limit, $t/\tau_\mathrm{micro}\to \infty$ with $t/\tau_\mathrm{macro}\to 0$. The first condition allows the integration bounds to be extended to infinity, while the second ensures that the slow relaxation of $Q$, and hence the corresponding long-time contribution to the correlator, remains negligible. At fixed nonzero $\lambda$, taking $t\to\infty$ instead would probe the eventual relaxation of $Q$, for which the mean-square displacement saturates rather than growing linearly.}
\begin{align}
    \braket[\rBig]{\cbig[Q_\mathrm{H}(t)-Q_\mathrm{H}(0)\cbig]^{\!\!\:2}\:\!}_{\mathrm{EQ}} &\,\xrightarrow{\,t/\tau_\mathrm{micro}\:\!\to\:\!\infty\,}\, t \,\mathlarger{\int}_{-\infty}^{\infty} \mathrm{d}t_\mathrm{rel} \: \braket[\cbig]{J_\mathrm{H}(t_\mathrm{rel}) J_\mathrm{H}(0)}_{\!\!\:\mathrm{EQ}} = t \lim_{\omega\to 0} S(\omega) \, .
\end{align}
In the last line, we once more employed definition~\eqref{eqn:power_spectrum_definition} of the power spectrum. Comparing the above late-time limit to the definition~\eqref{eqn:diffusion_rate_definition} of the diffusion rate $\gamma_\mathrm{diff}$, we can read off the diffusion rate as 
\begin{align}
    \gamma_\mathrm{diff}= \frac{1}{2V}\lim_{\omega\to 0} S(\omega) \, .
\end{align}
Most notably, we recover the Einstein relation connecting the relaxation rate $\Gamma_\mathrm{relax}$ and the diffusion rate $\gamma_\mathrm{diff}$, taking the form
\begin{align}
\label{eqn:relaxation_diffusion_relation}
    \tcboxmath{\Gamma_{\mathrm{relax}} = \frac{V}{\braket[\rbig]{Q_\mathrm{H}(0)^2}_\mathrm{EQ}}\,\gamma_\mathrm{diff}\, .}
\end{align}

\subsection{A brief interlude on entropy production}
In section~\ref{sec:Relation_Relaxation_Diffusion}, we have derived the fluctuation-dissipation relation, which maps between the relaxation and the diffusion rate. 
The relation holds despite the system being away from thermal equilibrium, and it does so because of the reduction to local thermal equilibrium.
Yet, we can still quantify the deviation from equilibrium via a different quantity: entropy production.
As the system relaxes towards equilibrium, the entropy increases and computing the rate of production is rather straightforward within the local equilibrium approach. To this end, we define the entropy of the system
\begin{equation}
    S(t) = -k_\mathrm{B}\, \mathrm{tr} \Big\{\rho(t) \log\!\big[\rho(t)\big] \!\!\:\Big\} \,, \label{eqn:shannon_entropy_definition}
\end{equation}
as was previously done in equation~\eqref{eq:EntropyFunctional},
and consider the entropy production rate $\mathrm{d}S(t)/\mathrm{d}t$.
Before turning to the computation, we need to address a legit worry. If we could compute the full density matrix $\rho(t)$ evolved via the von Neumann equation, we would find that the entropy production rate is trivially zero.
In fact, this is just a different way to state that unitary evolution produces no irreversibility and no dissipation.
Thus, we should really understand the entropy as a quantity that depends on the coarse-graining procedure. When integrating out the fast modes and focusing on the slow modes of the system, we are effectively coarse-graining the time evolution on macroscopic timescales of order $\tau_{\mathrm{macro}}$ to obtain an effective description. The coarse-graining step, which in the approach of section~\ref{sec:3_LEQ_Method} coincides with taking $\tau_{\mathrm{micro}}$ to zero while keeping $\tau_{\mathrm{macro}}$ finite, is the step at which irreversibility is introduced in the description of the system.
The system appears dissipative only because we are not tracking what is happening to the fast (microscopic) modes, but only include their effect in an average sense. By including faster and faster modes in the description of the system, the dissipation smoothly vanishes, and with it the production of entropy.
Given that, we should understand the steps that follow in this light: the entropy production gives a measure of how dissipative the system appears at the macroscopic scales, and it provides us with useful thermodynamic relations.\\

\noindent
Using the definition of equation~\eqref{eqn:shannon_entropy_definition}, together with the local equilibrium ansatz~\eqref{eq:rhoIrhoLEQ} in the interaction picture, the instantaneous entropy reads
\begin{align}
	S(t) &\approx -k_\mathrm{B}\, \mathrm{tr} \cBig\{\rho^{(\mathrm{LEQ})}_{\mathrm{I}}(t) \log\cbig[\rho^{(\mathrm{LEQ})}_{\mathrm{I}}(t)\cbig] \!\cBig\} \notag \\ 
    &= k_\mathrm{B}\beta(t) \braket[\cbig]{H^{\mathrm{(cons.)}}_\mathrm{I}}^{\!t}_{\!\!\:\mathrm{LEQ}}- k_\mathrm{B}\beta(t) \mu(t) \braket[\rbig]{ Q_\mathrm{I}}^t_{\mathrm{LEQ}} + k_\mathrm{B}\log \cbig[Z_\mathrm{I}^{(\mathrm{LEQ})}\!\!\:(t)\cbig]\:\! ,
    \label{eqn:entropy_definition}
\end{align}
from which it follows that\footnote{Note that, even being fully agnostic about the meaning of $\beta(t)$ and $\mu(t)$, which are introduced merely as Lagrange multipliers, these relations can be used to identify them as the (instantaneous) inverse temperature and chemical potential.}
\begin{equation}
	\frac{\partial S(t)}{\partial \braket[\rbig]{H\!\!\:\raisebox{-2.75pt}{\scalebox{0.7}{I}}\raisebox{4.25pt}{\scalebox{0.7}{\!\!\:$\mathrm{(cons.)}$}}}^{\!\!\;t}_{\mathrm{LEQ}}} =k_\mathrm{B}\beta(t) \,, \qquad \mathrm{and} \qquad \frac{\partial S(t)}{\partial \braket[\rbig]{ Q_\mathrm{I}}^{\!\!\;t}_{\mathrm{LEQ}}} =- k_\mathrm{B}\beta(t) \mu(t) \,.
\end{equation}
The entropy production rate is obtained by taking a total time derivative of equation~\eqref{eqn:entropy_definition}, which reads
\begin{align}
	\frac{\mathrm{d} S(t)}{\mathrm{d} t} &=
	\frac{\partial S(t)}{\partial \braket[\rbig]{H\!\!\:\raisebox{-2.75pt}{\scalebox{0.7}{I}}\raisebox{4.25pt}{\scalebox{0.7}{\!\!\:$\mathrm{(cons.)}$}}}^{t}_{\mathrm{LEQ}}} \frac{\mathrm{d} \braket[\rbig]{H\!\!\:\raisebox{-2.75pt}{\scalebox{0.7}{I}}\raisebox{4.25pt}{\scalebox{0.7}{\!\!\:$\mathrm{(cons.)}$}}}^{t}_{\mathrm{LEQ}}}{\mathrm{d}t} + \frac{\partial S(t)}{\partial \braket[\rbig]{ Q_\mathrm{I}}^{\!\!\;t}_{\mathrm{LEQ}}} \frac{\mathrm{d} \braket[\rbig]{ Q_\mathrm{I}}^{\!\!\;t}_{\mathrm{LEQ}}}{\mathrm{d} t} \,,
    \label{eq:Entropy_Prod_Rate_1}
\end{align}
where it is easy to check that the terms generated by the derivative acting on $\beta(t)$ and $\beta(t)\mu(t)$ cancel exactly with the time derivative of $\smash{\log\cbig[Z_\mathrm{I}^{(\mathrm{LEQ})}\!\!\:(t)\cbig]}$. The first term in equation~\eqref{eq:Entropy_Prod_Rate_1} is subleading in the coarse-grained dynamics and starts only at $\mathcal{O}(\lambda^3)$.\footnote{This can be seen by repeating the mesoscopic evolution leading to equation~\eqref{eq:Expansion_QDifference}, now replacing $Q_\mathrm{I}$ by $H^{\mathrm{(cons.)}}_\mathrm{I}$. The contribution linear in $\lambda$ vanishes identically because $\smash{H^{\mathrm{(cons.)}}_\mathrm{I}}$ commutes with $\smash{\rho^{(\mathrm{LEQ})}_{\mathrm{I}}(t)}$. At quadratic order, the commutator with $H^{\mathrm{(cons.)}}_\mathrm{I}$ may be written as a time derivative of $H^{\mathrm{(viol.)}}_\mathrm{I}$, such that the resulting relative-time integral reduces, after integration by parts, to boundary terms. The equal-time contribution vanishes identically, while the contribution at large relative times is suppressed by the decay of microscopic correlations. The remaining finite-$\Delta t$ correction is therefore suppressed by $\tau_{\mathrm{micro}}/\Delta t$ and, with the choice $\Delta t\sim\lambda^{-1}\tau_{\mathrm{micro}}$ adopted above, contributes only at $\mathcal{O}(\lambda^3)$.} Hence, the energy term does not contribute at the leading $\mathcal{O}(\lambda^2)$ order considered here. As for the second term, to leading order in $\lambda$ and in deviations from equilibrium, we can utilize the previously derived relaxation equation~\eqref{eq:Relax_Equation_LEQ_Method} together with the linear relation~\eqref{eq:Relation_mu_LEQ_approach} between $\braket[\rbig]{Q_\mathrm{I}}^t$ and $\mu(t)$ to relate the entropy production rate to the relaxation rate $\Gamma_\mathrm{relax}$. One therefore arrives at the leading-order expression for the entropy production rate in local equilibrium, given by
\begin{align}
	\!\frac{\mathrm{d} S(t)}{\mathrm{d} t} &= k_\mathrm{B}\beta(t) \mu(t) \Gamma_\mathrm{relax}\braket[\rbig]{Q_\mathrm{I}}^{\!\!\;t} \bigg\{1+\mathcal{O}\Big(\!\!\;\braket[\rbig]{Q_{\mathrm{I}}}^{\!\!\;t},\lambda\Big)\!\!\:\bigg\} \\ 
    &= k_\mathrm{B}\Gamma_\mathrm{relax}\,\frac{\beta(t)}{\beta_\mathrm{EQ}} \frac{\cbig(\!\!\;\braket[\rbig]{Q_\mathrm{I}}^{\!\!\;t}\cbig)^{\!2}}{\braket[\rbig]{Q_{\mathrm I}^2}_\mathrm{EQ,cons.}} \bigg\{1+\mathcal{O}\Big(\!\!\;\braket[\rbig]{Q_{\mathrm{I}}}^{\!\!\;t},\lambda\Big)\!\!\:\bigg\} = k_\mathrm{B}\Gamma_\mathrm{relax}\,\frac{\cbig(\!\!\;\braket[\rbig]{Q_\mathrm{I}}^{\!\!\;t}\cbig)^{\!2}}{\braket[\rbig]{Q_{\mathrm I}^2}_\mathrm{EQ}} \bigg\{1+\mathcal{O}\Big(\!\!\;\braket[\rbig]{Q_{\mathrm{I}}}^{\!\!\;t},\lambda\Big)\!\!\:\bigg\}  \,. \notag 
\end{align}
In the last line, we again worked to leading order in the deviation from equilibrium $\braket[\rbig]{Q_{\mathrm{I}}}^{\!\!\;t}$, allowing us to set $\beta(t)=\beta_\mathrm{EQ}$. We also replaced the charge-conserving equilibrium average by an equilibrium average involving the full Hamiltonian $H$, which is more readily evaluated in practice.
As expected, we find that the entropy production rate is positive, in accordance with the second law of thermodynamics, and, in the limit $\lambda\to 0$, it goes smoothly to zero with the relaxation rate $\Gamma_\mathrm{relax}$. This structure is a standard consequence of linear irreversible thermodynamics: the relation between thermodynamic forces, relaxation, and entropy production goes back to Onsager~\cite{Onsager1931}, while microscopic quantum expressions in terms of canonical rate-rate correlation functions were obtained by Kubo, Yokota \& Nakajima~\cite{KuboYokotaNakajima1957} and, for weak symmetry-breaking interactions particularly close to the present setting, by Peletminskii \& Yatsenko~\cite{Peletminskii1968}.

\section{Examples}
\label{sec:applications}

\noindent 
Having derived the general relaxation formula~\eqref{eq:Final_Relaxation_Rate}, and its relation to the diffusion rate~\eqref{eqn:relaxation_diffusion_relation} we now illustrate how it can be used in two representative examples. First, we study a simple perturbative model containing a charged scalar field weakly coupled to a light neutral scalar, so that any initial charge is slowly depleted via decays into the neutral particle.
Because the decay is a perturbative process, the system can also be studied via semiclassical kinetic theory, and the charge decay rate can be extracted from the Boltzmann equations.
It therefore provides a direct check that the correlator formula derived above reproduces the standard kinetic-theory result in a setting where both computations can be carried out explicitly.\\

\noindent
Next, we consider the relaxation of the baryon number in the early universe, the physical setting where much of Zubarev's framework presented in sections~\ref{sec:2_Zubarev_Method} and~\ref{sec:Zubarev_Calculation} found its application~\cite{KhlebnikovShaposhnikov1988,RabyMottola1990}.
We show how the relaxation rate of the anomalous $B+L$ charge can be related to the diffusion rate of the topological Chern--Simons number, also called the sphaleron rate.

\subsection{Charged scalar fields: comparison with kinetic theory}
\label{sec:boltzmann}

Here, we consider a perturbative example for which the relaxation rate can be computed in two independent ways. The purpose of this subsection is to apply the formula derived in the previous sections to a weakly charge-violating scalar theory and to compare the result with the one obtained from a linearized Boltzmann equation. This provides a useful check of the formalism in a setting where all quantities entering the relaxation rate can be evaluated explicitly.

\subsubsection{Setup}
Throughout this section and appendix \ref{app:scalar_model_details}, we work in natural units, $\hbar=c=k_{B}=1$. We consider a massive complex scalar field $\chi$ coupled weakly to a massless real scalar field $\phi$. The dynamics is described by
\begin{equation} \label{eq:scalar_model_lagrangian}
\mathcal L = (\partial_\mu \chi)^\dagger (\partial^\mu \chi) -
m_\chi^2 \chi^\dagger \chi +\frac{1}{2} (\partial_\mu \phi) (\partial^\mu \phi) - \frac{1}{2}\Big(\lambda_\chi \chi+\lambda_\chi^* \chi^\dagger\Big)\phi^2\, .
\end{equation}

\noindent 
The field $\phi$ is neutral under the charge carried by $\chi$. Therefore, in the limit $\lambda_\chi \to 0$, the theory has a global $U(1)_\chi$ symmetry under which
\begin{equation}
\chi \to e^{i\alpha}\chi\, , \qquad \chi^\dagger \to e^{-i\alpha}\chi^\dagger\,. 
\end{equation}
The corresponding Noether charge is
\begin{equation}
Q_\chi = i\mathlarger{\int}\mathrm d^3\vec{x}\:\! \left[ \chi^\dagger (\partial_0\chi) - \big(\partial_0\chi^\dagger\big)\chi \right].
\label{eq:scalar_model_Qchi}
\end{equation}
A $\chi$ particle carries charge $+1$ and a $\chi$ antiparticle charge $-1$, so that $Q_\chi$ measures the particle-antiparticle asymmetry. For nonzero $\lambda_\chi$, the interaction allows the charge-violating processes
\begin{equation}
\chi \leftrightarrow \phi\phi, \qquad \chi^\dagger \leftrightarrow \phi\phi .
\end{equation}
For $|\lambda_\chi|/m_\chi\ll1$, the charge $Q_\chi$ is therefore weakly non-conserved and constitutes the slow variable of interest. We define the corresponding charge density as
\begin{equation}
n_\chi(t) \equiv \frac{1}{V}\:\!\braket[\rbig]{Q_\chi}^t\, .
\end{equation}
In the late-time linear relaxation regime, its evolution takes the form
\begin{equation}
\frac{\mathrm d}{\mathrm dt}\,n_\chi(t)
=-\:\!\Gamma_{\mathrm{relax}}^{(\chi)}\,n_\chi(t)\, .
\end{equation}
Our goal is to compute $\Gamma_{\mathrm{relax}}^{(\chi)}$ from the equilibrium correlator of the charge-violating operator and then show that the same expression follows from the kinetic description.\\

\noindent 
For the purpose of the comparison below, we restrict to the case in which the small $\chi$ asymmetry is parametrized by a single time-dependent chemical potential $\mu_\chi(t)$, independent of momentum. Both the correlator and Boltzmann calculations are evaluated within this common ansatz.\\

\noindent In order to match the notation used in the general derivation, we pass to the Hamiltonian formulation. The Hamiltonian is obtained from the usual Legendre transform. Since the interaction in equation~\eqref{eq:scalar_model_lagrangian} contains no time derivatives, its contribution to the Hamiltonian is minus the corresponding contribution to the Lagrangian. We therefore write
\begin{equation}
H=H^{(\mathrm{cons.})}+\lambda H^{(\mathrm{viol.})}\,,
\end{equation}
where $\lambda_{\chi}\equiv\lambda m_\chi\eta_\chi$, $\lambda=\frac{|\lambda_\chi|}{m_\chi}\ll1$, $\eta_\chi=\frac{\lambda_\chi}{|\lambda_{\chi}|}$,  and $|\eta_\chi|=1$. With this convention, $\lambda$ is a dimensionless real positive number, and it measures the magnitude of the charge-violating coupling in units of $m_\chi$, while $\eta_\chi$ contains its phase. The charge-violating part of the Hamiltonian is then
\begin{equation}
H^{(\mathrm{viol.})} = \frac{m_{\chi}}{2} \mathlarger{\int} \mathrm d^3\vec{x}\, \Big(\eta_\chi \chi+\eta_\chi^* \chi^\dagger\Big)\phi^2\, ,
\end{equation}
containing the weak charge-violating interaction. As in section~\ref{sec:3_LEQ_Method}, we work in the interaction picture generated by $H^{(\text{cons.})}$. We remind the reader that in this picture the charge operator is time-independent
\begin{equation}
    Q_{\chi,\mathrm{S}}=Q_{\chi,\mathrm{I}}\, ,
\end{equation}
while the charge-violating Hamiltonian becomes time dependent through the free evolution of the fields. More explicitly, 
\begin{equation}
H_{\mathrm I}^{(\mathrm{viol.})}(t) =\frac{m_{\chi}}{2} \mathlarger{\int} \mathrm d^3\vec{x}\, \Big[\eta_\chi \chi_{\mathrm I}(\vec{x},t) +\eta_\chi^* \chi^\dagger_{\mathrm I}(\vec{x},t)\Big] \phi^2_{\mathrm I}(\vec{x},t)\, .
\end{equation}

\noindent where, following standard convention, spatial vectors are denoted in boldface. Following the notation introduced in equation ~\eqref{eq:ChargeViolationOperator_J_Interaction}, we define the charge-violation operator
\begin{equation}
J_{\chi,\mathrm I}(t) \equiv i\lambda\Big[H_{\mathrm I}^{(\mathrm{viol.})}(t),Q_{\chi,\mathrm I}\Big]\, .
\end{equation}
Using the commutators
\begin{equation}
\cbig[\chi_{\mathrm I}(x),Q_{\chi,\mathrm I}\cbig]\:\!=\chi_{\mathrm I}(x)\,,
\qquad
\cbig[\chi^\dagger_{\mathrm I}(x),Q_{\chi,\mathrm I}\cbig]\:\!=-\chi^\dagger_{\mathrm I}(x)\,,
\end{equation}
one obtains
\begin{equation} \label{eq:JIxixi}
J_{\chi,\mathrm I}(t) = i\, \frac{\lambda  m_\chi}{2}\mathlarger{\int}\mathrm d^3\vec{x}\,\Big[ \eta_\chi \chi_{\mathrm I}(\vec{x},t) - \eta_\chi^* \chi^\dagger_{\mathrm I}(\vec{x},t) \Big] \phi^2_{\mathrm I}(\vec{x},t)\, .
\end{equation}
This is the operator whose equilibrium two-point function enters the relaxation formula~\eqref{eq:Final_Relaxation_Rate}.

\subsubsection{Relaxation rate from the correlator formula}

We first compute the relaxation rate using the correlator formula derived above. Specializing equation~\eqref{eq:Final_Relaxation_Rate} to the charge $Q_\chi$, we obtain
\begin{equation}
\Gamma_{\mathrm{relax}}^{(\chi)} = \frac{1}{\braket[\rbig]{Q_{\chi,\mathrm I}^{2}}_{\mathrm{EQ}}}\mathlarger{\int}_0^\infty \mathrm dt \mathlarger{\int}_0^1 \mathrm ds\; \braket[\cBig]{e^{\scalebox{0.8}{$s\beta_\mathrm{EQ}H^{(\mathrm{cons.})}_\mathrm{I}$}} J_{\chi,\mathrm I}(0) e^{\scalebox{0.8}{$-s\beta_\mathrm{EQ}H^{(\mathrm{cons.})}_\mathrm{I}$}} J_{\chi,\mathrm I}(t)}_{\!\mathrm{EQ}}\, .
\label{eq:scalar_model_Gamma_corr}
\end{equation}
Note that the correlator is of order $\lambda^2$, since each $J_{\chi,\mathrm I}$ is linear in $\lambda$. At leading order, the remaining expectation value can therefore be evaluated in the charge-conserving theory.\\

\noindent The denominator in equation~\eqref{eq:scalar_model_Gamma_corr} is the equilibrium fluctuation of the charge $Q_\chi$. For the free complex scalar field at vanishing chemical potential, this gives 
\begin{equation}
\frac{1}{V} \:\!\braket[\cbig]{Q_{\chi,\mathrm I}^{2}}_{\mathrm{EQ}} = 2 \mathlarger{\int} \frac{\mathrm d^3\vec{k}}{(2\pi)^3}\, f_\chi^{\mathrm{EQ}}\big(E^\chi_{\vec{k}}\big) \left[ 1+f_\chi^{\mathrm{EQ}}\big(E^\chi_{\vec{k}}\big) \right], \qquad E^\chi_{\vec{k}}=\sqrt{\vec{k}^2+m_\chi^2}\, .
\label{eq:Q2_scalarfield}
\end{equation}
Here,
\begin{equation}
f_\chi^{\mathrm{EQ}}(E) = \frac{1}{e^{\beta_{\mathrm{EQ}}E}-1}
\end{equation}
is the equilibrium Bose--Einstein distribution. The analogous notation $f_\phi^{\mathrm{EQ}}(E)$ is used below for the $\phi$ field. The derivation of equation~\eqref{eq:Q2_scalarfield}, including the factor of two from particles and antiparticles, is given in appendix~\ref{app:scalar_model_details}.\\

\noindent The numerator is obtained by inserting the explicit expression for $J_{\chi,\mathrm I}$ into equation~\eqref{eq:scalar_model_Gamma_corr}. At leading order, the required correlator is evaluated in the charge-conserving equilibrium theory. Its calculation reduces to free thermal contractions and is presented in appendix~\ref{app:scalar_model_details}. The result can be written as
\begin{align}
\frac{\Gamma_{\mathrm{relax}}^{(\chi)}}{V} = \frac{\lambda^2m_\chi^2} {\braket[\rbig]{Q_{\chi,\mathrm I}^{2}}_{\mathrm{EQ}}} \mathlarger{\int}\mathrm d\Pi_\chi(\vec{k})\, \mathrm d\Pi_\phi(\vec{p})\, \mathrm d\Pi_\phi(\vec{q})\, (2\pi)^4\,
\delta\cbig(E_{\mathbf{k}}^\chi-E_{\mathbf{p}}^\phi-E_{\mathbf{q}}^\phi\cbig)\,
\delta^{(3)}\!\big(\mathbf{k}-\mathbf{p}-\mathbf{q}\big) \;\; \notag \\[-0.15cm]
\times \:\!f_\chi^{\mathrm{EQ}}\big(E^\chi_{\vec{k}}\big) \left[ 1+f_\chi^{\mathrm{EQ}}\big(E^\chi_{\vec{k}}\big) \right]\!\!\: \left[ 1+f_\phi^{\mathrm{EQ}}\big(E^\phi_{\vec{p}}\big) +f_\phi^{\mathrm{EQ}}\big(E^\phi_{\vec{q}}\big) \right] \:\!. \label{eq:scalar_model_Gamma_phasespace}
\end{align}
We have used the Lorentz-invariant phase space
\begin{equation}
\mathrm d\Pi_\chi(\vec{k}) \equiv \frac{\mathrm d^3\vec{k}}{(2\pi)^3\,2E^\chi_{\vec{k}}}\, , \qquad \mathrm d\Pi_\phi(\vec{p}) \equiv \frac{\mathrm d^3\vec{p}}{(2\pi)^3\,2E^\phi_{\vec{p}}}\,,
\end{equation}
and analogously for $\mathrm d\Pi_\phi(\vec{q})$, with
\begin{equation}
E^\chi_{\vec{k}}=\sqrt{\vec{k}^{\,2}+m_\chi^2}\: , \qquad E^\phi_{\vec{p}}=|\vec{p}|\, .
\end{equation}

\subsubsection{Comparison with the Boltzmann equation}

We now derive the same relaxation rate from a kinetic description. In this language,  the same charge density is expressed in terms of the phase-space distributions of the $\chi$ particles and antiparticles, namely
\begin{equation}
\label{eq:scalarmodel_nchi}
n_\chi(t) = \mathlarger{\int}\frac{\mathrm d^3\vec{k}}{(2\pi)^3} \Big[ f_\chi(\vec{k},t)-f_{\overline\chi}(\vec{k},t) \Big] \:\! ,
\end{equation}
where $f_\chi$ and $f_{\overline\chi}$ are the phase-space distributions of the $\chi$ particles and antiparticles, respectively. We write the distribution functions for all fields in terms of deviations from equilibrium, namely
\begin{subequations}
\begin{align}
f_\phi(\vec{p},t)&= f_\phi^{\mathrm{EQ}}\big(E^\phi_{\vec{k}}\big) + \delta f_\phi (\vec p, t)\,, \\
f_\chi(\vec k,t) &= f_\chi^\mathrm{EQ}\big(E^\chi_{\vec k}\big) + \delta f_\chi (\vec k, t) \,, \\
f_{\overline\chi}(\vec k,t ) &= f_{\overline\chi}^\mathrm{EQ}\big(E^\chi_{\vec k}\big) + \delta f_{\overline\chi}(\vec k, t) \,.
\end{align}
\end{subequations}
As we shall see, at linear order in deviations from equilibrium, any dependence on $\delta f_\phi$ drops out. As for the $\chi$ field, within the kinetic-equilibrium regime specified above, the particle and antiparticle distributions are Bose--Einstein distributions with opposite chemical potentials,
\begin{equation}
f_\chi(\vec{k},t) = \frac{1}{e^{\beta_{\mathrm{EQ}}(E^\chi_{\vec{k}}-\mu_\chi(t))}-1}, \qquad f_{\overline\chi}(\vec{k},t) = \frac{1}{e^{\beta_{\mathrm{EQ}}(E^\chi_{\vec{k}}+\mu_\chi(t))}-1}.
\end{equation}
Expanding to linear order in $\big\vert \beta_{\mathrm{EQ}}\mu_\chi(t)\big\rvert\ll1$, we have $f_\chi^{\mathrm{EQ}}=f_{\overline\chi}^{\mathrm{EQ}}$ and
\begin{equation}
\label{eq:scalar_model_linearized_distributions}
    \delta f_{\chi/\overline{\chi}} (\vec k, t) = \pm \beta_{\mathrm{EQ}}\,\mu_\chi(t)\, f_\chi^{\mathrm{EQ}}\big(E^\chi_{\vec{k}}\big) \left[ 1+f_\chi^{\mathrm{EQ}}\big(E^\chi_{\vec{k}}\big) \right] \,.
\end{equation}
Substituting equation~\eqref{eq:scalar_model_linearized_distributions} into equation~\eqref{eq:scalarmodel_nchi}, one obtains
\begin{equation}
n_\chi(t) = 2\beta_{\mathrm{EQ}}\mu_\chi(t)
\mathlarger{\int}\frac{\mathrm d^3\vec{k}}{(2\pi)^3}\:\!
f_\chi^{\mathrm{EQ}}\big(E^\chi_{\vec{k}}\big) \left[
1+f_\chi^{\mathrm{EQ}}\big(E^\chi_{\vec{k}}\big) \right] .
\label{eq:scalar_model_nchi_mu_relation_boltzmann}
\end{equation}
Using equation~\eqref{eq:Q2_scalarfield}, the equation above can be written as
\begin{equation} 
\label{eq:linear_relation_nchi}
n_\chi(t) = \frac{\beta_{\mathrm{EQ}}\,\mu_\chi(t)}{V} \braket[\rbig]{Q_{\chi,\mathrm I}^{2}}_{\mathrm{EQ}}\, .
\end{equation}

\noindent We now want to compute the time derivative of the charge density in the kinetic description. Since $n_\chi(t)$ is written in terms of the particle and antiparticle distributions, its evolution is determined by the Boltzmann equations for $f_\chi$ and $f_{\overline\chi}$. For the homogeneous perturbation considered here, there are no spatial-gradient terms, and the Boltzmann equations reduce to
\begin{equation}
\frac{\mathrm d f_\chi(\vec{k},t)}{\mathrm dt} = C_\chi(\vec{k},t)\, , \qquad \frac{\mathrm d f_{\overline\chi}(\vec{k},t)}{\mathrm dt} = C_{\overline\chi}(\vec{k},t)\, ,
\end{equation}
where $C_\chi$ and $C_{\overline\chi}$ are the collision terms. They describe the change of the occupation numbers due to microscopic reactions. Taking the time derivative of equation~\eqref{eq:scalarmodel_nchi} then gives
\begin{equation}
\frac{\mathrm d n_\chi}{\mathrm dt} = \mathlarger{\int}\frac{\mathrm d^3\vec{k}}{(2\pi)^3} \Big[ C_\chi(\vec{k},t) - C_{\overline\chi}(\vec{k},t) \Big] .
\label{eq:scalar_model_dndt_c_terms}
\end{equation}
 At leading order in the charge-violating coupling, the relevant tree-level processes are $\chi(\vec k)\leftrightarrow\phi (\vec p) \phi (\vec q)$ and  $\chi^\dagger (\vec k)\leftrightarrow\phi (\vec p) \phi (\vec q)$.\\

\noindent For the particle channel, the collision term can be written as
\begin{equation}
C_\chi(\vec{k},t) = -\frac{\lambda^2m_\chi^2}{4E_{\vec{k}}^{\chi}} \mathlarger{\int}\mathrm d\Pi_\phi(\vec{p})\, \mathrm d\Pi_\phi(\vec{q})\, (2\pi)^4\,
\delta\cbig(E_{\mathbf{k}}^\chi-E_{\mathbf{p}}^\phi-E_{\mathbf{q}}^\phi\cbig)\,
\delta^{(3)}\!\big(\mathbf{k}-\mathbf{p}-\mathbf{q}\big)\, \Delta_\chi\big(\vec{k},\vec{p},\vec{q};t\big)\, ,
\label{eq:scalar_model_Cchi}
\end{equation}
where $\Delta_\chi$ is the difference between the statistical factor for removing a $\chi$ particle and the statistical factor for producing one 
\begin{align}
\Delta_\chi\big(\vec{k},\vec{p},\vec{q};t\big) &= f_\chi(\vec{k},t) \Big[ 1+f_\phi(\vec p, t) \Big] \!\!\;\Big[1+f_\phi(\vec q, t) \Big] - f_\phi(\vec p, t) f_\phi(\vec q, t) \Big[ 1+f_\chi(\vec{k},t) \Big].
\label{eq:scalar_model_Deltachi}
\end{align}
The antiparticle collision term is obtained analogously.
Substituting the particle and antiparticle collision terms into equation~\eqref{eq:scalar_model_dndt_c_terms}, one obtains
\begin{align}
\frac{\mathrm d n_\chi}{\mathrm dt} ={}-\frac{\lambda^2 m_\chi^2}{2} \int \mathrm d\Pi_\chi(\vec{k})\,\mathrm d\Pi_\phi(\vec{p})\,\mathrm d\Pi_\phi(\vec{q})\, (2\pi)^4 \delta\cbig(E_{\mathbf{k}}^\chi-E_{\mathbf{p}}^\phi-E_{\mathbf{q}}^\phi\cbig)\,
\delta^{(3)}\!\big(\mathbf{k}-\mathbf{p}-\mathbf{q}\big) \; \notag\\ 
\times \Big[ \Delta_\chi\big(\vec{k},\vec{p},\vec{q};t\big) -\Delta_{\overline\chi}\big(\vec{k},\vec{p},\vec{q};t\big) \Big]\:\! .
\label{eq:scalar_model_boltzmann_delta_difference}
\end{align}
The detailed derivation of the collision terms and of the following linearization is given in appendix~\ref{app:scalar_model_details}. Using equation~\eqref{eq:scalar_model_linearized_distributions}, the difference of statistical factors becomes
\begin{align}
\Delta_\chi-\Delta_{\overline\chi} = 2\beta_{\mathrm{EQ}}\mu_\chi(t)\, f_\chi^{\mathrm{EQ}}\big(E^\chi_{\vec{k}}\big) \Big[
1+f_\chi^{\mathrm{EQ}}\big(E^\chi_{\vec{k}}\big) \Big]\!\!\; \Big[ 1+f_\phi^{\mathrm{EQ}}\big(E^\phi_{\vec{p}}\big) +f_\phi^{\mathrm{EQ}}\big(E^\phi_{\vec{q}}\big) \Big] + \mathcal O\big(\mu_\chi^2\big).
\label{eq:scalar_model_delta_linearized}
\end{align}

\noindent Finally, using the result above together with the linear relation between the chemical potential and the charge density~\eqref{eq:linear_relation_nchi}, the relaxation equation~\eqref{eq:scalar_model_dndt_c_terms} reduces to
\begin{equation}
\frac{\mathrm d n_\chi}{\mathrm dt} = -\:\!\Gamma_{\mathrm{relax}}^{(\chi,\mathrm{Boltzmann})} n_\chi(t)\,,
\end{equation}
with
\begin{align}
\frac{\Gamma_{\mathrm{relax}}^{(\chi,\mathrm{Boltzmann})}}{V} &= \frac{\lambda^2m_\chi^2} {\braket[\rbig]{Q_{\chi,\mathrm I}^{2}}_{\mathrm{EQ}}} \mathlarger{\int} \mathrm d\Pi_\chi(\vec{k})\, \mathrm d\Pi_\phi(\vec{p})\, \mathrm d\Pi_\phi(\vec{q})\, (2\pi)^4\,
\delta\cbig(E_{\mathbf{k}}^\chi-E_{\mathbf{p}}^\phi-E_{\mathbf{q}}^\phi\cbig)\,
\delta^{(3)}\!\big(\mathbf{k}-\mathbf{p}-\mathbf{q}\big) \notag\\[-0.15cm]
&\hspace{3.2cm}\times f_\chi^{\mathrm{EQ}}\big(E^\chi_{\vec{k}}\big) \Big[ 1+f_\chi^{\mathrm{EQ}}\big(E^\chi_{\vec{k}}\big) \Big] \!\!\;\Big[ 1+f_\phi^{\mathrm{EQ}}\big(E^\phi_{\vec{p}}\big) +f_\phi^{\mathrm{EQ}}\big(E^\phi_{\vec{q}}\big) \Big].
\label{eq:scalar_model_Gamma_boltzmann}
\end{align}
This agrees with the result obtained from the correlator formula in equation~\eqref{eq:scalar_model_Gamma_phasespace}.

\subsection{$B+L$ relaxation in the early universe}
\label{sec:application_sphaleron}

\noindent
In section~\ref{sec:Relation_Relaxation_Diffusion}, we derived a relation between the late-time relaxation of a slowly relaxing charge and the diffusion coefficient associated with that same charge. We now apply this result to the anomalous electroweak $\mathrm{B+L}$ charge, which is of particular interest in early-Universe cosmology. Electroweak sphaleron transitions violate $\mathrm{B+L}$ through the chiral anomaly, while preserving $\mathrm{B-L}$, and therefore play a central role in many scenarios of baryogenesis~\cite{Kuzmin:1985mm,Shaposhnikov:1986jp,Nelson:1991ab,Carena:1996wj,Morrissey:2012db,Garbrecht:2018mrp,Benso:2026csa}. In particular, they can redistribute a pre-existing $\mathrm{B-L}$ asymmetry between the baryon and lepton sectors, as occurs in leptogenesis, where an asymmetry initially stored in the lepton sector is partially converted into a baryon asymmetry. At the same time, they tend to erase any $\mathrm{B+L}$ asymmetry that is not protected by a non-vanishing $\mathrm{B-L}$ charge. The relaxation rate studied below quantifies how efficiently this washout proceeds during the electroweak epoch.
The remainder of this section is a review of conventional results that were formally derived decades ago in the seminal work by Khlebnikov and Shaposhnikov~\cite{KhlebnikovShaposhnikov1988}.\\

\noindent 
We denote the spatially integrated anomalous charge by $Q_{\mathrm{B+L}}(t)$, which is obtained by summing the baryon and lepton charges carried by the Standard Model chiral fermions. Explicitly, we have
\begin{equation}
Q_{\mathrm{B+L}} = \sum_{i=1}^{\:\!\scalebox{0.75}{$n_{\mathrm{f}}$}} \Bigg\{ \frac{1}{3} \sum_{c=1}^{3} \sum_{a=1}^{2} Q_{\scalebox{0.75}{$[Q_{L}]_i^{a,c}$}} + \frac{1}{3} \sum_{c=1}^{3} Q_{\scalebox{0.75}{$[u_{R}]_i^{c}$}}+ \frac{1}{3} \sum_{c=1}^{3} Q_{\scalebox{0.75}{$[d_{R}]_i^{c}$}} + \sum_{a=1}^{2} Q_{\scalebox{0.75}{$[L_{L}]_i^{a}$}} + Q_{\scalebox{0.75}{$[e_{R}]_i$}} \Bigg\} .
\label{eq:charge_decomposition}
\end{equation}
Here $i=1,\ldots,n_{\mathrm{f}}$ labels the fermion generation, $a=1,2$ labels the weak-isospin components of the left-handed doublets, and $c=1,2,3$ labels color. The operators appearing in equation~\eqref{eq:charge_decomposition} denote the net particle minus antiparticle number of each independent chiral species~\cite{Harvey:1990qw}. Its late-time relaxation can be written as 
\begin{equation}
\frac{\mathrm d}{\mathrm dt} \braket[\rbig]{Q_{\mathrm{B+L}}}^{\!\!\:t} = -\Gamma_{\mathrm{B+L}}\, \braket[\rbig]{Q_{\mathrm{B+L}}}^{\!\!\:t} \, ,
\label{eq:BplusL_relaxation_equation}
\end{equation}
where $\Gamma_{\mathrm{B+L}}$ is the relaxation rate of the physical $\mathrm{B+L}$ charge.\\

\noindent The quantity usually quoted in the sphaleron literature is not $\Gamma_{\mathrm{B+L}}$ itself. Instead, sphaleron dynamics are characterized by the diffusion of the electroweak Chern--Simons number, with diffusion rate $\gamma_{\mathrm{CS}}$~\cite{KhlebnikovShaposhnikov1988,Moore:1996qs}. This rate measures the real-time diffusion of the gauge fields between neighboring topological sectors, and it is also known as the sphaleron rate, a well-studied quantity which can be extracted from lattice simulations~\cite{Bodeker:1999gx,Moore2000,DOnofrio:2014rug}. In this subsection, we show how the formalism derived above relates this topological diffusion rate to the relaxation rate of the physical $\mathrm{B+L}$ charge. \\

\noindent  
We note that in the presence of additional exactly conserved charges, the equilibrium density matrix may in general contain the corresponding chemical potentials. In the following, we retain $\mathrm{B-L}$ as the only additional conserved constraint and restrict, for simplicity, to an equilibrium state with $\mu_{\mathrm{B-L}}=0$.
We must modify the ansatz for the LEQ density matrix accordingly, to include the chemical potential $\mu_\mathrm{B-L}(t)$ and the charge $Q_{\mathrm{B-L}}$.
Then, the derivation of section~\ref{sec:3_LEQ_Method} goes through untouched all the way to equation~\eqref{eq:CoarseGrained_EvolutionEquation_LO}.
What is affected, however, is the linear relation~\eqref{eq:Relation_mu_LEQ_approach} between $\mu_\mathrm{B+L}(t)$ and $\braket{Q_\mathrm{B+L}}^t$, which must be updated by including the expansion in the small parameter $\beta_\mathrm{EQ}\mu_\mathrm{B-L}(t)$, namely
\begin{equation}
    \braket[\rbig]{Q_\mathrm{B+L}}^t = \beta_\mathrm{EQ} \mu_\mathrm{B+L}(t) \braket[\rbig]{Q_\mathrm{B+L}^2}_\mathrm{EQ} + \beta_\mathrm{EQ}\mu_\mathrm{B-L}(t) \braket[\cbig]{Q_\mathrm{B+L}Q_\mathrm{B-L}}_\mathrm{EQ} \,.
\end{equation}
The chemical potential $\mu_\mathrm{B-L}(t)$ can in turn be fixed by computing the expectation value of the associated $\mathrm{B-L}$ charge, which is conserved and vanishes throughout the evolution of the system. A quick calculation yields
\begin{align}
    \braket[\rbig]{Q_\mathrm{B-L}}^t & =   \beta_\mathrm{EQ}\mu_\mathrm{B-L}(t) \braket[\rbig]{Q_\mathrm{B-L}^2}_\mathrm{EQ} + \beta_\mathrm{EQ} \mu_\mathrm{B+L}(t) \braket[\cbig]{Q_\mathrm{B+L}Q_\mathrm{B-L}}_\mathrm{EQ} \overset{!}{=} 0 \, ,\notag \\[0.1cm]
    \Longrightarrow \; 
    \beta_\mathrm{EQ}\mu_\mathrm{B-L}(t) &=  - \beta_\mathrm{EQ}\mu_\mathrm{B+L}(t) \frac{\braket[\rbig]{Q_\mathrm{B+L}Q_\mathrm{B-L}}_\mathrm{EQ}}{\braket[\rbig]{Q_\mathrm{B-L}^2}_\mathrm{EQ}} \,.
\end{align}
This results in a modified linear relation, namely
\begin{equation}
    \braket[\rbig]{Q_\mathrm{B+L}}^t = \beta_\mathrm{EQ} \mu_\mathrm{B+L}(t) \cBigg[ \braket[\rbig]{Q_\mathrm{B+L}^2}_\mathrm{EQ} - \frac{\braket[\rbig]{Q_\mathrm{B+L}Q_\mathrm{B-L}}_\mathrm{EQ}^2}{\braket[\rbig]{Q_\mathrm{B-L}^2}_\mathrm{EQ}} \cBigg] \, + \: \mathcal{O}\cbig(\mu_\mathrm{B\pm L}^2\cbig) \,,
\end{equation}
and the resulting relaxation rate must be modified accordingly. 
In practice, we can apply the result of equation~\eqref{eq:Relax_Rate_Alternative}, where the denominator must be replaced as follows~\cite{Bodeker:2014hqa}
\begin{equation}
    \label{eq:projectedQ2}\braket[\rbig]{Q_{\mathrm{B+L}}^2}_\mathrm{EQ} \longrightarrow \mathcal V_{\mathrm{B+L}} \equiv \braket[\rbig]{Q_{\mathrm{B+L}}^2}_\mathrm{EQ} - \frac{ \braket[\rbig]{Q_{\mathrm{B+L}}Q_{\mathrm{B-L}}}_\mathrm{EQ}^{2} }{ \braket[\rbig]{Q_{\mathrm{B-L}}^2}_\mathrm{EQ} }\,.
\end{equation}
This amounts to projecting the thermodynamic normalization onto the subspace of fixed $\mathrm{B-L}$ number.\\

\noindent 
Applying equation~\eqref{eqn:relaxation_diffusion_relation} to $Q_{\mathrm{B+L}}$ with the corresponding normalization, one obtains
\begin{equation}
\Gamma_{\mathrm{B+L}} = \frac{V\gamma_{\mathrm{B+L}}}{\mathcal{V}_{\mathrm{B+L}}}\, ,
\label{eq:relaxation_BplusL_diffusion}
\end{equation}
where
\begin{equation}
\gamma_{\mathrm{B+L}} \equiv \lim_{\substack{t/\tau_\mathrm{micro}\to\infty\\ t/\tau_\mathrm{macro}\to 0\;\,}} \frac{1}{2Vt} \braket[\cBig]{\!\!\:\cbig[Q_{\mathrm{B+L}}(t)-Q_{\mathrm{B+L}}(0) \cbig]^{\!\!\:2}}_{\!\!\:\mathrm{EQ}}
\label{eq:gamma_BplusL_def}
\end{equation}
is the diffusion coefficient of the anomalous charge itself. Therefore, in order to express $\Gamma_{\mathrm{B+L}}$ in terms of the standard sphaleron rate $\gamma_\mathrm{CS}$, two ingredients are needed. The first is the projected equilibrium fluctuation $\mathcal{V}_{\text{B+L}}$, which fixes the thermodynamic normalization in equation~\eqref{eq:relaxation_BplusL_diffusion}. The second is the electroweak anomaly, which relates changes in $Q_{\mathrm{B+L}}$ to changes in the Chern--Simons number and hence relates $\gamma_{\mathrm{B+L}}$ to $\gamma_{\mathrm{CS}}$.

\subsubsection{Equilibrium fluctuation of the anomalous charge}
\label{sec:sphaleron_charge_fluctuation}

We first evaluate the projected equilibrium fluctuation that appears in equation~\eqref{eq:relaxation_BplusL_diffusion}. In the approximation in which the thermal bath, which we refer to as plasma, is described by free relativistic fermionic species, the $\mathrm{B+L}$ charge is obtained by summing the baryon and lepton charges carried by the Standard Model chiral fermions as defined in equation \eqref{eq:charge_decomposition}. \\

\noindent
At vanishing chemical potentials, different species are uncorrelated in this approximation. Therefore the equilibrium fluctuation of $Q_{\mathrm{B+L}}$ is obtained by summing the equilibrium fluctuations of the individual charges, weighted by the square of their $\mathrm{B+L}$ charge. For one independent massless chiral fermionic species $X$ carrying unit charge, in the relativistic limit and at leading order in interactions, the equilibrium fluctuation of its net charge is~\cite{LeBellac:1996}
\begin{equation}
\braket[\rbig]{Q_X^2}_{\mathrm{EQ}} = \frac{VT^3}{6}\, .
\end{equation}
In the Standard Model, it follows that
\begin{align}
\braket[\cbig]{Q_{\mathrm{B+L}}^2}_{\mathrm{EQ}}= n_{\mathrm{f}} \cbigg[ 6\cBig(\frac{1}{3}\cBig)^{\!\!\!\:2} + 3\cBig(\frac{1}{3}\cBig)^{\!\!\!\:2} + 3\cBig(\frac{1}{3}\cBig)^{\!\!\!\:2} + 2 + 1 \cbigg] \frac{VT^3}{6}=
\frac{13n_{\mathrm{f}}}{18}\,VT^3 \, .
\label{eq:charge_BplusL_fluctuation}
\end{align}
The factors in the square bracket count, respectively, the two weak components and three colors of the left-handed quark doublets, the three colors of the right-handed up quarks, the three colors of the right-handed down quarks, the two weak components of the left-handed lepton doublets, and the right-handed charged leptons. The same counting can be applied to $Q_{\mathrm{B-L}}$ and to the mixed correlator between $Q_{\mathrm{B+L}}$ and $Q_{\mathrm{B-L}}$.  This gives
\begin{align}
\braket[\cbig]{Q_{\mathrm{B-L}}^2}_{\mathrm{EQ}} = \frac{13n_{\mathrm{f}}}{18}\,VT^3\,, \qquad 
\braket[\cbig]{Q_{\mathrm{B+L}}Q_{\mathrm{B-L}}}_{\mathrm{EQ}} = -\frac{5n_{\mathrm{f}}}{18}\,VT^3\,.
\label{eq:charge_other_fluctuations}
\end{align}
Therefore, the thermodynamic normalization that enters the relaxation-diffusion relation~\eqref{eq:projectedQ2} reads
\begin{equation}
\mathcal V_{\mathrm{B+L}}= \frac{8n_{\mathrm{f}}}{13}\,VT^3\,.
\end{equation}

\subsubsection{Electroweak anomaly and Chern--Simons diffusion rate}
\label{sec:anomaly_CS_diffusion}

We now turn to the real-time ingredient entering equation~\eqref{eq:relaxation_BplusL_diffusion}. In the electroweak plasma, the microscopic processes responsible for the violation of $\mathrm{B+L}$ are transitions of the $SU(2)_L$ gauge fields between topologically distinct configurations~\cite{Manton:1983nd,Klinkhamer:1984di,ArnoldMcLerran1987}. These transitions are conventionally described in terms of the Chern--Simons number. Our goal is to relate the diffusion coefficient of the physical charge, $\gamma_{\mathrm{B+L}}$, to the Chern--Simons diffusion rate, $\gamma_{\mathrm{CS}}$.\\

\noindent 
We use the convention
\begin{equation}
q(x) = \frac{g^2}{32\pi^2}
F^A_{\!\!\:\mu\nu}(x)\widetilde F^{A\mu\nu}(x)\,,
\label{eq:sphaleron_topological_density}
\end{equation}
where $q(x)$ is the topological charge density of the electroweak $SU(2)_L$ gauge fields. Note that here and in the following the four-vector $x=(t,\vec x)$ denotes the spacetime coordinate. The electroweak chiral anomaly relates this density to the non-conservation of the $\mathrm{B+L}$ current ~\cite{tHooft:1976rip,Kuzmin:1985mm},
\begin{equation}
\partial_{\!\!\;\mu} \:\!j^\mu_{\mathrm{B+L}}(x) = 2n_{\mathrm{f}}\:\!q(x)\, .
\label{eq:sphaleron_anomaly_equation}
\end{equation}
 This relation allows us to rewrite the diffusion of $Q_{\mathrm{B+L}}$ in terms of a topological diffusion rate. The same topological density can be written as a total derivative,
\begin{equation}
q(x)=\partial_{\!\!\;\mu} K^\mu(x)\,,
\label{eq:sphaleron_topological_density_total_derivative}
\end{equation}
where $K^\mu$ is the Chern--Simons current. The corresponding Chern--Simons number is defined by
\begin{equation}
N_{\mathrm{CS}}(t) = \int \mathrm d^3\vec{x}\,K^0(\vec{x},t)\, ,
\label{eq:sphaleron_NCS_definition}
\end{equation}
Working in a finite spatial volume with periodic boundary conditions, akin to what is done on the lattice, the integral of the spatial divergence of $K^\mu$ vanishes. Therefore,  equation~\eqref{eq:sphaleron_topological_density_total_derivative} implies
\begin{equation}
\dot N_{\mathrm{CS}}(t) = \int \mathrm d^3\vec{x}\:q(\vec{x},t)\, ,
\label{eq:sphaleron_NCS_dot}
\end{equation}
where the overhead dot denotes a time derivative. On the other hand, integrating equation~\eqref{eq:sphaleron_anomaly_equation} over space gives
\begin{equation}
\dot Q_{\mathrm{B+L}}(t) = 2n_{\mathrm{f}} \int\mathrm d^3\vec{x}\:q(\vec{x},t)\, .
\label{eq:sphaleron_Qdot_BplusL}
\end{equation}
Combining equations~\eqref{eq:sphaleron_NCS_dot} and~\eqref{eq:sphaleron_Qdot_BplusL}, we obtain
\begin{equation}
\dot Q_{\mathrm{B+L}}(t) = 2n_{\mathrm{f}}\,\dot N_{\mathrm{CS}}(t)\qquad \Longrightarrow \qquad \Delta Q_{\mathrm{B+L}} = 2n_{\mathrm{f}}\,\Delta N_{\mathrm{CS}} \, .
\label{eq:sphaleron_deltaQ_deltaNCS}
\end{equation}
At finite temperature, sphaleron transitions make $N_{\mathrm{CS}}$ diffuse in real time ~\cite{Moore:1996qs,Arnold:1996dy,Bodeker:1998hm}. We define the Chern--Simons diffusion coefficient, in analogy with equation~\eqref{eqn:diffusion_rate_definition}, as the late-time growth rate of the mean squared change of $N_{\mathrm{CS}}$~\cite{Bodeker:1999gx}, 
\begin{equation}
\gamma_{\mathrm{CS}} \equiv \lim_{t\to\infty} \frac{1}{2Vt} \braket[\rBig]{\!\!\;\cbig[ N_{\mathrm{CS}}(t)-N_{\mathrm{CS}}(0) \cbig]^{\!\!\:2}}_{\mathrm{EQ}} \, .
\label{eq:sphaleron_gamma_CS_def}
\end{equation}
 Using equation~\eqref{eq:sphaleron_NCS_dot}, this rate may equivalently be written as the real-time spacetime integral of the topological charge-density correlator,
\begin{equation}
\gamma_{\mathrm{CS}} = \frac{1}{2} \int_{-\infty}^{\infty}\mathrm dt \int \mathrm d^3\vec{x}\, \braket[\cbig]{q(\vec{x},t)\,q(\vec{0},0)}_{\mathrm{EQ}} \, .
\end{equation}
In terms of the electroweak field strength this becomes\footnote{Our result differs from the one of references~\cite{Bodeker:1999gx,Moore2000,Burnier:2005hp} by a factor of $1/2$, which can be traced back to the definition of $\gamma_{\mathrm{CS}}$ in equation~\eqref{eq:sphaleron_gamma_CS_def}, which we choose to adhere to the conventional definition of a diffusion rate, as in equation~\eqref{eqn:diffusion_rate_definition}.}
\begin{align}
\gamma_{\mathrm{CS}} = \frac{1}{2}\left(\frac{g^2}{32\pi^2}\right)^{\!\!2} \mathlarger{\int}_{-\infty}^{\infty}\mathrm dt \mathlarger{\int}\mathrm d^3\vec{x}\, \left\langle F^A_{\!\!\:\mu\nu}(\vec{x},t)\widetilde F^{A\mu\nu}(\vec{x},t)
F^B_{\!\!\:\rho\sigma}(\vec{0},0)\widetilde F^{B\rho\sigma}(\vec{0},0)
\right\rangle_{\mathrm{EQ}} .
\end{align}
We note that the sphaleron rate is a real-time dynamical quantity and it must be obtained from the real-time dynamics of the electroweak gauge fields (see reference~\cite{Moore:2010jd} for a detailed discussion on this point), for instance through effective descriptions or non-perturbative methods that have been used in the sphaleron literature~\cite{DOnofrio:2014rug}.\\

\noindent We can now translate this topological diffusion rate into the diffusion coefficient of the anomalous charge. Comparing the definition of the $\mathrm{B+L}$ diffusion rate~\eqref{eq:gamma_BplusL_def} and of the topological diffusion rate~\eqref{eq:sphaleron_gamma_CS_def}, and using the linear relation between $\Delta Q$ and $\Delta N$ given in equation~\eqref{eq:sphaleron_deltaQ_deltaNCS}, we obtain
\begin{equation} \gamma_{\mathrm{B+L}} = 4n_{\mathrm{f}}^2\,\gamma_{\mathrm{CS}} .
\label{eq:gamma_BplusL_gammaCS}
\end{equation}
Within the simplified treatment adopted here, with $\gamma_{\mathrm{B+L}}$ being the intermediate-time diffusion coefficient defined in equation~\eqref{eq:gamma_BplusL_def}, the substitution of equation~\eqref{eq:gamma_BplusL_gammaCS} and the projected equilibrium fluctuation obtained above into equation~\eqref{eq:relaxation_BplusL_diffusion} yields
\begin{equation}
\Gamma_{\mathrm{B+L}} =  \frac{13 n_{\mathrm{f}}}{2 T^3}\, \gamma_{\mathrm{CS}} .
\label{eq:sphaleron_final_relaxation_rate}
\end{equation}
This is the desired relation between the relaxation rate of the physical $\mathrm{B+L}$ charge and the Chern--Simons diffusion rate conventionally used to characterize electroweak sphaleron transitions~\cite{Bochkarev:1987wf,Rubakov:1996,Cline:2000nw}.  More complete treatments of the chemical-equilibrium constraints replace this constant by a slightly more general thermodynamic coefficient, for which the prefactor derived here provides an excellent approximation~\cite{Burnier:2005hp}. Intrinsically, $\Gamma_{\mathrm{B+L}}$ is defined in terms of the real-time dynamics of the $\mathrm{B+L}$ charge, normalized by its projected equilibrium fluctuation. The anomaly relation allows the corresponding dynamical correlator to be expressed in terms of Chern--Simons number diffusion. Thus, $\gamma_{\mathrm{CS}}$ provides the non-perturbative real-time gauge dynamics, while $\mathcal V_{\mathrm{B+L}}$ determines how this diffusion translates into the physical relaxation rate.

\section{Transport in inhomogeneous systems}
\label{sec:transport}

\noindent
The local equilibrium framework discussed in section~\ref{sec:3_LEQ_Method} is not limited to the evolution of global quantities, such as the total charge, but can also be employed to derive dynamical equations for local densities.
In practice, this is equivalent to constructing an effective description of the slow macroscopic variables of the system which, as we shall see, results in a hydrodynamical description. 
This derivation is closely related to microscopic approaches to hydrodynamics based on local equilibrium~\cite{Hayata:2015lga,Mabillard2023}, and is complementary to modern Schwinger--Keldysh formulations of hydrodynamics~\cite{Liu:2018kfw,Akyuz:2023lsm,Firat:2025upx}.
In the following, we show how the local-equilibrium construction leads to a hydrodynamic description of the system, reproducing the equations of a perfect fluid in the energy--momentum sector and a diffusion--relaxation equation for an approximately conserved charge.

\subsection{Energy transport}
We start with an even simpler setting than what we have analyzed so far.
We consider a closed system, invariant under spatial translations, whose total energy is conserved at all times, and study the evolution of the energy density as it starts from a collection of inhomogeneous fluctuations and relaxes towards the equilibrium homogeneous configuration. 
The Hamiltonian of the system is
\begin{equation}
    H = \int \mathrm{d}^3\vec{x} \, h_\mathrm{H}(\vec{x}, t) \,,
\end{equation}
where $h_\mathrm{H}$ is the Hamiltonian density and its expectation value is the energy density $\varepsilon(\vec{x}, t) = \braket[\rbig]{h_\mathrm{H}(\vec{x}, t)}$. As before, spatial vectors are denoted in boldface. Although we restrict our discussion to three spatial dimensions, the derivation carries over essentially unchanged to an arbitrary spatial dimension $d$. We will briefly discuss the modifications to the final result in $d$ dimensions at the end of the computation. For convenience, we work in the Heisenberg picture, so that the density matrix $\rho_\mathrm{H}$ is fixed and does not evolve, allowing us to drop time superscripts on all matrix elements. 
The Hamiltonian density satisfies the following continuity equation
\begin{equation}
    \label{eqn:continuity_eq_h}
    \partial_t h_\mathrm{H}(\vec{x}, t) =\frac{i}{\hbar} \cbig[ H, h_\mathrm{H}(\vec{x}, t) \cbig] =- \nabla \cdot \vec{p}_\mathrm{H}(\vec{x}, t) \,,
\end{equation}
where $\vec{p}_\mathrm{H}$ is the energy density flux, or momentum density. The time derivative of the Hamiltonian density must be a total divergence exactly because the total energy is conserved, namely
\begin{equation}
    \frac{\mathrm{d}}{\mathrm{d}t} H = \frac{\mathrm{d}}{\mathrm{d}t} \int \mathrm{d}^3\vec{x} \, h_\mathrm{H}(\vec{x}, t) = - \int \mathrm{d}^3\vec{x}\;\nabla\cdot \vec{p}_\mathrm{H}(\vec{x}, t) =0 \,,
\end{equation}
where we used the fact that the integral of a total divergence over the whole space vanishes by Gauss' law.
Integrating over the momentum density we obtain the total three-momentum, which is conserved as a consequence of spatial translation invariance,
\begin{equation}
    \vec \Pi_\mathrm{H} = \int \mathrm{d}^3\vec x \, \vec p_\mathrm{H}(\vec x, t) \,.
\end{equation}
This implies that the momentum density $\vec p_\mathrm{H}$ is also a slow variable, just as the energy density $h_\mathrm{H}$, and it also satisfies a continuity equation
\begin{equation}
    \label{eqn:continuity_eq_p}
    \partial_t p^i_\mathrm{H}(\vec x, t) = - \partial_j T^{ij}_\mathrm{H}(\vec x, t) \,,
\end{equation}
where $T^{ij}_\mathrm{H}$ is a rank-two tensor that can be identified with the spatial components of the energy-momentum tensor.
In fact, these quantities can be conveniently collected into a symmetric energy-momentum tensor operator $T_{\mathrm H}^{\mu\nu}$,
\begin{equation}
    T^{\mu\nu}_\mathrm{H} =
    \begin{pmatrix}
        h_\mathrm{H} & p^i_\mathrm{H} \\
        p^j_\mathrm{H} & T^{ij}_\mathrm{H}
    \end{pmatrix},
\end{equation}
which satisfies the conservation law
\begin{equation}
    \partial_\mu T^{\mu\nu}_\mathrm{H}(\vec x, t) = 0 \,.
\end{equation}
It is easy to verify that this conservation law yields the two continuity equations~\eqref{eqn:continuity_eq_h} and~\eqref{eqn:continuity_eq_p} for the energy and momentum density.
Note that we can also define the four-momentum density $p^\mu_\mathrm{H} = T^{0\mu}_\mathrm{H} = \big(h_\mathrm{H}, \vec p_\mathrm{H}\big)$.\\

\noindent
Our goal is to extract the time-evolution equations for the four slow variables $\braket[\rbig]{h_\mathrm{H}(\vec x, t)}$ and $\braket[\rbig]{\vec p_\mathrm{H}(\vec x, t)}$, at linear order in deviations from equilibrium.
Because the density matrix in the Heisenberg picture is time-independent, both equations take on a very simple form 
\begin{subequations}
\begin{align}
    \partial_t \braket[\cbig]{h_\mathrm{H}(\vec x, t)} &= - \nabla \cdot \braket[\cbig]{\vec p_\mathrm{H}(\vec x, t)} \,, \label{eqn:energy_density_evolution}\\
    \partial_t \braket[\cbig]{\mathrm p_\mathrm{H}^i(\vec x, t)} &= - \:\! \partial_j \braket[\cbig]{T^{ij}_\mathrm{H} (\vec x, t)} \,. \label{eqn:momentum_density_evolution}
\end{align}
\end{subequations}
At this point, we introduce a local-equilibrium ensemble to evaluate the expectation values of the slow local observables, namely
\begin{align}
\rho_\mathrm{H}^{(\mathrm{LEQ})}(t) &= \frac{1}{Z^{(\mathrm{LEQ})}(t)} \exp \bigg\{\!- \!\int \mathrm{d}^3\vec{x} \, \beta(\vec{x}, t) u_\mu(\vec x, t) p^\mu_\mathrm{H}(\vec{x}, t) \bigg\} \notag \\
    &= \frac{1}{Z^{(\mathrm{LEQ})}(t)} \exp \bigg\{\!- \!\int \mathrm{d}^3\vec{x} \, \beta(\vec{x}, t) \gamma_\mathrm{pl}(\vec x, t) \Big[ h_\mathrm{H}(\vec{x}, t) - \vec v_\mathrm{pl}(\vec x,t ) \cdot \vec p_\mathrm{H}(\vec x, t) \Big] \bigg\} \,,
    \label{eqn:leq_density_matrix_local_relativistic}
\end{align}
where we have introduced the local (inverse) temperature $\beta(\vec x, t)$ as well as the plasma four-velocity $u_\mu$, which we parametrize in the inertial frame in which the homogeneous equilibrium state is at rest as $u^\mu = \gamma_{\mathrm{pl}} (1, \vec v_\mathrm{pl})$.
Here, $\vec{v}_{\mathrm{pl}}(\vec{x},t)$ is the local collective velocity of the medium; the subscript ``pl'' anticipates the plasma applications relevant to this work. Accordingly, $\gamma_\mathrm{pl} = \big(1 - \vec v_\mathrm{pl}^2\big)^{-1/2}$ is the plasma Lorentz factor. As in section~\ref{sec:3_LEQ_Method}, the use of the local-equilibrium ensemble relies on a separation between microscopic and macroscopic time scales. Over an interval $\tau_{\mathrm{micro}}\ll\Delta t\ll\tau_{\mathrm{macro}}$, the microscopic correlations that are not retained in $\smash{\rho_\mathrm{H}^{(\mathrm{LEQ})}(t)}$ lose their influence, while the local energy and momentum densities change only weakly. At each time, the local thermodynamic parameters $\beta(\vec x,t)$ and $\vec v_\mathrm{pl}(\vec x,t)$ are fixed by requiring the local-equilibrium ensemble to reproduce the instantaneous expectation values of the energy and momentum densities via the constraints
\begin{subequations}
\begin{align}
    \braket[\cbig]{h_\mathrm{H}(\vec x,t)}^t_\mathrm{LEQ}  &\overset{!}{=} \braket[\cbig]{h_\mathrm{H}(\vec x,t)} \,,  \label{eqn:constraint_energy_density}   \\
    \braket[\cbig]{\vec p_\mathrm{H}(\vec x,t)}^t_\mathrm{LEQ}  &\overset{!}{=} \braket[\cbig]{\vec p_\mathrm{H}(\vec x,t)} \,. \label{eqn:constraint_momentum_density}
\end{align}
\end{subequations}
In the Heisenberg picture, the exact density matrix remains time-independent; the time dependence of $\rho_\mathrm{H}^{(\mathrm{LEQ})}(t)$ reflects instead the slow evolution of these parameters.\\

\noindent Next, we proceed with the expansion around equilibrium. We define the deviation from the equilibrium (inverse) temperature $\delta\beta(\vec x, t) = \beta(\vec x, t) - \beta_\mathrm{EQ}$, and we expand for small deviations away from thermal equilibrium, namely $|\delta\beta(\vec x, t)|\ll\beta_\mathrm{EQ}$ and $|\vec v_\mathrm{pl}(\vec x, t)|\ll 1$, where $\vec v_{\mathrm{pl}}$ itself measures the velocity perturbation since the reference equilibrium state is at rest. Note that, to linear order in the plasma velocity, we can write $\gamma_\mathrm{pl}\simeq 1$. 
Expanding the density matrix, we have
\begin{align}
    \scalebox{0.965}{$\displaystyle{\rho_\mathrm{H}^{(\mathrm{LEQ})}(t) =  \rho_\mathrm{H}^{(\mathrm{EQ})} \Bigg\{ 1 }$}\,&\scalebox{0.965}{$\displaystyle{\:- \mathlarger{\int} \mathrm{d}^3\vec{y} \: \delta\beta(\vec{y}, t) \mathlarger{\int}_0^1\mathrm{d}s\, e^{s\beta_{\mathrm{EQ}}H} \delta h_\mathrm{H}(\vec{y}, t) e^{-s\beta_{\mathrm{EQ}}H} }$} \label{eq:rhoexpanded-transport}\\
    & \scalebox{0.965}{$\displaystyle{+ \:\beta_\mathrm{EQ} \mathlarger{\int} \mathrm{d}^3\vec y \mathlarger{\int}_0^1 \mathrm{d}s \: e^{s\beta_{\mathrm{EQ}}H} \Big[\vec v_\mathrm{pl}(\vec y, t) \cdot \vec p_\mathrm{H}(\vec{y}, t)\Big] e^{-s\beta_{\mathrm{EQ}}H} + \mathcal{O}\cbig(\delta\beta^2, \vec v_\mathrm{pl}^2,\delta\beta\,\vec v_\mathrm{pl}\cbig) \Bigg\}\,, }$} \notag
\end{align}
where we defined the deviation from equilibrium energy density $\delta h_\mathrm{H}(\vec x, t) = h_\mathrm{H}(\vec x, t) - \braket[\rbig]{h_\mathrm{H}(\vec{x}, t)}_{\mathrm{EQ}}$, and the equilibrium density matrix $\smash{\rho_\mathrm{H}^{(\mathrm{EQ})}=\exp(-\beta_{\mathrm{EQ}}H)/Z^{(\mathrm{EQ})}}$.\\

\noindent
Looking back at equations~\eqref{eqn:energy_density_evolution} and~\eqref{eqn:momentum_density_evolution}, we have three expectation values that we must express within the local equilibrium approach.
Evaluating the local-equilibrium expectation values in the matching conditions~\eqref{eqn:constraint_energy_density} and~\eqref{eqn:constraint_momentum_density} using the expansion~\eqref{eq:rhoexpanded-transport} gives, to linear order,
\begin{subequations}
\begin{align}
	\braket[\cbig]{\delta h_\mathrm{H}(\vec x, t)} &= - \mathlarger{\int} \mathrm{d}^3\vec y \: \delta\beta(\vec y, t) \mathlarger{\int}_0^1 \mathrm{d}s \, \braket[\rBig]{ e^{s\beta_{\mathrm{EQ}}H} \delta h_\mathrm{H}(\vec{y}, t) \:\! e^{-s\beta_{\mathrm{EQ}}H} \delta h_\mathrm{H}(\vec x, t)}_{\!\!\:\mathrm{EQ}} + \mathcal{O}\cbig(\delta\beta^2, \vec v_\mathrm{pl}^2\cbig) \,, \\
	\braket[\cbig]{\vec p_\mathrm{H}(\vec x, t)} &= \beta_\mathrm{EQ} \mathlarger{\int} \mathrm{d}^3\vec y \mathlarger{\int}_0^1 \mathrm{d}s \, \braket[\cBig]{ e^{s\beta_{\mathrm{EQ}}H} \Big[\vec v_\mathrm{pl}(\vec y, t) \cdot \vec p_\mathrm{H}(\vec{y}, t)\Big] e^{-s\beta_{\mathrm{EQ}}H} \vec p_\mathrm{H}(\vec x, t) }_{\!\mathrm{EQ}} + \mathcal{O}\cbig(\delta\beta^2, \vec v_\mathrm{pl}^2\cbig) ,
\end{align}
\end{subequations}
where we have already used that cross-terms of the form $\smash{\braket[\rbig]{\delta h_\mathrm{H}(\vec x, t) \vec p_\mathrm{H}(\vec y, t)}_\mathrm{EQ}}$ vanish since the momentum is odd under time reversal while the energy density is even. \\

\noindent To further simplify these relations, let us recognize that at equilibrium the system has a finite correlation length $\ell_{\mathrm{mfp}}$, which corresponds to the mean-free-path of a particle. It follows that both two-point correlators above are localized within a box of linear size $|\vec y - \vec x|\sim \ell_{\mathrm{mfp}}$.
This justifies expanding both hydrodynamic parameters around $\vec{x}$. Exemplifying for the temperature fluctuations, we have
\begin{equation}
    \delta\beta\big(\vec y, t\big) =\delta \beta(\vec{x}, t) + (\vec y- \vec x) \cdot \nabla \delta\beta (\vec{x}, t) + \mathcal{O}\big(\nabla^2 \delta\beta\big) \,.
    \label{eqn:derivative_exp_delta_beta}
\end{equation}
By assumption, $\delta\beta$ is a long wavelength mode, which means that it changes over a distance $\ell_{\delta\beta}$ much larger than the mean-free path $\ell_\mathrm{mfp}$. Consequently, its relative gradient scales roughly as $|\nabla\delta\beta|/|\delta \beta|\sim \ell_{\delta\beta}^{-1}$, and the first gradient term in the expansion is of order $\ell_{\mathrm{mfp}}/\ell_{\delta\beta}$, while higher-order derivative terms are further suppressed by higher powers of the same factor.
The same holds for the plasma velocity $\vec v_\mathrm{pl}$, which
varies over a characteristic spatial scale $\ell_\mathrm{v}$, with higher-order derivative terms suppressed by powers of $\ell_\mathrm{mfp}/\ell_\mathrm{v}$.
The scales $\ell_{\delta\beta}$ and $\ell_\mathrm{v}$ are set by the macroscopic perturbations and, as we shall see at the end of this subsection, determine the validity of the fluid description.
Working at leading order in the derivative expansion, we obtain the following two linear relations
\begin{subequations}
\begin{align}
	\braket[\cbig]{\delta h_\mathrm{H}(\vec x, t)} &= - \chi_{hh} \delta\beta(\vec x, t) + \mathcal{O}\cbig( \delta\beta^2, \vec v_\mathrm{pl}^2,\ell_\mathrm{mfp}/\ell_{\delta\beta} \cbig) \,, \label{eqn:relation_deltah_deltabeta}\\
	\braket[\cbig]{\vec p_\mathrm{H}(\vec x, t)} &= \chi_{\vec p \vec p} \vec v_\mathrm{pl}(\vec x, t) + \mathcal{O}\cbig(\delta\beta^2, \vec v_\mathrm{pl}^2, \ell_\mathrm{mfp}/\ell_\mathrm{v}\cbig) \,,
\end{align}
\end{subequations}
where we have introduced the energy and momentum susceptibilities,
\begin{subequations}
\begin{align}
	\chi_{hh} &= \frac{1}{V} \:\!\braket[\cbig]{\!\!\;\big(H-E\big)^{\!\!\:2}}_\mathrm{EQ} \,, \label{eqn:def_chi_hh} \\
	\chi_{\vec p \vec p} &= \frac{\beta_\mathrm{EQ}}{3V} \:\!\braket[\rbig]{\vec \Pi_\mathrm{H}^2}_\mathrm{EQ}\,.
\end{align}
\end{subequations}
In obtaining the expression for the conductivity, we have used the isotropy of the equilibrium ensemble to simplify $\braket[\cbig]{\Pi_\mathrm{H}^i \Pi_\mathrm{H}^j}_\mathrm{EQ} = \frac{1}{3}\:\!\delta^{ij} \braket[\rbig]{\vec \Pi_\mathrm{H}^2}_\mathrm{EQ}$.\\

\noindent
To close the momentum equation~\eqref{eqn:momentum_density_evolution}, we are left with evaluating the expectation value of the energy-momentum tensor
\begin{align}
	\braket[\cbig]{T^{ij}_\mathrm{H}(\vec x, t)} = 
    \braket[\cbig]{T^{ij}_\mathrm{H}(\vec x, t)}_\mathrm{EQ} &- \mathrm{\int} \mathrm{d}^3 \vec y \: \delta\beta(\vec y, t) \mathrm{\int}_0^1 \mathrm{d}s \, \braket[\rBig]{e^{s\beta_\mathrm{EQ}H}\delta h_\mathrm{H}(\vec y, t) e^{-s\beta_\mathrm{EQ}H} T^{ij}_\mathrm{H}(\vec x, t)}_\mathrm{EQ} \notag \\[0.05cm]
    & + \beta_\mathrm{EQ} \mathrm{\int} \mathrm{d}^3 \vec y \mathrm{\int}_0^1 \mathrm{d}s \, \braket[\cBig]{e^{s\beta_\mathrm{EQ}H} \Big[\vec v_\mathrm{pl}(\vec y, t) \cdot \vec p_\mathrm{H}(\vec y, t)\Big] e^{-s\beta_\mathrm{EQ}H} T^{ij}_\mathrm{H}(\vec x, t)}_{\!\mathrm{EQ}}  \notag \\[0.15cm] 
    &+ \mathcal{O} \big(\delta\beta^2,\vec v_\mathrm{pl}^2 \big)  \,. 
\end{align}
The last term vanishes identically because the momentum operator is odd under time-reversal, while $T^{ij}_\mathrm{H}$ is even.
As for the first term, it amounts to $\smash{\braket[\rbig]{T^{ij}_\mathrm{H}(\vec x, t)}_\mathrm{EQ} = \delta^{ij} \braket[\rbig]{P_\mathrm{H}(\vec x, t)}_\mathrm{EQ} = \delta^{ij} \mathcal{P}_\mathrm{EQ}}$, where we used isotropy of the equilibrium ensemble. Here, $P_\mathrm{H}(\vec x,t)\equiv \frac{1}{3}\:\!T^{ii}_\mathrm{H}(\vec x,t)$ denotes one third of the spatial trace of the energy-momentum tensor, and its equilibrium expectation value $\mathcal{P}_\mathrm{EQ}\equiv \braket[\rbig]{P_\mathrm{H}(\vec x,t)}_\mathrm{EQ}$ is the equilibrium pressure.
Because $\mathcal{P}_\mathrm{EQ}$ is spatially homogeneous, it drops out when acted on by the spatial derivative $\partial_j$ and does not contribute to the dynamics.
We can then focus on the second term, which contains the leading contribution to the momentum transport equation.
Working at leading order, we can drop all derivative terms, and $\delta\beta(\vec x, t)$ can be pulled out of the $\vec y$-integral, which now only acts onto $\delta h$,
\begin{align}
    \mathlarger{\int} \mathrm{d}^3 \vec y \: \delta\beta(\vec y, t) & \mathlarger{\int}_0^1 \mathrm{d}s \, \braket[\cBig]{e^{s\beta_\mathrm{EQ}H}\delta h_\mathrm{H}(\vec y, t) e^{-s\beta_\mathrm{EQ}H} T^{ij}_\mathrm{H}(\vec x, t)}_{\!\!\:\mathrm{EQ}}  = \notag \\
    &= \delta\beta(\vec x, t) \mathlarger{\int}_0^1 \mathrm{d}s \, \braket[\rbigg]{e^{s\beta_\mathrm{EQ}H} \bigg[\!\!\:\int \mathrm{d}^3 \vec y \: \delta h_\mathrm{H}(\vec y, t) \bigg] e^{-s\beta_\mathrm{EQ}H} T^{ij}_\mathrm{H}(\vec x, t)\!\!\:}_{\!\mathrm{EQ}} \notag \\
    &= \delta\beta(\vec x, t) \braket[\rBig]{\!\!\:\big(H-E\big) T^{ij}_\mathrm{H}(\vec x, t)}_{\!\!\:\mathrm{EQ}}
    = \delta\beta(\vec x, t) \:\!\delta^{ij} \braket[\rBig]{\!\!\:\big(H-E\big)P_\mathrm{H}(\vec x, t)}_{\!\!\:\mathrm{EQ}} \,.
\end{align}
We recall that $E=\braket{H}_\mathrm{EQ}$ is the total energy of the system. Collecting the equilibrium and linear contributions, we obtain
\begin{equation}
    \braket[\cbig]{T_\mathrm{H}^{ij}(\vec x,t)} =    \delta^{ij}\mathcal{P}_\mathrm{EQ} -     \delta\beta(\vec x,t)\,\delta^{ij} \braket[\rBig]{\!\!\:\big(H-E\big)P_\mathrm{H}(\vec x,t)}_{\!\!\:\mathrm{EQ}} + \cdots \,.
\end{equation}
The remaining quantity is an equilibrium expectation value of a composite operator computed at a single space-time coordinate, and by space and time-translational invariance, it is independent of the coordinate.
Let us massage it into a more familiar form
\begin{align}
    \braket[\cbig]{\big(H-E\big)P_\mathrm{H}}_\mathrm{EQ} &=  -
    \frac{\partial \braket{P_\mathrm{H}}_\mathrm{EQ}}{\partial \beta_\mathrm{EQ}} = - \underbrace{\frac{\partial \braket{P_\mathrm{H}}_\mathrm{EQ}}{\partial \braket{h_\mathrm{H}}_\mathrm{EQ}}}_{\displaystyle{=c_s^2}} \underbrace{\frac{\partial \braket{h_\mathrm{H}}_\mathrm{EQ}}{\partial \beta_\mathrm{EQ} \vphantom{\braket{}_\mathrm{EQ}}}}_{\displaystyle{=-\chi_{hh}\vphantom{c_s^2}}} \,,
\end{align}
where $c_s^2$ is the plasma sound-speed and the energy susceptibility $\chi_{hh}$ has been defined in equation~\eqref{eqn:def_chi_hh}. As we show below, $c_s$ is precisely the speed at which perturbations of energy and momentum travel through the medium.
Using the momentum continuity equation~\eqref{eqn:momentum_density_evolution}, we can then write the linear dynamical equation for the transport of momentum density
\begin{equation}
    \partial_t \:\! \braket[\cbig]{\vec p_\mathrm{H}(\vec x, t)} = c_s^2 \!\:\chi_{hh} \nabla \delta\beta(\vec x, t) = - \nabla \braket[\cbig]{P_\mathrm{H}(\vec x, t) } \,.
\end{equation}
Using the linear relation~\eqref{eqn:relation_deltah_deltabeta} to eliminate $\delta\beta$ from the momentum equation above, and recalling the energy continuity equation~\eqref{eqn:energy_density_evolution}, we arrive at the two transport equations for the system
\begin{subequations}
\begin{align}
    \partial_t \!\:\braket[\cbig]{\delta h_\mathrm{H}(\vec x, t)} &= - \nabla \cdot \braket[\cbig]{\vec p_\mathrm{H}(\vec x, t)} \,, \\
    \partial_t \!\:\braket[\cbig]{\vec p_\mathrm{H}(\vec x, t)} &= - c_s^2 \nabla \braket[\cbig]{\delta h_\mathrm{H}(\vec x, t)} \,. \label{eqn:linear_eq_momentum}
\end{align}
\end{subequations}
Differentiating both equations with respect to time, we obtain two decoupled second-order equations
\begin{subequations}
\begin{align}
    \partial_t^2 \braket[\cbig]{\delta h_\mathrm{H}(\vec x, t)} &= c_s^2 \!\:\nabla^2 \braket[\cbig]{\delta h_\mathrm{H}(\vec x, t)} \,, \label{eqn:sound_eq_energy_density}\\
    \partial_t^2 \braket[\cbig]{\vec p_\mathrm{H}(\vec x, t)} &= c_s^2 \!\:\nabla \Big[ \nabla\cdot \braket[\cbig]{\vec p_\mathrm{H}(\vec x, t)} \Big] \,.
\end{align}
\end{subequations}
From the first equation, we learn that energy density fluctuations propagate in waves at the speed of sound $c_s$.
As for the second one, further manipulations are needed.
First, we can write 
\begin{equation}
    \nabla \Big[ \nabla\cdot \braket[\cbig]{\vec p_\mathrm{H}(\vec x, t)} \Big] = \nabla^2 \braket[\cbig]{\vec p_\mathrm{H} (\vec x, t)} + \nabla \times \Big[\nabla \times \braket[\cbig]{\vec p_\mathrm{H}(\vec x, t)}\Big] \,.
\end{equation}
Then, let us split the momentum density into a purely longitudinal and purely transverse part, namely
\begin{align}
    \braket[\cbig]{\vec p_\mathrm{H} (\vec x, t)} = \braket[\cbig]{\vec p_\mathrm{H}^\mathrm{L}(\vec x, t)} + \braket[\cbig]{\vec p_\mathrm{H}^\mathrm{T}(\vec x, t) } \,, \quad \mathrm{with} \quad \cBigg\{\:\!\begin{matrix}
        \nabla\times \braket[\cbig]{\vec p_\mathrm{H}^\mathrm{L}(\vec x, t)} = 0 \\[0.2cm]
        \;\;\nabla \cdot \braket[\cbig]{\vec p_\mathrm{H}^\mathrm{T}(\vec x, t) } = 0
    \end{matrix} \; .
\end{align}
From the first-order equation~\eqref{eqn:linear_eq_momentum} for the momentum, we observe that the time-varying part of the momentum density is purely longitudinal.
Two equations follow
\begin{subequations}
\begin{align}
    \partial_t^2 \:\!\braket[\cbig]{\vec p_\mathrm{H}^\mathrm{L}(\vec x, t)} &= c_s^2 \!\:\nabla^2 \braket[\cbig]{\vec p_\mathrm{H}^\mathrm{L}(\vec x, t)} \,, \\
    \partial_t \,\braket[\cbig]{\vec p_\mathrm{H}^\mathrm{T}(\vec x, t) } &= 0 \,.
\end{align}
\end{subequations}
We learn that, at this order, transverse momentum excitations are frozen and do not propagate, while longitudinal momentum modes carry all the dynamics. The longitudinal modes describe compression-decompression waves, namely sound waves, and the equation is redundant with equation~\eqref{eqn:sound_eq_energy_density} for the energy density fluctuations.\\

\noindent
The general solution of the wave equation can be written as a sum over plane-waves of momentum $|\vec k|=2\pi/\lambda_\mathrm{k}$, where $\lambda_\mathrm{k}$ is the wavelength.
Because our derivation relies on an expansion in spatial gradients, this fluid description is only valid for wavelengths $\lambda_\mathrm{k}\gg \ell_\mathrm{mfp}$, namely as the wavelength approaches the microscopic correlation scale, the gradient expansion breaks down and the current description is no longer valid.
A posteriori, this fixes constraints on the allowed values of $\ell_{\delta\beta}$ and $\ell_\mathrm{v}$, which must be large enough for the fluid description to be valid.
Finally, let us observe that the equations we obtained describe linear hydrodynamics in an \emph{ideal} fluid.
At the next order in derivatives we would see dissipation and viscosity arising, yielding the description of a real fluid.

\subsection{Charge transport}

\noindent 
Having illustrated how the local equilibrium approach can be employed to obtain dynamical equations for local conserved densities, we now turn to the case of interest of this work, namely the dynamics of an approximately conserved charge density $q$. For this calculation, it is convenient to return to the interaction picture used in section~\ref{sec:3_LEQ_Method}, since the charge-conserving dynamics is carried by the operators, while charge violation enters through the evolution of the density matrix. This allows us to treat transport and washout separately. In this picture, we write the global charge operator $Q_\mathrm{I}$ in terms of the charge density as
\begin{equation}
    Q_\mathrm{I} = \int \mathrm{d}^3 \vec{x} \: q_\mathrm{I}(\vec{x}, t)\,.
\end{equation}
Because in the interaction picture the total charge is time-independent, the charge density satisfies a continuity equation, just as the Hamiltonian density did in the Heisenberg picture. We have
\begin{equation}
    \frac{\mathrm{d}}{\mathrm{d}t} \:\! q_\mathrm{I}(\vec{x},t) = \frac{i}{\hbar} \Big[ H_\mathrm{I}^{(\mathrm{cons.})},q_{{\mathrm{I}}}(\vec{x},t) \Big] \equiv - \nabla \cdot \vec{j}_\mathrm{I}(\vec{x}, t)\,,
\end{equation}
with $\vec{j}_\mathrm{I}$ the charge current in the interaction picture.
We can quickly check
\begin{align}
    \frac{\mathrm{d}}{\mathrm{d}t} \:\!Q_\mathrm{I} &= \mathlarger{\int} \mathrm{d}^3\vec{x} \: \frac{\mathrm{d}}{\mathrm{d}t}\:\! q_\mathrm{I}(\vec{x},t) = - \mathlarger{\int} \mathrm{d}^3 \vec{x}\;\nabla\cdot\vec{j}_\mathrm{I}(\vec{x},t) =0 \,.
\end{align}
To obtain the transport-relaxation equation for the charge density, we consider once again the change over a mesoscopic time interval $\big[t,t+\Delta t\big]$, with $\Delta t$ chosen to obey the scale separation relation~\eqref{eq:TimeScales} previously introduced in section~\ref{sec:3_LEQ_Method}. As before, we study the time evolution of the expectation value $\smash{\braket[\rbig]{q_\mathrm{I}(\vec{x},t)}^t}$ over the given time interval, with the goal of deriving a finite-difference quotient analogous to that obtained in equation~\eqref{eq:Finite_Difference_Quotient}. \\
 
\noindent 
There are two crucial differences from the case of energy transport: first, the density matrix $\rho_\mathrm{I}(t)$ is not a constant over the time interval $\Delta t$, because of the charge-violating effects; second, the integrated charge current is not a conserved quantity, as opposed to the total momentum of the fluid. Thus, we obtain
\begin{align}
	\label{eqn:transport-relaxation_before_expansion}
	\braket[\cbig]{q_\mathrm{I}\big(\vec{x}, t+\Delta t\big)}^{\!\!\:t+\Delta t} - \braket[\cbig]{q_\mathrm{I}(\vec{x}, t)}^{\!\!\:t} &= \mathrm{tr} \bigg\{ q_\mathrm{I}(\vec{x}, t) \Big[ \rho_\mathrm{I}\big(t+\Delta t\big) - \rho_\mathrm{I}(t) \Big] \!\!\:\bigg\} \notag \\ 
    &\phantom{=}\; - \mathlarger{\int}_t^{t+\Delta t} \mathrm{d}t' \: \mathrm{tr} \cBig\{ \nabla \cdot \vec{j}_\mathrm{I}(\vec{x}, t') \:\! \rho_\mathrm{I}\big(t+\Delta t\big)\!\!\;\cBig\} \,.
\end{align}
This first term on the right-hand side represents the relaxation of the charge density, which we have already investigated previously in this work, whereas the second term encodes transport, namely the spatial spreading of a local charge overdensity, analogous to the transport of energy density studied in the previous section. For later convenience, we denote the two contributions to the finite change in equation above by
\begin{subequations}
\label{eqn:transport-relaxation_split}
\begin{align}
\mathrm{tr} \bigg\{ q_\mathrm{I}(\vec{x}, t) \Big[ \rho_\mathrm{I}\big(t+\Delta t\big) - \rho_\mathrm{I}(t) \Big] \!\!\:\bigg\} &\equiv \cBig[ \braket[\cbig]{q_\mathrm{I}\big(\vec{x}, t+\Delta t\big)}^{\!\!\:t+\Delta t} - \braket[\cbig]{q_\mathrm{I}(\vec{x}, t)}^{\!t}\:\! \cBig]_{\mathrm{washout}}\,, \label{eqn:transport-relaxation_washout} \\[-0.1cm]
- \mathlarger{\int}_t^{t+\Delta t} \mathrm{d}t' \: \mathrm{tr} \cBig\{ \nabla \cdot \vec{j}_\mathrm{I}(\vec{x}, t') \: \rho_\mathrm{I}\big(t+\Delta t\big) \!\!\;\cBig\} &\equiv \cBig[ \braket[\cbig]{q_\mathrm{I}\big(\vec{x}, t+\Delta t\big)}^{\!\!\:t+\Delta t} - \braket[\cbig]{q_\mathrm{I}(\vec{x}, t)}^{\!t}\:\! \cBig]_{\mathrm{transport}}\,.
\label{eqn:transport-relaxation_transport}
\end{align}
\end{subequations}

\noindent
At this point, we again employ a local equilibrium ansatz of the form
\begin{align} \label{eq:rhoIrhoLEQ_Transport}  \rho^{(\mathrm{LEQ})}_{\mathrm{I}}(t)= \frac{1}{Z_\mathrm{I}^{(\mathrm{LEQ})}\!\!\:(t)}\,\exp\cbigg\{ - \mathlarger{\int}\mathrm{d}^3\vec{x} \;\beta(\vec{x},t) \gamma_\mathrm{pl}(\vec x, t) \Big[ h^{\mathrm{(cons.)}}_\mathrm{I}(\vec{x},t) &- \vec v_\mathrm{pl}(\vec x,t) \cdot \vec p_\mathrm{I}(\vec x, t) \notag \\[-0.15cm]
& -\mu(\vec{x}, t) \:\!q_\mathrm{I}(\vec{x}, t)\Big]\!\cbigg\}\, ,
\end{align}
having introduced the charge-conserving Hamiltonian density $\smash{h^{\mathrm{(cons.)}}_\mathrm{I}(\vec{x},t)}$, as well as the local chemical potential $\mu(\vec{x}, t)$.  As in the previous subsection, $\beta(\vec{x},t)$ and $\vec v_\mathrm{pl}(\vec{x},t)$ are fixed by the energy  and momentum density matching conditions. The local chemical potential is fixed at each time by the matching
condition
\begin{equation} \label{eqn:matching_qxt}
    \braket[\cbig]{q_\mathrm{I}(\vec{x},t)}_{\mathrm{LEQ}}^{t}    \overset{!}{=}    \braket[\cbig]{q_\mathrm{I}(\vec{x},t)}^{t}\,.
\end{equation}
Inserting the local equilibrium ansatz into equation~\eqref{eqn:transport-relaxation_before_expansion} and expanding at leading order in both deviations from equilibrium and the small coupling $\lambda$ lets us arrive at the sought-after linear transport-relaxation equation. We will illustrate this procedure in the following. \\

\noindent
As a final note before proceeding to the calculation, let us mention that, at linear order in deviation from equilibrium, at leading order in $\lambda$, and in the ideal fluid limit (namely leading order in the gradients), the dynamics of the momentum density (or equivalently of the energy density) and that of the charge density decouple.
Thus, we will avoid deriving the dynamical equations for the energy and momentum density again.
At higher order in derivatives, dissipative effects such as fluid viscosity and charge diffusion become relevant.

\subsubsection{Local washout}
Let us start our analysis with the washout term \eqref{eqn:transport-relaxation_washout}, with the aim of showing that charge density relaxes with the same rate as the global charge. Just as for a global charge, the first genuine local relaxation appears at $\mathcal{O}(\lambda^2)$. 
Following the derivation from section~\ref{sec:3_LEQ_Method}, all steps until equation~\eqref{eq:EDOQ-leq00} follow, upon replacing $Q_\mathrm{I}$ by the local charge density operator $q_\mathrm{I}(\vec{x}, t)$, and we can write
\begin{align}
   \bigg[\frac{\mathrm{d}}{\mathrm{d}t} \braket[\rbig]{q_{\mathrm I}(\vec{x},t )}^{\!\!\:t} \bigg]_{\mathrm{washout}}= -\frac{\lambda^2}{\hbar^2} \mathlarger{\int}_0^\infty \mathrm{d}t_{\mathrm{rel}}\;\mathrm{tr}\bigg\{\!\!\:\Big[\rho^{(\mathrm{LEQ})}_{\mathrm{I}}(t),H^{(\mathrm{viol.})}_{\mathrm I}(0)\Big]\!\!\:\Big[H^{(\mathrm{viol.})}_{\mathrm I}(t_\mathrm{rel}) , q_{\mathrm I}(\vec{x},t ) \Big]\!\!\:\bigg\} \phantom{\, .}\notag \\ 
   \times \cbigg\{1+\mathcal{O}\cBig(\lambda\:\!,\frac{\tau_{\mathrm{micro}}}{\Delta t},\frac{\Delta t}{\tau_{\mathrm{macro}}}\cBig)\!\!\:\cbigg\}\, ,
\end{align}
Here, the derivation departs slightly from what is presented in section~\ref{sec:3_LEQ_Method}, and we shall show the steps explicitly. Inserting the ansatz~\eqref{eq:rhoIrhoLEQ_Transport}, the commutator of the density matrix with the charge-violating part of the Hamiltonian reads
\begin{align}
    \cBig[\rho_{\mathrm{I}}^{(\mathrm{LEQ})}(t),H_{\mathrm{I}}^{(\mathrm{viol.})}(0)\cBig] &=- \mathlarger{\int}_0^1 \mathrm{d}s \mathlarger{\int}\mathrm{d}^3\vec{y}  \, \beta(\vec{y}, t) \gamma_\mathrm{pl}(\vec y,t) \Big(\rho_{\mathrm{I}}^{(\mathrm{LEQ})}\!\!\;(t)\!\!\:\Big)^{\!1-s} \notag\\ 
    &\qquad\qquad\qquad\;\; \times \Big[h^{\mathrm{(cons.)}}_\mathrm{I}(\vec{y}, t) - \vec v_\mathrm{pl}(\vec y, t)\cdot \vec p_{{\mathrm{I}}}(\vec y, t) \\ 
    &\qquad\qquad\qquad\qquad\qquad\qquad\;\;\;\;\: - \mu(\vec{y}, t) \:\! q_\mathrm{I}(\vec{y}, t),H_{\mathrm{I}}^{(\mathrm{viol.})}(0)\Big] \Big(\rho_{\mathrm{I}}^{(\mathrm{LEQ})}\!\!\;(t)\!\!\:\Big)^{\!s}\,. \notag 
\end{align}
We linearize around the equilibrium density matrix, which, as in section~\ref{sec:3_LEQ_Method}, is defined with respect to the charge-conserving part $H^{(\mathrm{cons.})}_\mathrm{I}$ of the Hamiltonian, granting
\begin{align}
    \rho_{\mathrm{I}}^{\,(\mathrm{LEQ})}(t)= \rho^{(\mathrm{EQ})}_\mathrm{I} \Bigg\{1&+\beta_\mathrm{EQ}\mathlarger{\int}_0^1 \mathrm{d}s\; e^{\scalebox{0.8}{$s\beta_\mathrm{EQ}H^{(\mathrm{cons.})}_\mathrm{I}$}} \!\!\:\cbigg[\int\mathrm{d}^3\vec{x}\: \mu(\vec{x},t) \!\: q_{\mathrm{I}}(\vec{x},t)\cbigg] e^{\scalebox{0.8}{$-s\beta_\mathrm{EQ}H^{(\mathrm{cons.})}_\mathrm{I}$}} \notag \\[-0.1cm]
    &-\mathlarger{\int}_0^1 \mathrm{d}s \; e^{\scalebox{0.8}{$s\beta_\mathrm{EQ}H^{(\mathrm{cons.})}_\mathrm{I}$}}  \cbigg[\int\mathrm{d}^3\vec{x}\: \delta\beta(\vec{x}, t) \delta h^{(\mathrm{cons.})}_\mathrm{I}(\vec{x}, t) \cbigg] e^{\scalebox{0.8}{$-s\beta_\mathrm{EQ}H^{(\mathrm{cons.})}_\mathrm{I}$}}  \notag \\[-0.1cm]
    &+{\beta_{\text{EQ}}}\mathlarger{\int}_0^1 \mathrm{d}s \; e^{\scalebox{0.8}{$s\beta_\mathrm{EQ}H^{(\mathrm{cons.})}_\mathrm{I}$}}  \cbigg[\int\mathrm{d}^3\vec{x}\: \vec v_\mathrm{pl} (\vec{x}, t) \cdot \vec p_\mathrm{I}(\vec{x}, t) \cbigg] e^{\scalebox{0.8}{$-s\beta_\mathrm{EQ}H^{(\mathrm{cons.})}_\mathrm{I}$}}  \notag \\[-0.1cm]
    & + \mathcal{O}\Big[\mu(\vec{x},t)^2, \delta\beta(\vec{x},t)^2, \vec v_\mathrm{pl}(\vec x, t)^2\Big]\!\!\:\Bigg\} \, .
\end{align}
When plugging this expression inside the equation for the washout and truncating at leading order in the gradients, the contributions involving $h^{\mathrm{(cons.)}}_\mathrm{I}$ and $\vec v_\mathrm{pl}\cdot\vec p_\mathrm{I}$ vanish.
The only non-vanishing contribution at zeroth order in derivatives is the term proportional to the local chemical potential, which reads
\begin{align}
    \!\scalebox{0.985}{$\displaystyle{\bigg[\frac{\mathrm{d}}{\mathrm{d}t} \braket[\rbig]{q_{\mathrm I}(\vec{x},t )}^{\!\!\:t} \bigg]_{\mathrm{washout}} }$} &\: \scalebox{0.985}{$\displaystyle{= 
    -\frac{\lambda^2\beta_\mathrm{EQ}}{\hbar^2} \mathlarger{\int}_0^\infty \mathrm{d}t_{\mathrm{rel}}\mathlarger{\int}_0^1 \mathrm{d}s \mathlarger{\int}\mathrm{d}^3\vec{y} \: \mu(\vec{y}, t) }$}\notag \\
    &\: \scalebox{0.985}{$\displaystyle{\quad \times \braket[\rbigg]{e^{\scalebox{0.8}{$s\beta_\mathrm{EQ}H^{(\mathrm{cons.})}_\mathrm{I}$}} \!\!\:\Big[q_\mathrm{I}(\vec{y}, t) ,H^{(\mathrm{viol.})}_{\mathrm I}(0)\Big] e^{\scalebox{0.8}{$-s\beta_\mathrm{EQ}H^{(\mathrm{cons.})}_\mathrm{I}$}} \!\!\:\Big[H^{(\mathrm{viol.})}_{\mathrm I}(t_\mathrm{rel}) , q_{\mathrm I}(\vec{x},t)\Big]\!\!\;}_{\!\mathrm{EQ}} }$}\notag \\ 
    &\: \scalebox{0.985}{$\displaystyle{\quad \times \cBig\{1+\mathcal{O}\cbig(\lambda, \mu^2, \delta\beta\:\!\mu\cbig)\!\!\:\cBig\}\,.}$}
\end{align}
Again, the two-point correlator localizes the $y$-integral in a region $|\vec{y}-\vec{x}| \sim \ell_{\mathrm{mfp}}$, and we can expand the local chemical potential in spatial gradients, where we call $\ell_\mu$ the typical length over which $\mu$ varies.
At leading order, $\mu$ can be taken out of the $y$-integral, which then only acts onto the charge density $\int_y q_\mathrm{I}(\vec{y}, t) = Q_\mathrm{I}$. The resulting correlator only depends on single spatial coordinate, and by translational invariance, it must be independent of the position. We can divide by the volume and integrate over the remaining spatial coordinate, so that we arrive at
\begin{align}
    \bigg[\frac{\mathrm{d}}{\mathrm{d}t} \braket[\cbig]{q_\mathrm{I}(\vec{x},t)\!\!\;}^{\!\!\:t} \:\!\bigg]_{\mathrm{washout}} &= 
    \frac{\lambda^2\beta_\mathrm{EQ}}{V \hbar^2} \:\!\mu(\vec{x}, t) \mathlarger{\int}_0^\infty \mathrm{d}t_{\mathrm{rel}} \mathlarger{\int}_0^1 \mathrm{d}s \notag \\
    & \quad \times \braket[\rbigg]{e^{\scalebox{0.8}{$s\beta_\mathrm{EQ}H^{(\mathrm{cons.})}_\mathrm{I}$}} \Big[H^{(\mathrm{viol.})}_{\mathrm I}(0),Q_\mathrm{I}\Big] e^{\scalebox{0.8}{$-s\beta_\mathrm{EQ}H^{(\mathrm{cons.})}_\mathrm{I}$}} \!\!\:\Big[H^{(\mathrm{viol.})}_{\mathrm I}(t_\mathrm{rel}) , Q_\mathrm{I} \Big]\!\!\;}_{\!\mathrm{EQ}}\notag \\ 
    & \quad \times \cBig\{1+\mathcal{O}\cbig(\lambda, \mu^2, \delta\beta\:\!\mu\cbig)\!\!\:\cBig\}\,.
    \label{eqn:wahsout_in_terms_of_mu}
\end{align}
Finally, we only need the linear relation between the charge density and the local chemical potential, which follows by expanding the matching condition \eqref{eqn:matching_qxt} to linear order in $\mu$, and to leading order in derivatives. It reads 
\begin{align}
    \braket[\cbig]{q_\mathrm{I}(\vec{x},t)\!\!\;}^{\!\!\:t} &= \beta_{\mathrm{EQ}}\mathlarger{\int}_0^1\mathrm{d}s \mathlarger{\int}\mathrm{d}^3\vec{y}\:\mu(\vec{y},t)\braket[\rBig]{e^{\scalebox{0.8}{$s\beta_\mathrm{EQ}H^{(\mathrm{cons.})}_\mathrm{I}$}}  q_\mathrm{I}(\vec{y},t) e^{\scalebox{0.8}{$-s\beta_\mathrm{EQ}H^{(\mathrm{cons.})}_\mathrm{I}$}} q_\mathrm{I}(\vec{x},t) }_{\mathrm{EQ}} \notag \\
    &= \beta_\mathrm{EQ} \mu(\vec{x},t)\:\!\chi_{qq}\:\! \cbigg\{1+ \mathcal{O}\bigg(\mu^2, \frac{\ell_{\mathrm{mfp}}}{\ell_\mu}\bigg)\!\cbigg\}\,,
    \label{eq:transport_susceptibility}
\end{align}
where the static charge susceptibility is given by
\begin{align}
    \chi_{qq}&= \mathlarger{\int}_0^1\mathrm{d}s \mathlarger{\int}\mathrm{d}^3\vec{y} \: \braket[\rBig]{e^{\scalebox{0.8}{$s\beta_\mathrm{EQ}H^{(\mathrm{cons.})}_\mathrm{I}$}}  q_\mathrm{I}(\vec{y},0) e^{\scalebox{0.8}{$-s\beta_\mathrm{EQ}H^{(\mathrm{cons.})}_\mathrm{I}$}} q_\mathrm{I}(\vec{0},0)}_{\mathrm{EQ}} 
    =\frac{1}{V}\:\!\braket[\cbig]{Q_\mathrm{I}^2}_\mathrm{EQ} \,.
\end{align}
Inserting the linear relation~\eqref{eq:transport_susceptibility} into equation~\eqref{eqn:wahsout_in_terms_of_mu}, we find
\begin{equation}
    \cBig[\partial_t\:\!\braket[\cbig]{q_\mathrm{I}(\vec{x},t)\!\!\;}^{\!t}\,\cBig]_{\mathrm{washout}} = -\Gamma_{\mathrm{relax}}\,\braket[\cbig]{q_\mathrm{I}(\vec{x},t)\!\!\;}^{\!\!\:t}\,\cbigg\{1+\mathcal{O}\cBig[\lambda,\braket[\cbig]{q_\mathrm{I}(\vec{x},t)\!\!\;}^{\!\!\:t}\cBig]\!\!\:\cbigg\}\,,
\end{equation}
with $\Gamma_\mathrm{relax}$ the relaxation rate of equation~\eqref{eq:Relaxation_Rate_InteractionPic}, identical to the one obtained for the global charge in section~\ref{sec:3_LEQ_Method}.

\subsubsection{Transport current} 
We now evaluate the transport contribution defined in equation~\eqref{eqn:transport-relaxation_transport}, with the aim of deriving the diffusive part of the local charge evolution. The charge transport is encoded in the term
\begin{align}
	- \mathlarger{\int}_t^{t+\Delta t} \mathrm{d}t' \: \mathrm{tr} \Big[\nabla \cdot \vec{j}_\mathrm{I}(\vec{x}, t') \:\! \rho_\mathrm{I}(t+\Delta t) \Big] &= - \mathlarger{\int}_t^{t+\Delta t} \mathrm{d}t' \: \mathrm{tr} \Big[\nabla \cdot \vec{j}_\mathrm{I}(\vec{x}, t') \rho_\mathrm{I}(t) \Big] \Big[ 1 + \mathcal{O}(\lambda) \Big] \notag \\
	&= - \nabla \cdot \mathlarger{\int}_t^{t+\Delta t}\mathrm{d}t' \: \braket[\cbig]{\,\!\!\;\vec{j}_\mathrm{I}(\vec{x}, t')\!\!\;}^{\!\!\: t}\,,
\end{align}
where we replace $\rho_\mathrm{I}(t+\Delta t)$ by $\rho_\mathrm{I}(t)$, since their difference is of order $\lambda$. Such corrections can be neglected here because the diffusion term is determined at leading order by the charge-conserving dynamics.  
Using the linearly expanded density matrix to evaluate the correlator on the RHS, we note that all but one term vanish via $\mathcal{CPT}$ arguments, and we are left with
\begin{align}
    \mathlarger{\int}_t^{t+\Delta t}\mathrm{d}t' \: \braket[\cbig]{\,\!\!\;\vec{j}_\mathrm{I}(\vec{x}, t')\!\!\;}^{\!\!\: t} &= 
    \beta_\mathrm{EQ} \mathlarger{\int}_t^{t+\Delta t}\mathrm{d}t' \: \mathlarger{\int}\mathrm{d}^3\vec y \: \mu(\vec y, t) \mathlarger{\int}_0^1 \mathrm{d}s \label{eqn:int_j_intermediate_expr} \\ 
    &\qquad\qquad\qquad\qquad\quad\times \braket[\cBig]{e^{s\beta_\mathrm{EQ}H_\mathrm{I}^{(\mathrm{cons.})}} q_\mathrm{I}(\vec y, t) e^{-s\beta_\mathrm{EQ}H_\mathrm{I}^{(\mathrm{cons.})}} \vec j_\mathrm{I}(\vec x, t')\!\!\:}_\mathrm{EQ} \,. \notag 
\end{align}
Once again, we employ the gradient expansion for $\mu(\vec y, t)$.
When plugging the term with no derivatives inside equation~\eqref{eqn:int_j_intermediate_expr}, the $\vec y$-integral only hits the charge density $q_\mathrm{I}$, and it evaluates to the time-independent global charge $Q_\mathrm{I}$. The resulting correlator $\braket[\rbig]{Q_\mathrm{I} \, \vec j_\mathrm{I}(\vec x, t')}_\mathrm{EQ}$ vanishes because the current $\vec j_\mathrm{I}$ is odd under time reversal.
The first non-trivial contribution then appears at first order in derivatives, namely
\begin{align}
    \mathlarger{\int}_t^{t+\Delta t}\mathrm{d}t' \: \braket[\cbig]{\,\!\!\;\vec{j}_\mathrm{I}(\vec{x}, t')\!\!\;}^{\!\!\: t} &= 
    \beta_\mathrm{EQ} \cbig[\nabla \mu(\vec x, t) \cbig]^{\!\!\;i} \mathlarger{\int}_t^{t+\Delta t} \mathrm{d}t' \mathlarger{\int} \mathrm{d}^3\vec y \mathlarger{\int}_0^1 \mathrm{d}s \: \big(\vec y - \vec x\big)^{\!\!\:i} \label{eqn:int_j_intermediate_2}\\
    & \qquad\quad\qquad\qquad\qquad \times \braket[\cBig]{e^{s\beta_\mathrm{EQ}H_\mathrm{I}^{(\mathrm{cons.})}} q_\mathrm{I}(\vec y, t) e^{-s\beta_\mathrm{EQ}H_\mathrm{I}^{(\mathrm{cons.})}} \vec j_\mathrm{I}(\vec x, t')\!\!\:}_\mathrm{EQ} \,, \notag 
\end{align}
where the repeated index $i$ is summed over.
Define the charge dipole moment  
\begin{equation}
    \vec I_{{\mathrm{I}}}( t) = \int \mathrm{d}^3 \vec y \;\vec y \:\! q_\mathrm{I}(\vec y, t) \,.
\end{equation}
Computing its time-derivative, we have
\begin{equation}
    \frac{\mathrm{d}}{\mathrm{d} t} \:\!\vec I_{\mathrm{I}}(t) = - \int \mathrm{d}^3\vec y \; \vec y \cbig[ \nabla \cdot \vec j_\mathrm{I}(\vec y, t) \cbig]= \int \mathrm{d}^3\vec y \;\vec j_\mathrm{I}(\vec y, t) \,,
\end{equation}
where we used integration by parts.
It follows, that we can write
\begin{equation}
    \vec I_{{ \mathrm{I}}}(t) = \vec I_{{\mathrm{I}}}(t_1) - \int_t^{t_1} \mathrm{d}t_2 \int \mathrm{d}^3\vec y \; \vec j_\mathrm{I}(\vec y, t_2) \,.
\end{equation}
Back to equation~\eqref{eqn:int_j_intermediate_2}, we rename $t'$ to $t_1$, and use the identity above. Then, the equal time correlator at $t_1$ vanishes because of $\vec j_\mathrm{I}$ being odd under time reversal.
Finally, we are left with a single unequal-time correlator
\begin{align}
    \mathlarger{\int}_t^{t+\Delta t}\mathrm{d}t' \: \braket[\cbig]{\,\!\!\;\vec{j}_\mathrm{I}(\vec{x}, t')\!\!\;}^{\!\!\: t}  &= 
    -\beta_\mathrm{EQ} \cbig[\nabla \mu(\vec x, t) \cbig]^{\!\!\;i} \mathlarger{\int}_t^{t+\Delta t}\mathrm{d}t_1 \mathlarger{\int}_t^{t_1} \mathrm{d}t_2 \mathlarger{\int}\mathrm{d}^3 \vec y\mathlarger{\int}_0^1 \mathrm{d}s \notag \\
    &\;\qquad\qquad\qquad\qquad \times \braket[\cBig]{e^{s\beta_\mathrm{EQ}H_\mathrm{I}^{(\mathrm{cons.})}} j_{\mathrm{I}}^{\:\!i}(\vec y, t_2) e^{-s\beta_\mathrm{EQ}H_\mathrm{I}^{(\mathrm{cons.})}} \vec j_\mathrm{I}(\vec x, t_1)\!\!\;}_{\!\!\:\mathrm{EQ}} \notag \\
    &= -\frac{1}{3}\:\!\beta_\mathrm{EQ} \nabla \mu(\vec x, t) \mathlarger{\int}_0^{\Delta t}\mathrm{d}t_1 \mathlarger{\int}_0^{t_1} \mathrm{d}t_2 \mathlarger{\int} \mathrm{d}^3 \vec y \mathlarger{\int}_0^1 \mathrm{d}s  \\
    &\;\qquad\qquad\qquad\qquad \times \braket[\rbigg]{\!\!\;\cBig[e^{s\beta_\mathrm{EQ}H_\mathrm{I}^{(\mathrm{cons.})}} \vec j_{\mathrm{I}}(\vec y, t_2-t_1) e^{-s\beta_\mathrm{EQ}H_\mathrm{I}^{(\mathrm{cons.})}}\cBig]\:\!\cdot \:\vec j_\mathrm{I}(\vec 0, 0)\!\!\:}_{\!\mathrm{EQ}}, \notag
\end{align}
where in the last step we have exploited the time- and space-translational invariance of the equilibrium density matrix, as well as its isotropy to write, schematically, $\braket[\rbig]{j^i j^\ell}_\mathrm{EQ} = \frac{1}{3}\:\!\delta^{i\ell} \braket[\rbig]{\:\!\vec j^2}_\mathrm{EQ}$.\\

\noindent Now is the time to employ the hierarchy of time-scales: the unequal-time correlator is localised over a short time-interval $|t_\mathrm{rel}|=|t_2-t_1|\sim\tau_\mathrm{micro}$. By assumption, $\Delta t\gg\tau_\mathrm{micro}$, so that following the same treatment as for the washout, the double integral  reduces to a single integral over the relative time times a factor of $\Delta t$.
Then, dividing by $\Delta t$ and approximating the finite difference with the instantaneous time derivative, we finally arrive at 
\begin{equation}
    \Big[\partial_t\:\!\braket[\rbig]{q_\mathrm{I}(\vec{x},t)\!\!\;}^{\!\!\:t}\Big]_{\mathrm{transport}} =D_q \nabla^2 \:\!\braket[\rbig]{q_\mathrm{I}(\vec{x},t)\!\!\;}^{\!\!\:t} \,,
\end{equation}
with the charge diffusion being
\begin{align} \label{eq:Charge_Diffusion_Coefficient}
    D_q = 
    \frac{1}{3\chi_{qq}} \mathlarger{\int}_0^1\mathrm{d}s \mathlarger{\int}_{-\infty}^0 \mathrm{d}t_{\mathrm{rel}} \mathlarger{\int} \mathrm{d}^3\vec{y} \: \braket[\rbigg]{\!\!\:\cBig[e^{\scalebox{0.8}{$s\beta_\mathrm{EQ}H^{(\mathrm{cons.})}_\mathrm{I}$}} \vec{j}_\mathrm{I}(\vec{y}, t_{\mathrm{rel}}) \!\: e^{\scalebox{0.8}{$-s\beta_\mathrm{EQ}H^{(\mathrm{cons.})}_\mathrm{I}$}}\cBig] \cdot \:\vec{j}_\mathrm{I}(\vec{0}, 0)\!\!\:}_{\!\mathrm{EQ}} \;\notag \\ 
    \times \Big[ 1 + \mathcal{O}\cbig(\lambda, \ell_{\mathrm{mfp}}^2/\ell_\mu^2\cbig) \Big] .
\end{align}
Again, changing from $3$ spatial dimensions to $d$ ones simply sees the factor $\frac{1}{3}$ replaced by $\frac{1}{d}$.
Just as shown in section~\ref{sec:Relation_Relaxation_Diffusion} for the relaxation rate, also the charge diffusion constant $D_q$ can be written in terms of a rate. In particular, define the charge diffusion rate 
\begin{equation}
    \gamma_q = \lim_{t\to\infty} \frac{1}{6Vt} \braket[\rBig]{\cbig[ \vec I_\mathrm{I}(t) - \vec I_\mathrm{I}(0) \cbig]^{\!\!\:2}\:\!}_\mathrm{EQ} \,.
\end{equation}
Along the same lines as in section~\ref{sec:Relation_Relaxation_Diffusion}, and at leading order in $\lambda$ and in spatial gradients, one can show that
\begin{equation}
    \gamma_q = \chi_{qq} D_q \,.
\end{equation}

\subsubsection{Diffusion--relaxation equation}
Collecting both contributions, the late-time evolution of the charge density obeys\footnote{We have derived this diffusion--relaxation equation in the rest-frame of the plasma. Boosting it to an inertial reference frame where the plasma has a stationary four-velocity $u_{0,\mathrm{pl}}$, the left-hand side of equation~\eqref{eq:Diffusion_Relaxation_Equation} must be modified by replacing $\partial_t\to u_{0,\mathrm{pl}}^{\mu}\partial_\mu $, and we recover the usual charge convection term.}
\begin{equation} \label{eq:Diffusion_Relaxation_Equation}
    \tcboxmath{\partial_t\:\! \braket[\rbig]{q_\mathrm{I}(\vec{x},t)\!\!\;}^{\!\!\:t} = D_q\:\!\nabla^2 \braket[\rbig]{q_\mathrm{I}(\vec{x},t)\!\!\;}^{\!\!\:t}- \Gamma_\mathrm{relax}\:\!\braket[\rbig]{q_\mathrm{I}(\vec{x},t)\!\!\;}^{\!\!\:t}\, ,}
\end{equation}
valid up to corrections of order $\lambda$ and $\ell_{\mathrm{mfp}}^2/\ell_\mu^2$. The charge density spreads diffusively on the scale set by  $D_q$ while being slowly washed out at the rate $\Gamma_{\mathrm{relax}}$. Integrating over the volume, the diffusion term becomes a boundary term that vanishes for a closed system, and one recovers the global relaxation law $\mathrm{d}\langle Q\rangle/\mathrm{d}t = -\Gamma_{\mathrm{relax}}\langle Q\rangle$ of section~\ref{sec:3_LEQ_Method}, a useful consistency check.\\

\noindent
To understand the dynamics described by the diffusion--relaxation equation above, let us go to momentum space and define
\begin{equation}
    \widetilde{q}_\mathrm{I}(\vec{k}, t) = \int \mathrm{d}^3\vec{x} \: e^{i \vec{k} \cdot \vec{x}}\, q_\mathrm{I}(\vec{x}, t)\,.
\end{equation}
In momentum space, the equation becomes
\begin{equation} \label{eq:Diffusion_Relaxation_Momentum}
    \partial_t \:\! \braket[\rbig]{\widetilde{q}_\mathrm{I}(\vec{k},t)\!\!\;}^{\!\!\:t} = - \big( D_q\:\!\vec{k}^2 + \Gamma_\mathrm{relax}\big) \braket[\rbig]{\widetilde{q}_\mathrm{I}(\vec{k},t)\!\!\;}^{\!\!\:t} \,.
\end{equation}
From it, we learn that for modes with momentum $\vec{k}^2\gtrsim \Gamma_\mathrm{relax}/D_q$ any charge overdensity is quickly spread out by the diffusion before it has any chance to decay via charge-violating effects of order $\lambda^2$. Instead, for low momentum modes such that $\vec{k}^2\lesssim \Gamma_\mathrm{relax}/D_q$, charge-violating effects become dominant. In the weak charge-violation limit this justifies only studying the decay of the homogeneous, i.e. $k=0$, mode, which is nothing but the total charge.\\

\noindent
Just as for the transport of energy and momentum, our derivation relies on the expansion in spatial derivatives, implying that the diffusive description we obtained is only valid for smooth enough densities, and it breaks down for perturbations with $|\vec{k}|\,\ell_\mathrm{mfp}\gtrsim 1$.

\section{Conclusions}
\label{sec:conclusions}

\noindent
The relaxation of an approximately conserved charge provides a particularly simple setting in which irreversible macroscopic dynamics can be connected directly to microscopic equilibrium fluctuations. The essential ingredients are not tied to a specific statistical construction: the charge violation must be weak, the system must remain close to local equilibrium, and microscopic correlations must decay on a timescale parametrically shorter than the one governing the evolution of the charge. Once this hierarchy is established, the microscopic memory of the state becomes irrelevant for the late-time dynamics, and the charge obeys a time-local relaxation law, with the resulting leading-order relaxation rate fixed entirely by equilibrium properties of the underlying theory. The fact that such a law exists, and that its rate reduces to the form of a Green--Kubo expression, has of course been appreciated for a long time. The contribution of the present work lies elsewhere, namely in the route by which this result may be reached, in the assumptions on which it actually rests, and in what remains ambiguous once one attempts to proceed beyond leading order.\\

\noindent
Our primary result is a construction that yields this relaxation law through a particularly direct approach. It shares with Zubarev's NSO formalism the local-equilibrium ansatz, which represents the instantaneous macroscopic state at late times, but dispenses with the non-equilibrium statistical operator constructed upon it, since the latter retains substantially more information than the evolution of the slow charge requires. The local-equilibrium state is instead evolved over an intermediate time interval, long enough for microscopic correlations to decay, yet short enough that the slow charge changes only perturbatively, such that the resulting finite change of the charge yields the desired rate. At leading non-trivial order, the two constructions agree, so that the retarded averaging over the history of the system proves inessential at this order. The virtue of this shorter route is not merely its economy. Because the separation of timescales is implemented in a single, explicit step, each of the assumptions listed above enters at an identifiable point of the derivation, such that the resulting relaxation law is accompanied by an explicit statement of its domain of validity rather than by a tacit one.\\ %The same reasoning delivers the associated fluctuation--dissipation relation without any additional machinery.\\

\noindent
This perspective also sharpens the fluctuation--dissipation interpretation of charge relaxation. The macroscopic decay rate may be obtained either from an equilibrium two-point function of the charge-violating dynamics, or from the equilibrium diffusion of the corresponding charge fluctuations, with the equilibrium variance of the charge fixing the conversion between the two. These are complementary manifestations of the same physics, and our derivation makes explicit the minimal assumptions under which the connection emerges. The same reasoning extends naturally from a homogeneous charge to slowly varying charge densities, where diffusion redistributes the charge while weak symmetry violation removes it. Relaxation, diffusion, and spatial transport can therefore be understood within a common near-equilibrium framework.\\

%\noindent
%Our second result is of a more cautionary nature and concerns the NSO formalism itself. As discussed in section~\ref{sec:ZubarevDiscussion}, the local-equilibrium operator underlying the construction is not unique: alternatives such as equation~\eqref{eq:LEQ_Operator_altered} satisfy the same defining requirements while differing at order $\lambda$, such that distinct yet seemingly equally justified choices generate distinct contributions at next-to-leading order. Independently of this, the self-consistency relations~\eqref{eq:Constraints_Schrödinger} accomplish more than fixing the time-dependent Lagrange multipliers. By demanding equality of the relevant averages evaluated in the local-equilibrium and non-equilibrium states, they effectively select the uniform relaxation regime that the method is intended to describe, and we have not been able to establish this selection from scale separation and the weakening of initial correlations alone. Neither observation affects the results obtained above, all of which are derived at leading non-trivial order in $\lambda$. Taken together, however, they delimit the regime in which the greater generality of the NSO framework can presently be exploited: at leading order it agrees with the far more economical local-equilibrium construction, whereas at the order at which the two could differ, its own predictions are not unambiguous. We regard this as an open problem of the formalism, and one that deserves to be settled before either construction is pushed to next-to-leading order in $\lambda$.\\

\noindent
Finally, we have illustrated how these tools apply to physically relevant observables in two rather different limiting regimes. The perturbative scalar theory serves as a check of the general result in a regime where a perturbative kinetic description is viable: the equilibrium correlation function and the linearized Boltzmann equation yield one and the same phase-space integral, and since the two calculations share no intermediate steps, their agreement is a non-trivial one. The electroweak $\mathrm{B+L}$ charge lies in the opposite regime, its violation being non-perturbative and admitting no kinetic description. There, the formalism serves instead to re-derive the known relation between the relaxation rate of the physical $\mathrm{B+L}$ charge and the Chern--Simons diffusion rate, also known as the sphaleron rate, the latter being the quantity accessible to lattice simulations~\cite{Bodeker:1999gx,Moore2000,DOnofrio:2014rug}. In both cases the general framework reproduces results established by other means, while providing a common perspective on the relaxation of approximately conserved charges.\\

\noindent
Our analysis has been restricted throughout to small departures from local equilibrium, to a single approximately conserved charge, to a pronounced separation between microscopic and macroscopic timescales, and to leading non-trivial order in the charge-violating interaction. Two of these restrictions deserve comment, since the physics lying beyond them is not merely conceivable, but well motivated.\\

\noindent
The first is the restriction to a single slow charge, which is only rarely realized in the applications of interest. In electroweak baryogenesis, the transport problem involves an entire network of chemical potentials, coupled through Yukawa interactions as well as through strong and weak sphalerons, with rates spread over several orders of magnitude~\cite{Nelson:1991ab,Cline:2000nw,Morrissey:2012db,Garbrecht:2018mrp,vandeVis:2025efm}. In leptogenesis, spectator processes redistribute a generated asymmetry among charges which equilibrate at widely separated temperatures~\cite{Harvey:1990qw,Bodeker:2019ajh}. Within the present framework, the required modification is clear in principle: the local-equilibrium ansatz must be extended to include the operator associated with each slow variable, together with its conjugate Lagrange multiplier, and the resulting relaxation law will be modified accordingly. What matters for the validity of the construction is not that the slow variables share a common timescale, but only that each of them evolves slowly compared to the microscopic fluctuations. A partial hierarchy, in which the retained variables themselves relax at widely separated rates, is therefore admissible, provided that all of them are included in the ansatz. Which variables must be retained, and by what criterion they are to be identified as slow, is a question the construction poses explicitly rather than tacitly, and we regard it as the natural next step.\\

\noindent
The second is the assumed hierarchy $\tau_{\mathrm{micro}}\ll\tau_{\mathrm{macro}}$, together with the assumption of a single, homogeneous equilibrium state. A cosmological first-order phase transition~\cite{Linde:1978px,Kibble:1980mv,Witten:1984rs,Hindmarsh:2020hop} strains both, and it is worth separating what the formalism can accommodate from what it cannot. The passage of a bubble wall breaks the isotropy and the translational invariance of the equilibrium state, so that the momentum flux across the wall acquires the status of a further relevant operator, with the plasma velocity field as its conjugate Lagrange multiplier. This is precisely an instance of the extension just described, and thus lies within reach of the construction, even though it is excluded by the charge-neutral linearization of section~\ref{sec:transport}. Less readily accommodated is the absence of a single equilibrium phase, a supercooled state being metastable, and the circumstance that the relaxation and transport coefficients are themselves functions of the wall profile. Across a wall, which in a strongly first-order transition is typically both thin and rapidly moving relative to the characteristic velocities in the plasma~\cite{Moore:1995ua,Moore:1995si,Ai:2025bjw}, the $\mathrm{B+L}$ relaxation rate falls from its unsuppressed symmetric-phase value to one exponentially suppressed by the sphaleron energy~\cite{Moore:1998swa}. Conventional treatments of electroweak baryogenesis accommodate this by promoting the relaxation rate to a prescribed function of position~\cite{Joyce:1994zn,Cline:2000nw,Konstandin:2013caa,Morrissey:2012db}. Deriving this local rate, rather than postulating it, constitutes a well-posed question for the construction developed here.\\

\noindent
Higher orders in the deviation from equilibrium and in the gradient expansion constitute a further natural extension. Both are viable within the construction presented here, and both are well motivated, being precisely what is required in order to proceed beyond the perfect-fluid limit obtained in section~\ref{sec:transport}. We do not claim that the simplified local-equilibrium construction supersedes the more general NSO framework. Its merit lies rather in the circumstance that every approximation it makes is visible at the very step at which it is made, which is what turns the restrictions listed above into statements about the physics rather than about the formalism.

\subsection*{Acknowledgments}

N.W. gratefully acknowledges support from the German Academic Scholarship Foundation, the Marianne-Plehn-Program of the Elite Network of Bavaria, and the International Max Planck Research School on Elementary Particle Physics (IMPRS EPP). M. C. acknowledges support from the Fundamental Research Funds for the Central Universities of China (Grant No. E5ER6601A2). The authors thank Gerd Röpke for initial discussions on Zubarev's NSO method. We acknowledge the assistance of LLMs, specifically ChatGPT-5.6 Sol and Claude Sonnet 5, to support the critical assessment of our derivations, identify errors and typographical mistakes, and improve the clarity of explanations throughout the manuscript.\\

\appendix 

\section{Expansion of exponential-type operators}
\label{sec:Dyson_Expansion} 

Let us briefly discuss the expansion of operators of the form $\exp\!\big(A+\delta B\big)$, where $A$ and $B$ are assumed to be non-commuting operators. Our goal is to expand this exponential for small $\delta\ll 1$, with the resulting series being reminiscent of the Dyson series familiar from QFT. Defining the shorthand 
\begin{align}
    B(s)\equiv e^{-sA}\:\!B\:\! e^{sA}\, ,
\end{align}
we utilize the operator equality
\begin{align}
    \exp\!\big(A+\delta B\big) = e^{A}\,\mathcal{T}_s \exp\!\bigg\{\delta\int_0^1\mathrm{d}s\, B(s)\bigg\}\, ,
\end{align}
where $\mathcal{T}_s$ denotes an ordering operator, indicating that the exponential is $s$-ordered: operators are arranged such that those with larger values of $s$ appear to the left. Expanding the $s$-ordered exponential then directly leads to the Dyson series 
\begin{align}
    \exp\!\big(A+\delta B\big) = e^{A}\,\Bigg\{1&+\delta\int_0^1 \mathrm{d}s\, B(s) + \delta^2 \int_0^1\mathrm{d}s\int_0^s\mathrm{d}s'\: B(s) B(s') \notag \\[-0.05cm] 
    &+ \delta^3 \int_0^1\mathrm{d}s\int_0^s\mathrm{d}s'\int_0^{s'} \mathrm{d}s''\:B(s) B(s') B(s'')+\mathcal{O}\big(\delta^4\big)\Bigg\} \, . 
\end{align}
In our case for which we desire to expand the LEQ operator $\smash{\rho_\mathrm{H}^{(\mathrm{LEQ})}\!\!\:(t)}$ around its equilibrium counterpart $\smash{\rho_\mathrm{H}^{(\mathrm{EQ})}}$, we plug in
\begin{subequations}
    \begin{align}
        A&=-\mathrm{\beta}_\mathrm{EQ}H\, , \\
        B&=-\Delta\beta(t) H+\beta_\mathrm{EQ}\:\!\mu(t)Q_\mathrm{H}(t)\, ,
    \end{align}
\label{eq:Substitutions_Expansion}%
\end{subequations}
while setting the formal expansion parameter $\delta=1$. Note that in this case one must divide by the trace of the operator, which should be expanded in the same manner. We have
\begin{align}
    \!\mathrm{tr}\cBig\{\exp\!\big(A+\delta B\big)\!\!\:\cBig\} = \mathrm{tr}\cbig\{e^{A}\cbig\}\,&+\delta\int_0^1 \mathrm{d}s\: \mathrm{tr}\Big\{e^{A}B(s)\Big\} + \delta^2 \int_0^1\mathrm{d}s\int_0^s\mathrm{d}s'\: \mathrm{tr}\Big\{e^{A}B(s) B(s')\Big\} \notag \\[-0.05cm] 
    &+ \delta^3 \int_0^1\mathrm{d}s\int_0^s\mathrm{d}s'\int_0^{s'} \mathrm{d}s''\: \mathrm{tr}\cBig\{e^{A}B(s) B(s') B(s'')\cBig\}\,+\: \mathcal{O}\big(\delta^4\big)\, , 
\end{align}
where we can readily simplify the first-order term, as $\mathrm{tr}\big\{e^{A}B(s)\big\}$ is $s$-independent due to cyclicity of the trace. Therefore, one arrives at the combined expansion 
\begin{align}
    \frac{\exp\!\big(A+\delta B\big)}{\mathrm{tr}\cBig\{\exp\!\big(A+\delta B\big)\!\!\:\cBig\}} = \frac{e^A}{\mathrm{tr}(e^A)} \Bigg\{1&+\delta \mathlarger{\int}_0^1 \mathrm{d}s\, \Big[B(s) -  \raisebox{0.3pt}{$\braket{\raisebox{-0.3pt}{$B$}}_{\!\!\:A}$}\Big] \notag \\[-0.25cm] 
    &+ \delta^2 \mathlarger{\int}_0^1\mathrm{d}s \mathlarger{\int}_0^s\mathrm{d}s'\, \cBig[B(s) B(s')-\braket[\cbig]{B(s) B(s')}_{\!\!\:\!A}\cBig] \notag \\[0.05cm]
    &- \delta^2 \mathlarger{\int}_0^1 \mathrm{d}s\, \raisebox{0.3pt}{$\braket{\raisebox{-0.3pt}{$B$}}_{\!\!\:A}$}\Big[B(s) -  \raisebox{0.3pt}{$\braket{\raisebox{-0.3pt}{$B$}}_{\!\!\:A}$}\Big] \notag \\
    & + \delta^3 \mathlarger{\int}_0^1 \mathrm{d}s \mathlarger{\int}_0^s \mathrm{d}s' \mathlarger{\int}_0^{s'} \mathrm{d}s'' \, \cBig[ B(s)B(s')B(s'') - \braket[\cbig]{B(s) B(s') B(s'')}_{\!\!\:\!A} \cBig] \notag \\
    & - \delta^3 \mathlarger{\int}_0^1 \mathrm{d}s \mathlarger{\int}_0^s \mathrm{d}s' \mathlarger{\int}_0^1 \mathrm{d}s'' \, \braket[\cbig]{B(s) B(s')}_{\!\!\:\!A} \Big[B(s'') - \raisebox{0.3pt}{$\braket{\raisebox{-0.3pt}{$B$}}_{\!\!\:A}$} \Big] \notag\\
    & - \delta^3 \mathlarger{\int}_0^1 \mathrm{d}s  \mathlarger{\int}_0^s \mathrm{d}s' \, \raisebox{0.3pt}{$\braket{\raisebox{-0.3pt}{$B$}}_{\!\!\:A}$} \cBig[ B(s)B(s') - \braket[\cbig]{B(s) B(s')}_{\!\!\:\!A} \cBig] \notag \\
    & + \delta^3 \mathlarger{\int}_0^1 \mathrm{d}s \, \cbig[\raisebox{0.3pt}{$\braket{\raisebox{-0.3pt}{$B$}}_{\!\!\:A}$}\cbig]^{\!\!\:2}\Big[ B(s) -  \raisebox{0.3pt}{$\braket{\raisebox{-0.3pt}{$B$}}_{\!\!\:A}$} \Big]
    + \mathcal{O}\big(\delta^4\big)\Bigg\}\, ,
    \label{eq:Combined_Expansion_Dyson}
\end{align}
where we introduced the shorthand
\begin{align}
    \braket{\raisebox{-0.45pt}{$O$}}_{\!\!\: A}\equiv \frac{\mathrm{tr}\big(e^AO\big)}{\mathrm{tr}(e^A)}\, .
\end{align}
Instating the substitutions~\eqref{eq:Substitutions_Expansion} and retaining only the first non-trivial order, we arrive at relation~\eqref{eq:Expanded_LEQ_Operator}.

\section{Consequences of $\mathcal{CPT}$-symmetry}
\label{sec:CPT_Arguments}

In the present section, we review some aspects of the discrete symmetry transformations $\mathcal{C}$ (charge conjugation), $\mathcal{P}$ (parity) and $\mathcal{T}$ (time reversal) and demonstrate how they lead to the vanishing of a specific set of traces entering the derivation of the relaxation rate. A consistent way of implementing the discrete symmetry transformations is to specify their action both on $c$-numbers as well as operators.\footnote{When spinorial or other tensor structures are present, they may also be affected nontrivially. For the time being, we focus on the scalar case.} Requiring both $\mathcal{C}$ and $\mathcal{P}$ to be unitary operators implies that they act trivially on $c$-numbers. By contrast, $\mathcal{T}$ is anti-unitary and thus acts by complex conjugation on $c$-numbers, i.e. it obeys $\mathcal{T} i \mathcal{T}^{-1} = -i$. To determine the transformation properties of operators constructed from field operators under $\mathcal{C}$, $\mathcal{P}$ and $\mathcal{T}$, it suffices to specify their action on the set of creation and annihilation operators. \\

\noindent 
We will make this explicit in the case of a complex scalar field $\Phi(x)$, for which the Heisenberg field operator reads 
\begin{align}
    \!\!\Phi(\vec{x},t) = \mathlarger{\mathlarger{\int}}\frac{\mathrm{d}^3\vec{p}}{(2\pi)^3} \frac{1}{2E(\vec{p})}\Bigg(a(\vec{p}) \exp\!\bigg\{\!\!\!\:-\!\!\:\frac{i}{\hbar}\Big[E(\vec{p})\:\!t-\vec{x}\cdot \vec{p}\Big]\!\!\:\bigg\}+b^\dagger\!\!\:(\vec{p}) \exp\!\bigg\{\frac{i}{\hbar}\Big[E(\vec{p})\:\!t-\vec{x}\cdot \vec{p}\Big]\!\!\:\bigg\}\!\!\:\Bigg)\, .
\end{align}
In the present case, we have omitted the label H for the Heisenberg picture in favor of notational simplicity. The action of the discrete symmetry transformations on the creation and annihilation operators is then defined as follows: 
\begin{subequations}
    \begin{align}
    \text{Charge conjugation $\mathcal{C}$:} && \mathcal{C} \,\!\!\; a(\vec{p})\,\mathcal{C}^{-1} &= b(\vec{p})\, , & \mathcal{C} \,\!\!\; b(\vec{p})\,\mathcal{C}^{-1} &= a(\vec{p}) \, , \\ 
    \text{Parity $\mathcal{P}$:} && \mathcal{P} \:\! a(\vec{p})\:\!\mathcal{P}^{-1} &= a(-\vec{p})\, , & \mathcal{P} \:\! b(\vec{p})\:\!\mathcal{P}^{-1} &= b(-\vec{p}) \, , \\ 
    \text{Time reversal $\mathcal{T}$:} && \mathcal{T} \:\! a(\vec{p})\:\!\mathcal{T}^{-1} &= a(-\vec{p})\, , & \mathcal{T} \:\! b(\vec{p})\:\!\mathcal{T}^{-1} &= b(-\vec{p}) \, .
\end{align}
\end{subequations}
Note that while $\mathcal{P}$ and $\mathcal{T}$ act identically on operators, they differ in their transformation behavior on $c$-numbers. It is then straightforward to verify that this transformation behavior reproduces the expected action on the full field operator, namely
\begin{subequations}
\begin{align}
    \scalebox{0.86}{$\displaystyle{ \mathcal{C}\:\!\Phi(\vec{x},t)\:\!\mathcal{C}^{-1} }$} \: & \scalebox{0.86}{$\displaystyle{ = \mathlarger{\mathlarger{\int}}\frac{\mathrm{d}^3\vec{p}}{(2\pi)^3} \frac{1}{2E(\vec{p})}\Bigg(\mathcal{C}\:\!a(\vec{p})\:\!\mathcal{C}^{-1} \exp\!\bigg\{\!\!\!\:-\!\!\:\frac{i}{\hbar}\Big[E(\vec{p})\:\!t-\vec{x}\cdot \vec{p}\Big]+\mathcal{C}\:\!b^\dagger\!\!\:(\vec{p})\:\!\mathcal{C}^{-1} \exp\!\bigg\{\frac{i}{\hbar}\Big[E(\vec{p})\:\!t-\vec{x}\cdot \vec{p}\Big]\!\!\:\bigg\}\!\!\:\Bigg) }$}\notag \\ 
    &\scalebox{0.86}{$\displaystyle{ = \mathlarger{\mathlarger{\int}}\frac{\mathrm{d}^3\vec{p}}{(2\pi)^3} \frac{1}{2E(\vec{p})}\Bigg(b(\vec{p}) \exp\!\bigg\{\!\!\!\:-\!\!\:\frac{i}{\hbar}\Big[E(\vec{p})\:\!t-\vec{x}\cdot \vec{p}\Big]+a^\dagger\!\!\;(\vec{p}) \exp\!\bigg\{\frac{i}{\hbar}\Big[E(\vec{p})\:\!t-\vec{x}\cdot \vec{p}\Big]\!\!\:\bigg\}\!\!\:\Bigg) }$} \notag \\ 
    &\scalebox{0.86}{$\displaystyle{ = \Phi^\dagger\!\!\:(\vec{x},t)\, ,}$} \\[0.3cm] 
    \scalebox{0.86}{$\displaystyle{ \mathcal{P}\:\!\Phi(\vec{x},t)\mathcal{P}^{-1} }$} \:& \scalebox{0.86}{$\displaystyle{ = \mathlarger{\mathlarger{\int}}\frac{\mathrm{d}^3\vec{p}}{(2\pi)^3} \frac{1}{2E(\vec{p})}\Bigg(\mathcal{P}\:\!a(\vec{p})\:\!\mathcal{P}^{-1} \exp\!\bigg\{\!\!\!\:-\!\!\:\frac{i}{\hbar}\Big[E(\vec{p})\:\!t-\vec{x}\cdot \vec{p}\Big]+\mathcal{P}\:\!b^\dagger\!\!\:(\vec{p})\:\!\mathcal{P}^{-1} \exp\!\bigg\{\frac{i}{\hbar}\Big[E(\vec{p})\:\!t-\vec{x}\cdot \vec{p}\Big]\!\!\:\bigg\}\!\!\:\Bigg) }$} \notag \\ 
    &\scalebox{0.86}{$\displaystyle{ = \mathlarger{\mathlarger{\int}}\frac{\mathrm{d}^3\vec{p}}{(2\pi)^3} \frac{1}{2E(\vec{p})}\Bigg(a(-\vec{p}) \exp\!\bigg\{\!\!\!\:-\!\!\:\frac{i}{\hbar}\Big[E(\vec{p})\:\!t-\vec{x}\cdot \vec{p}\Big]+b^\dagger\!\!\:(-\vec{p}) \exp\!\bigg\{\frac{i}{\hbar}\Big[E(\vec{p})\:\!t-\vec{x}\cdot \vec{p}\Big]\!\!\:\bigg\}\!\!\:\Bigg) }$} \notag \\ 
    &\scalebox{0.86}{$\displaystyle{ = \mathlarger{\mathlarger{\int}}\frac{\mathrm{d}^3\vec{p}}{(2\pi)^3} \frac{1}{2E(-\vec{p})}\Bigg(a(\vec{p}) \exp\!\bigg\{\!\!\!\:-\!\!\:\frac{i}{\hbar}\Big[E(-\vec{p})\:\!t+\vec{x}\cdot \vec{p}\Big]+b^\dagger\!\!\:(\vec{p}) \exp\!\bigg\{\frac{i}{\hbar}\Big[E(-\vec{p})\:\!t+\vec{x}\cdot \vec{p}\Big]\!\!\:\bigg\}\!\!\:\Bigg) }$} \notag \\ 
    &\scalebox{0.86}{$\displaystyle{ = \Phi(-\vec{x},t)\, , }$} \\[0.3cm] 
    \mathcal{T}\:\!\Phi(\vec{x},t)\mathcal{T}^{-1} \:&\scalebox{0.86}{$\displaystyle{ = \mathlarger{\mathlarger{\int}}\frac{\mathrm{d}^3\vec{p}}{(2\pi)^3} \frac{1}{2E(\vec{p})}\Bigg(\mathcal{T}\:\!a(\vec{p})\:\!\mathcal{T}^{-1} \exp\!\bigg\{\frac{i}{\hbar}\Big[E(\vec{p})\:\!t-\vec{x}\cdot \vec{p}\Big]+\mathcal{T}\:\!b^\dagger\!\!\:(\vec{p})\:\!\mathcal{T}^{-1} \exp\!\bigg\{\!\!\!\:-\!\!\:\frac{i}{\hbar}\Big[E(\vec{p})\:\!t-\vec{x}\cdot \vec{p}\Big]\!\!\:\bigg\}\!\!\:\Bigg) }$} \notag \\ 
    &\scalebox{0.86}{$\displaystyle{ = \mathlarger{\mathlarger{\int}}\frac{\mathrm{d}^3\vec{p}}{(2\pi)^3} \frac{1}{2E(\vec{p})}\Bigg(a(-\vec{p}) \exp\!\bigg\{\frac{i}{\hbar}\Big[E(\vec{p})\:\!t-\vec{x}\cdot \vec{p}\Big]+b^\dagger\!\!\:(-\vec{p}) \exp\!\bigg\{\!\!\!\:-\!\!\:\frac{i}{\hbar}\Big[E(\vec{p})\:\!t-\vec{x}\cdot \vec{p}\Big]\!\!\:\bigg\}\!\!\:\Bigg) }$} \notag \\ 
    &\scalebox{0.86}{$\displaystyle{ = \mathlarger{\mathlarger{\int}}\frac{\mathrm{d}^3\vec{p}}{(2\pi)^3} \frac{1}{2E(-\vec{p})}\Bigg(a(\vec{p}) \exp\!\bigg\{\!\!\!\:-\!\!\:\frac{i}{\hbar}\Big[E(-\vec{p})\:\!t-\vec{x}\cdot \vec{p}\Big]+b^\dagger\!\!\:(\vec{p}) \exp\!\bigg\{\frac{i}{\hbar}\Big[E(-\vec{p})\:\!t-\vec{x}\cdot \vec{p}\Big]\!\!\:\bigg\}\!\!\:\Bigg) }$} \notag \\ 
    &\scalebox{0.86}{$\displaystyle{ = \Phi(\vec{x},-t)\, .}$}
\end{align}%
\end{subequations}
Here, we have used the fact that $E(\vec{p})$ is real and satisfies $E(\vec{p})=E(-\vec{p})$. With the (conserved) charge operator $Q_\Phi$ in this scalar case given by 
\begin{align}
    Q_\Phi = \mathlarger{\int}\frac{\mathrm{d}^3\vec{p}}{(2\pi)^3} \frac{1}{2E(\vec{p})}\Big[a^\dagger \!\!\;(\vec{p})\:\! a(\vec{p})-b^\dagger\!\!\:(\vec{p})\:\! b(\vec{p})\Big]\, ,
\end{align}
it is found to be invariant under both $\mathcal{P}$ and $\mathcal{T}$, while changing its sign under $\mathcal{C}$, as is expected. \\ 

\noindent 
Let us return to the minimal setup considered throughout most of this work, involving a single approximately conserved charge $Q$ and the full Hamiltonian $H$ governing the system. We assume that $H$ is invariant under the combined $\mathcal{CPT}$ transformation, whereas the charge operator $Q_\mathrm{H}(t)$, much like the charge operator $Q_\Phi$ discussed previously, satisfies the transformation properties
\begin{align}
    \mathcal{C}\:\!Q_\mathrm{H}(t)\,\mathcal{C}^{-1}&=-Q_\mathrm{H}(t)\, , & \mathcal{P}Q_\mathrm{H}(t)\:\!\mathcal{P}^{-1}&=Q_\mathrm{H}(t)\, , & \mathcal{T}Q_\mathrm{H}(t)\:\!\mathcal{T}^{-1}&=Q_\mathrm{H}(-t)\, .
\end{align}
The nontrivial behavior under time reversal stems solely from the weak non-conservation of the operator $Q_\mathrm{H}(t)$, which reverses the time argument. Under a combined $\mathcal{CPT}$ transformation, one would therefore find the transformation behaviors
\begin{align}
    \big(\mathcal{CPT}\big)H\big(\mathcal{CPT}\big)^{\!-1} = H\, , \qquad \text{and}\qquad \big(\mathcal{CPT}\big)Q_\mathrm{H}(t)\big(\mathcal{CPT}\big)^{\!-1} = -Q_\mathrm{H}(-t)\, .
\end{align}
The above relations are sufficient for what follows. However, because $\mathcal{T}$ is anti-unitary, care must be taken when manipulating traces, as the standard cyclicity property requires modification. Indeed, one has the altered relation
\begin{align}
    \mathrm{tr}\cbig[\mathcal{T}O(t)\mathcal{T}^{-1}\cbig]=\mathrm{tr}\big[O(t)\big]^* = \mathrm{tr}\big[O^\dagger\!\!\:(t)\big] \, ,
\end{align}
with the star denoting complex conjugation, which arises directly from the defining relation of anti-unitary operators, namely that $\braket[\rbig]{\mathcal{T}\psi\!\!\;|\mathcal{T}\phi}=\braket{\psi\!\!\:|\phi}^*$. We can now apply this directly to the desired expressions. Note that due to $H$ being invariant under $\mathcal{CPT}$, so is $e^{-\beta_\mathrm{EQ}H}$ and thus the equilibrium density matrix $\smash{\rho^{(\mathrm{EQ})}}$. Using the fact that both $H$ and $Q_\mathrm{S}$, and therefore similarly $Q_\mathrm{H}(t)$, are Hermitian operators, we find 
\begin{align}
    \braket[\rbig]{Q_\mathrm{H}(t)}_\mathrm{EQ} = \mathrm{tr}\Big[Q_\mathrm{H}(t)\rho^{(\mathrm{EQ})}\Big] &= \mathrm{tr}\Big[Q_\mathrm{H}^\dagger(t)\rho^{(\mathrm{EQ}),\dagger}\Big] \notag \\ 
    &= \mathrm{tr}\Big[\big(\mathcal{CPT}\big)Q_\mathrm{H}(t)\rho^{(\mathrm{EQ})}\big(\mathcal{CPT}\big)^{\!-1}\Big] \notag \\ 
    &= - \,\mathrm{tr}\Big[Q_\mathrm{H}(-t)\rho^{(\mathrm{EQ})}\Big] = -\,\braket[\rbig]{Q_\mathrm{H}(t)}_\mathrm{EQ} = 0\, .
\end{align}
In the last line, we utilized that the time-evolution operators entailed in the Heisenberg operator $Q_\mathrm{H}(t)$ can be commuted through the equilibrium density matrix. As a result, the equilibrium expectation value becomes time-independent, and the negative sign in the time argument is therefore immaterial. An identical conclusion can be drawn for equilibrium traces entailing any odd number of number operators $Q_\mathrm{H}(t)$, as for example $\braket[\rbig]{HQ_\mathrm{H}(t)}_\mathrm{EQ}=0$. Moreover, one finds that the matrix element
\begin{align}
    \braket[\rbigg]{\exp\cbig(s\beta_\mathrm{EQ}H\cbig)\!\: Q_{\mathrm{H}}(t) \exp\cbig(-s\beta_\mathrm{EQ}H\cbig)\:\!\frac{\mathrm{d}Q_{\mathrm{H}}(t)}{\mathrm{d}t}}_{\!\!\mathrm{EQ}}
\end{align}
entering equation~\eqref{eq:LEQ_Average_DerivativeN_Expanded} vanishes identically, which becomes evident upon utilizing the identity 
\begin{align}
    \frac{\mathrm{d}Q_{\mathrm{H}}(t)}{\mathrm{d}t} = \frac{i}{\hbar}\Big[H,Q_\mathrm{H}(t)\Big]\, .
\end{align}
Even though we have an even number of $Q_\mathrm{H}(t)$-insertions, the factor $i$ is also odd under a $\mathcal{CPT}$-transformation. Therefore, we again obtain an overall minus sign and the given equilibrium average vanishes.\footnote{Beware that the Hermitian conjugation entering when distributing the factor $\mathds{1}=\big(\mathcal{CPT}\big)\big(\mathcal{CPT}\big)^{\!-1}$ changes the order of the commutator and flips the sign of $i$. However, the two sign changes cancel each other.} Crucially, the described simplifications only arise when all factors of $Q_\mathrm{H}(t)$ are evaluated at the same time $t$. Only in this case do the time-evolution operators cancel, rendering the expectation value time-independent and allowing one to conclude that the matrix element vanishes.

\section{Details of the charged scalar computation}
\label{app:scalar_model_details}

In this appendix we collect the technical steps used in the charged scalar example of Section~\ref{sec:boltzmann}. The main text presents the relaxation rate directly in phase-space form and compares it with the result obtained from the linearized Boltzmann equation. Here we give the intermediate derivations. We first evaluate the equilibrium correlator appearing in the relaxation formula, including the fluctuation of the charge \(Q_\chi\). We then give the details of the kinetic calculation. To simplify the notation in this appendix, we use four-vector notation $\big[$e.g., $p=(E_{\vec{p}},\vec{p})\big]$ throughout and omit the field labels on the single-particle energies. Thus, $E_{\vec{k}}$, $E_{\vec{p}}$, and $E_{\vec{q}}$ denote the energies associated with the corresponding momenta and fields, as determined by the context.

\subsection{Correlator calculation}
\label{app:scalar_model_correlator}

Here we give the details leading to equations~\eqref{eq:Q2_scalarfield} and~\eqref{eq:scalar_model_Gamma_phasespace}. Throughout this subsection the equilibrium average is taken with respect to the charge-conserving theory, and all fields are evolved with $H^{(\mathrm{cons.})}$. \\

\noindent We use the standard mode expansion for the massive complex scalar,
\begin{equation}
\chi_{\mathrm I}(x) = \mathlarger{\int} \frac{\mathrm d^3\vec{k}}{(2\pi)^3\,2E_{\vec{k}}} \left[ a_{\vec{k}}\,e^{-ik\cdot x} + b^\dagger_{\vec{k}}\,e^{ik\cdot x} \right]\,, \end{equation}
with $E_{\vec{k}}=\sqrt{\vec{k}^2+m_\chi^2}\,$. Substituting this expansion into \eqref{eq:scalar_model_Qchi}, one obtains
\begin{equation}
Q_{\chi,\mathrm I} = \mathlarger{\int} \frac{\mathrm d^3\vec{k}}{(2\pi)^3\,2E_{\vec{k}}} \left[ a^\dagger_{\vec{k}}a_{\vec{k}} - b^\dagger_{\vec{k}}b_{\vec{k}} \right].
\label{eq:app_scalar_Q_operator}
\end{equation}
The creation and annihilation operators satisfy
\begin{equation}
\cbig[ a_{\vec{k}},a^\dagger_{\vec{p}} \cbig] = \cbig[ b_{\vec{k}},b^\dagger_{\vec{p}} \cbig] = (2\pi)^3\,2E_{\vec{k}}\, \delta^{(3)}\!\!\:\big(\vec{k}-\vec{p}\big), \end{equation}
and the non-vanishing thermal averages at vanishing chemical potential are
\begin{align} \label{app:eq:daggeraa-aadagger}
\cbig\langle a^\dagger_{\vec{k}}a_{\vec{p}}
\cbig\rangle_{\mathrm{EQ}} &= (2\pi)^3\,2E_{\vec{k}}\, \delta^{(3)}\!\!\:\big(\vec{k}-\vec{p}\big)\, f_\chi^{\mathrm{EQ}}(E_{\vec{k}}), \notag\\ 
\cbig\langle a_{\vec{k}}a^\dagger_{\vec{p}} \cbig\rangle_{\mathrm{EQ}} &= (2\pi)^3\,2E_{\vec{k}}\, \delta^{(3)}\!\!\:\big(\vec{k}-\vec{p}\big)\, \left[ 1+f_\chi^{\mathrm{EQ}}(E_{\vec{k}}) \right],
\end{align}
with identical relations for the antiparticle operators $b_{\vec{k}}$. Therefore the equilibrium fluctuation of $Q_\chi$ receives equal contributions from particles and antiparticles. One obtains
\begin{equation}
\frac{1}{V} \braket[\rbig]{Q_{\chi,\mathrm I}^2
}_{\mathrm{EQ}} = 2 \mathlarger{\int} \frac{\mathrm d^3\vec{k}}{(2\pi)^3}\, f_\chi^{\mathrm{EQ}}(E_{\vec{k}}) \left[ 1+f_\chi^{\mathrm{EQ}}(E_{\vec{k}}) \right].
\label{eq:app_scalar_Q_variance}
\end{equation}
This is the result given in equation~\eqref{eq:Q2_scalarfield}.\\

\noindent 
The numerator of the relaxation formula~\eqref{eq:scalar_model_Gamma_corr} is
\begin{equation} 
\mathcal N_\chi \equiv \int_0^\infty \mathrm dt \int_0^1 \mathrm ds\, \mathcal G_\chi(s,t),
\end{equation}
where 
\begin{equation}
\mathcal G_\chi(s,t) \equiv \left\langle e^{s\beta_{\mathrm{EQ}}H^{(\mathrm{cons.})}} J_{\chi,\mathrm I}(0) e^{-s\beta_{\mathrm{EQ}}H^{(\mathrm{cons.})}} J_{\chi,\mathrm I}(t) \right\rangle_{\mathrm{EQ}} ,
\end{equation}
Since each $J_{\chi,\mathrm I}$ is linear in $\lambda$, the correlator is of order $\lambda^2$. At leading order, the remaining expectation values can therefore be evaluated at zeroth order in the charge-violating interaction, with the fields evolving freely under $H^{(\mathrm{cons.})}$.\\

\noindent We use 
\begin{equation}
 x_s \equiv \big(\!\!\:-\!\!\;is\beta_{\mathrm{EQ}},\vec{x}\big),
\end{equation}
so that
\begin{equation}
e^{s\beta_{\mathrm{EQ}}H^{(\mathrm{cons.})}} \mathcal O_{\mathrm I}(0,\vec{x}) e^{-s\beta_{\mathrm{EQ}}H^{(\mathrm{cons.})}} = \mathcal O_{\mathrm I}(x_s).
\end{equation}
Inserting equation~\eqref{eq:JIxixi} the previous expression reduces to
\begin{align}
\mathcal G_\chi(s,t) = \frac{\lambda^2m_\chi^2}{4}\int \mathrm d^3\vec{x}\,\mathrm d^3\vec{y}\, \Big[  \left\langle \chi_{\mathrm I}(x_s) \chi^\dagger_{\mathrm I}(y) \right\rangle_{\mathrm{EQ}} + \left\langle \chi^\dagger_{\mathrm I}(x_s) \chi_{\mathrm I}(y) \right\rangle_{\mathrm{EQ}} \Big] \left\langle \phi^2_{\mathrm I}(x_s) \phi^2_{\mathrm I}(y)
\right\rangle_{\mathrm{EQ}} .
\label{eq:app_scalar_corr_factorized}
\end{align}
where $y=(t,\vec{y})$. Note that the correlator factorizes because, at the order considered here, the $\chi$ and $\phi$ sectors evolve independently and the equilibrium density operator takes the form
$\rho_{\mathrm{EQ}}=\rho_{\chi,\mathrm{EQ}}\otimes\rho_{\phi,\mathrm{EQ}}$. Furthermore, we have used that the $U(1)_\chi$ invariance of the charge-conserving equilibrium state implies $\big\langle
\chi_{\mathrm I}(x_s)\chi_{\mathrm I}(y) \big\rangle_{\mathrm{EQ}} = \big\langle\chi^\dagger_{\mathrm I}(x_s)\chi^\dagger_{\mathrm I}(y) \big\rangle_{\mathrm{EQ}} = 0$ and also $|\eta_\chi|^2=1$.\\

\noindent  
For the free real field $\phi$, Wick's theorem gives
\begin{equation}
\braket[\cbig]{\phi_{\mathrm I}^2(x_s)\phi_{\mathrm I}^2(y)}_{\mathrm{EQ}}
=\braket[\rbig]{\phi_{\mathrm I}^2(x_s)}_{\mathrm{EQ}}
\braket[\rbig]{\phi_{\mathrm I}^2(y)}_{\mathrm{EQ}}
+
2\braket[\cbig]{\phi_{\mathrm I}(x_s)\phi_{\mathrm I}(y)}_{\mathrm{EQ}}^{\!2} \, .
\end{equation}
Because  $\braket[\rbig]{\phi_{\mathrm I}^2(x)}_\mathrm{EQ}$ is independent of the space-time coordinate $x$, the disconnected contribution factors out of the integral in $\vec x$ and $\vec y$. Then, the integral only acts over the $\chi$ two-point function, selecting its zero-momentum limit, which vanishes since the two-point function is evaluated on shell and $\chi$ is massive. Therefore, we can omit the disconnected term and retain only the connected contraction, which amounts to the replacement
\begin{equation}
\braket[\cbig]{\phi_{\mathrm I}^2(x_s)\phi_{\mathrm I}^2(y)}_{\mathrm{EQ}}
\;\longrightarrow\;
2\:\!\braket[\cbig]{\phi_{\mathrm I}(x_s)\phi_{\mathrm I}(y)}_{\mathrm{EQ}}^{\!\!\:2} \, .
\end{equation}
Using the thermal contractions in equation~\eqref{app:eq:daggeraa-aadagger} and the same relations for the $\phi$ creation and annihilation operators, the relevant correlators give, 
\begin{align}
\braket[\cbig]{\chi_{\mathrm I}(x_s) \chi^\dagger_{\mathrm I}(y)}_{\mathrm{EQ}} &= \mathlarger{\int} \frac{\mathrm d^3\vec{k}}{(2\pi)^3\,2E_{\vec{k}}} \bigg\{\!\left[ 1+f_\chi^{\mathrm{EQ}}(E_{\vec{k}}) \right] e^{-ik\cdot(x_s-y)}+ f_\chi^{\mathrm{EQ}}(E_{\vec{k}}) \:\! e^{ik\cdot(x_s-y)} \bigg\}.
\end{align}
and 
\begin{align}
\braket[\cbig]{\phi_{\mathrm I}(x_s) \phi_{\mathrm I}(y) }_{\mathrm{EQ}} &= \mathlarger{\int} \frac{\mathrm d^3\vec{q}}{(2\pi)^3\,2E_{\vec{q}}}\bigg\{\!
\left[ 1+f_\phi^{\mathrm{EQ}}(E_{\vec{q}}) \right] e^{-iq\cdot(x_s-y)}+ f_\phi^{\mathrm{EQ}}(E_{\vec{q}}) \:\!e^{iq\cdot(x_s-y)} \bigg\}\,,
\end{align}
where $E_{\vec{q}}$ denotes the energy of the $\phi$ mode. The expression for $\big\langle \chi^\dagger_{\mathrm I}(x_s) \chi_{\mathrm I}(y) \big\rangle_{\mathrm{EQ}}$ is identical to its Hermitian-conjugate counterpart. \\

\noindent Substituting the above correlators  gives eight products of phases. They are all of the same perturbative order. However, after the spatial and time integrations, each product would be proportional to
\begin{equation}
\delta^{(3)} \cbig(\sigma_k\vec{k}+\sigma_p\vec{p}+\sigma_q\vec{q} \cbig) \:\! \delta\cbig( \sigma_k E_{\vec{k}} + \sigma_p E_{\vec{p}} + \sigma_q E_{\vec{q}} \cbig), \qquad \sigma_i=\pm 1. \end{equation}
For a massive $\chi$ mode and massless $\phi$ modes, the only sign
choices that have solutions are
$(\sigma_k,\sigma_p,\sigma_q)=(+,-,-)$ and
$(\sigma_k,\sigma_p,\sigma_q)=(-,+,+)$.\footnote{For example, $(+,+,-)$ would require $ \vec{q}=\vec{k}+\vec{p}$, and $E_{\vec{q}}=E_{\vec{k}}+E_{\vec{p}}$. However, $ E_{\vec{q}} = |\vec{q}| = |\vec{k}+\vec{p}| \leq |\vec{k}|+|\vec{p}| < E_{\vec{k}}+E_{\vec{p}} $, where the last inequality follows from $E_{\vec{k}}=\sqrt{\vec{k}^2+m_\chi^2}>|\vec{k}|$.} These correspond to the decay and inverse-decay contributions. Therefore, we can safely keep the relevant contribution and ignore the rest. After spatial integration we get
\begin{align}
\frac{\mathcal G_\chi(s,t)}{V} &=
\lambda^2 m_\chi^2\mathlarger{\int} \frac{\mathrm d^3\vec{k}}{(2\pi)^3\,2E_{\vec{k}}}\, \frac{\mathrm d^3\vec{p}}{(2\pi)^3\,2E_{\vec{p}}}\, \frac{\mathrm d^3\vec{q}}{(2\pi)^3\,2E_{\vec{q}}}\, (2\pi)^3 \delta^{(3)}\!\!\:\big(\vec{k}-\vec{p}-\vec{q}\big) \notag\\[0.05cm]
&\qquad\qquad\quad  \times \bigg\{f_\chi^{\mathrm{EQ}}(E_{\vec{k}}) \left[ 1+f_\phi^{\mathrm{EQ}}(E_{\vec{p}}) \right] \left[ 1+f_\phi^{\mathrm{EQ}}(E_{\vec{q}}) \right] e^{-i(t+i s \beta_{\text{EQ}})\Delta E}  \notag\\[-0.125cm] 
&\qquad\qquad\qquad\qquad\qquad\!\!\!\; + \left[ 1+f_\chi^{\mathrm{EQ}}(E_{\vec{k}}) \right] f_\phi^{\mathrm{EQ}}(E_{\vec{p}}) f_\phi^{\mathrm{EQ}}(E_{\vec{q}}) e^{i(t+i s \beta_{\text{EQ}})\Delta E }  \bigg\}\,,
\label{eq:app_scalar_corr_intermediate}
\end{align}
where
\begin{equation}
\Delta E \equiv E_{\vec{k}} - E_{\vec{p}} - E_{\vec{q}}.
\end{equation}
Performing now the integral in $s$, and using the equilibrium KMS relation, which for Bose-Einstein distributions implies $1+f_a^{\mathrm{EQ}}(E)=e^{\beta_{\mathrm{EQ}}E} f_a^{\mathrm{EQ}}(E)$ the statistical factors can be rearranged as
\begin{equation}
\left[ 1+f_\chi^{\mathrm{EQ}}(E_{\vec{k}}) \right] f_\phi^{\mathrm{EQ}}(E_{\vec{p}}) f_\phi^{\mathrm{EQ}}(E_{\vec{q}}) = e^{\beta_{\mathrm{EQ}}\Delta E} f_\chi^{\mathrm{EQ}}(E_{\vec{k}}) \left[ 1+f_\phi^{\mathrm{EQ}}(E_{\vec{p}}) \right] \!\left[ 1+f_\phi^{\mathrm{EQ}}(E_{\vec{q}}) \right] .
\end{equation}
We then obtain
\begin{align}
    \mathlarger{\int}_{0}^{1}\text{d}s\, \frac{\mathcal{G}_\chi(s,t)}{V} = \lambda^2m_\chi^2 \mathlarger{\int}&\frac{\mathrm d^3\vec{k}}{(2\pi)^3\,2E_{\vec{k}}}\, \frac{\mathrm d^3\vec{p}}{(2\pi)^3\,2E_{\vec{p}}}\, \frac{\mathrm d^3\vec{q}}{(2\pi)^3\,2E_{\vec{q}}}\, (2\pi)^3 \delta^{(3)}\!\!\:\big(\vec{k}-\vec{p}-\vec{q}\big) \\
    &f_\chi^{\mathrm{EQ}}(E_{\vec{k}}) \left[ 1+f_\phi^{\mathrm{EQ}}(E_{\vec{p}}) \right] \left[ 1+f_\phi^{\mathrm{EQ}}(E_{\vec{q}}) \right] \frac{ e^{\beta_{\mathrm{EQ}}\Delta E}-1 }{ \beta_{\mathrm{EQ}}\Delta E } \left[ e^{-i\Delta E t} + e^{i\Delta E t}\right]. \notag 
\end{align}
The remaining time integral can be evaluated in the distributional sense. Using
\begin{equation}
\int_0^\infty \mathrm dt\, e^{-i\Delta E t} = \pi\delta(\Delta E) - i\,\mathcal P\frac{1}{\Delta E}, \qquad \int_0^\infty \mathrm dt\, e^{i\Delta E t} = \pi\delta(\Delta E) + i\,\mathcal P\frac{1}{\Delta E},
\end{equation}
the principal-value parts cancel in the sum. Therefore
\begin{equation}
\int_0^\infty \mathrm dt\, \left[
e^{-i\Delta E t} + e^{i\Delta E t} \right] = 2\pi\delta(\Delta E).
\end{equation}
The factor multiplying the delta function is regular at $\Delta E=0$, so that
\begin{equation}
\frac{ e^{\beta_{\mathrm{EQ}}\Delta E}-1 }{ \beta_{\mathrm{EQ}}\Delta E } \delta(\Delta E) = \delta(\Delta E).
\end{equation}
Substituting these results into the definition of $\mathcal N_\chi$, we get
\begin{align} \frac{\mathcal N_\chi}{V} = \lambda^2 m_\chi^2 \mathlarger{\int}
&\frac{\mathrm d^3\vec{k}}{(2\pi)^3\,2E_{\vec{k}}}\, \frac{\mathrm d^3\vec{p}}{(2\pi)^3\,2E_{\vec{p}}}\, \frac{\mathrm d^3\vec{q}}{(2\pi)^3\,2E_{\vec{q}}}\, (2\pi)^4  \delta^{(4)}\!\!\:\big(k-p-q\big)\notag\\
&\hspace{2.6cm}\times f_\chi^{\mathrm{EQ}}(E_{\vec{k}}) \!\left[ 1+f_\chi^{\mathrm{EQ}}(E_{\vec{k}}) \right] \!\left[ 1 + f_\phi^{\mathrm{EQ}}(E_{\vec{p}}) + f_\phi^{\mathrm{EQ}}(E_{\vec{q}}) \right].
\label{eq:app_scalar_corr_result}
\end{align}
Combining all previous results, the explicit expression for the relaxation rate then gives 
\begin{align}
\frac{\Gamma_{\mathrm{relax}}^{(\chi)}}{V} &= \frac{\lambda^2 m_\chi^2} {\langle Q_{\chi,\mathrm I}^{2}\rangle_{\mathrm{EQ}}} \mathlarger{\int} \mathrm d\Pi_\chi(\vec{k})\, \mathrm d\Pi_\phi(\vec{p})\, \mathrm d\Pi_\phi(\vec{q})\, (2\pi)^4 \delta^{(4)}\!\!\:\big(k-p-q\big)\notag \\
&\hspace{2.6cm}\times f_\chi^{\mathrm{EQ}}(E_{\vec{k}}) \!\left[ 1+f_\chi^{\mathrm{EQ}}(E_{\vec{k}}) \right] \!\left[ 1+f_\phi^{\mathrm{EQ}}(E_{\vec{p}}) +f_\phi^{\mathrm{EQ}}(E_{\vec{q}}) \right] ,
\label{app:scalar_model_Gamma_phasespace}
\end{align}
where
\begin{equation}
\mathrm d\Pi_a(\vec{p}) \equiv \frac{\mathrm d^3\vec{p}}{(2\pi)^3\,2E_{\vec{p}}^{a}}\, , \qquad \text{with }\quad E_{\vec{k}}=\sqrt{\vec{k}^2+m_\chi^2} \quad \text{and}\quad  E_{\vec{p}}=|\vec{p}|\, ,
\end{equation}
is the Lorentz-invariant phase space.

\subsection{Boltzmann calculation}
\label{app:scalar_model_boltzmann}

We now give the details of the collision terms~\eqref{eq:scalar_model_Deltachi} and of the linearization~\eqref{eq:scalar_model_delta_linearized} used in Section~\ref{sec:boltzmann}. We begin by explaining the statistical
factors $\Delta_\chi$ and $\Delta_{\overline\chi}$ entering the collision terms~\eqref{eq:scalar_model_Cchi}.\\

\noindent As recalled in equation~\eqref{app:eq:daggeraa-aadagger}, the phase-space distribution of a bosonic mode is related to the expectation value of its number operator. The same relations apply to the creation and annihilation operators of the $\phi$ field. It can be then shown that a bosonic mode in the initial state contributes a factor $f$, whereas a bosonic mode in the final state contributes a factor $1+f$ to the statistical part of the collision term; see, for example, references~\cite{Berges:2004yj,Enomoto:2023cun}. For an explicit derivation of the collision term within the Schwinger--Keldysh formalism, including its extension to a complex scalar field, see reference~\cite{Weinstock:2005jw}.\\

\noindent Consider first the decay
\begin{equation}
\chi(k)\to\phi(p)\phi(q).
\end{equation}
Since the two final-state $\phi$ particles are identical, the corresponding phase-space integral carries the usual symmetry factor $1/2!$. The initial $\chi$ mode contributes a factor $f_\chi(\vec{k},t)$, while the two final-state bosons contribute the Bose enhancement factors $1+f_\phi(E_{\vec{p}})$ and
$1+f_\phi(E_{\vec{q}})$. The statistical factor associated with this process is therefore
\begin{equation}
f_\chi(\vec{k},t) \cbig[ 1+f_\phi(E_{\vec{p}}) \cbig] \!\!\:\cbig[ 1+f_\phi(E_{\vec{q}}) \cbig].
\end{equation}

\noindent For the inverse process
\begin{equation}
\phi(p)\phi(q)\to\chi(k),
\end{equation}
the two initial $\phi$ modes contribute the factors
$f_\phi(E_{\vec{p}})$ and $f_\phi(E_{\vec{q}})$, while the final-state $\chi$ boson contributes the Bose-enhancement factor $1+f_\chi(\vec{k},t)$.
\begin{equation}
f_\phi(E_{\vec{p}}) f_\phi(E_{\vec{q}}) \cbig[ 1+f_\chi(\vec{k},t) \cbig].
\end{equation}

\noindent The first process removes particles from the $\chi$ distribution, whereas the inverse process produces them. The collision term gives the net change of $f_\chi(\vec{k},t)$ and therefore contains the statistical factor for the inverse process minus that for the decay. In the convention adopted in equation~\eqref{eq:scalar_model_Cchi}, this difference is written with an overall minus sign by defining
\begin{align}
\Delta_\chi(\vec{k},\vec{p},\vec{q};t)
=f_\chi(\vec{k},t)
\cbig[
1+f_\phi(E_{\vec{p}})
\cbig]\!\!\:
\cbig[
1+f_\phi(E_{\vec{q}})
\cbig]-
f_\phi(E_{\vec{p}})
f_\phi(E_{\vec{q}})
\cbig[
1+f_\chi(\vec{k},t)
\cbig].
\end{align}
Thus, $\Delta_\chi$ is the statistical factor for the decay minus that for the inverse process, while the collision term itself is proportional to the inverse-process contribution minus the decay contribution. The same reasoning applies to the antiparticle process. $\Delta_{\overline \chi}$ can be obtained from the previous equation with the replacement  $f_\chi\to f_{\overline\chi}$.\\

\noindent Subtracting the two statistical factors, we find
\begin{align}
\Delta_\chi-\Delta_{\overline\chi} = \cbig[ f_\chi(\vec{k},t)-f_{\overline\chi}(\vec{k},t) \cbig] \!\!\:\cbig[ 1+f_\phi(E_{\vec{p}}) +f_\phi(E_{\vec{q}}) \cbig].
\end{align}
Finally, using the linearized distributions in equation~\eqref{eq:scalar_model_linearized_distributions}, we obtain
\begin{equation}
f_\chi(\vec{k},t)-f_{\overline\chi}(\vec{k},t) = 2\beta_{\mathrm{EQ}}\mu_\chi(t)\, f_\chi^{\mathrm{EQ}}(E_{\vec{k}}) \left[ 1+f_\chi^{\mathrm{EQ}}(E_{\vec{k}}) \right] +\mathcal O\big(\mu_\chi^2\big),
\end{equation}
and hence
\begin{align}
\Delta_\chi-\Delta_{\overline\chi} \:\!&= 2\beta_{\mathrm{EQ}}\mu_\chi(t)\, f_\chi^{\mathrm{EQ}}(E_{\vec{k}}) \left[ 1+f_\chi^{\mathrm{EQ}}(E_{\vec{k}}) \right] \notag\\
&\qquad\qquad\qquad\qquad\quad\!\times \left[ 1+f_\phi^{\mathrm{EQ}}(E_{\vec{p}}) +f_\phi^{\mathrm{EQ}}(E_{\vec{q}}) \right] +\mathcal O\big(\mu_\chi^2\big),
\end{align}
in agreement with the result given in the main text equation~\eqref{eq:scalar_model_delta_linearized}.
Note that at linear order around equilibrium, any term proportional to $\delta f_\phi = f_\phi - f_\phi^\mathrm{EQ}$ drops out because $f_\chi^\mathrm{EQ} = f_{\overline\chi}^{\mathrm{EQ}}$.

\bibliographystyle{CustomBibliography}  
\addcontentsline{toc}{section}{\protect\numberline{}References}
\bibliography{Literature}

@article{RabyMottola1990,
    author = "Mottola, Emil and Raby, Stuart",
    title = "{Baryon number dissipation at finite temperature in the standard model}",
    doi = "10.1103/PhysRevD.42.4202",
    journal = "Phys. Rev. D",
    volume = {42},
    number = {12},
    pages = "4202--4208",
    year = "1990",
    month = {Dec},
    publisher = {American Physical Society}
}

@article{ArnoldMcLerran1987,
    author = "Arnold, Peter Brockway and McLerran, Larry D.",
    title = {Sphalerons, small fluctuations, and baryon-number violation in electroweak theory},
    doi = "10.1103/PhysRevD.36.581",
    journal = "Phys. Rev. D",
    volume = {36},
    number = {2},
    pages = {581--595},
    year = "1987",
    month = {Jul},
    publisher = {American Physical Society}
}

@book{ZubarevBook1971,
    title={{Nonequilibrium Statistical Thermodynamics}},
    author={Zubarev, Dmitry Nikolayevich},
    year={1974},
    publisher={Springer US}, 
    note="[originally published as ``Neravnovesnaya Statisticheskaya Termodinamika'' in Russian (1971), Nauka Press]"
}

@book{ZubarevBook1996,
    title={{Statistical Mechanics of Nonequilibrium Processes. Volume 1: Basic Concepts, Kinetic Theory}},
    author={Zubarev, Dmitry Nikolayevich and Röpke, Gerd and Morozov, Vladimir},
    year={1996},
    publisher={Akademie Verlag, Berlin}
}

@book{ZubarevBook1997,
    title={{Statistical Mechanics of Nonequilibrium Processes. Volume 2: Relaxation and Hydrodynamic Processes}},
    author={Zubarev, Dmitry Nikolayevich and Röpke, Gerd and Morozov, Vladimir},
    year={1997},
    publisher={Akademie Verlag, Berlin}
}

@article{ZubarevArticle1970,
    author={Zubarev, D. N.},
    title={Boundary conditions for statistical operators in the theory of nonequilibrium processes and quasiaverages},
    journal={Theor. Math. Phys.},
    year={1970},
    month={Jun},
    day={01},
    volume={3},
    number={2},
    pages={505--512},
    doi={10.1007/BF01046515},
    url={https://doi.org/10.1007/BF01046515}
}

@article{RoepkeArticle2018,
    author = "Röpke, Gerd",
    title={Electrical Conductivity of Charged Particle Systems and Zubarev's Nonequilibrium Statistical Operator Method},
    journal={Theor. Math. Phys.},
    volume={194},
    number={1},
    pages={74--104},
    doi={10.1134/S0040577918010063},
    eprint = "1809.03357",
    archivePrefix = "arXiv",
    primaryClass = "physics.plasm-ph",
    year = "2018"
}

@book{RoepkeBook2013,
    title={Nonequilibrium Statistical Physics},
    author={Röpke, Gerd},
    year={2013},
    publisher={Wiley-VCH}
}

@article{BecattiniArticle2019,
    author = "Becattini, F. and Buzzegoli, M. and Grossi, E.",
    title = "{Reworking Zubarev’s Approach to Nonequilibrium Quantum Statistical Mechanics}",
    eprint = "1902.01089",
    archivePrefix = "arXiv",
    primaryClass = "cond-mat.stat-mech",
    doi = "10.3390/particles2020014",
    journal = "Particles",
    volume = "2",
    number = "2",
    pages = "197--207",
    year = "2019"
}

@article{KhlebnikovShaposhnikov1988,
    author = "Khlebnikov, S. Yu. and Shaposhnikov, M. E.",
    title = "{The statistical theory of anomalous fermion number non-conservation}",
    doi = "10.1016/0550-3213(88)90133-2",
    journal = "Nucl. Phys. B",
    volume = "308",
    number = {4},
    pages = "885--912",
    year = "1988"
}

@inproceedings{Moore2000,
    author = "Moore, Guy D.",
    title = "{Do we understand the sphaleron rate?}",
    booktitle = "{Strong and Electroweak Matter 2000}",
    eprint = "hep-ph/0009161",
    archivePrefix = "arXiv",
    doi = "10.1142/9789812799913_0007",
    pages = "82--94",
    month = "6",
    year = "2000"
}

@article{Moore:2010jd,
    author = "Moore, Guy D. and Tassler, Marcus",
    title = "{The sphaleron rate in SU(N) gauge theory}",
    eprint = "1011.1167",
    archivePrefix = "arXiv",
    primaryClass = "hep-ph",
    doi = "10.1007/JHEP02(2011)105",
    journal = "JHEP",
    volume = "2011",
    number = "2",
    pages = "105",
    year = "2011",
    month={Feb}
}

@article{Hatano:2026tdm,
    author = "Hatano, Naomichi and Ordonez, Gonzalo",
    title = "{Non-Hermitian Quantum Mechanics of Open Quantum Systems: Revisiting The One-Body Problem}",
    eprint = "2602.14105",
    archivePrefix = "arXiv",
    primaryClass = "quant-ph",
    month = "2",
    year = "2026"
}

@article{GamowAlphaDecay,
    author={Gamow, G.},
    title={Zur Quantentheorie des Atomkernes},
    journal={Z. Phys.},
    year={1928},
    month={Mar},
    day={01},
    volume={51},
    number={3},
    pages={204--212},
    issn={0044-3328},
    doi={10.1007/BF01343196}
}

@article{SiegertRadiativeStates,
    author = "Siegert, A. J. F.",
    title = "{On the Derivation of the Dispersion Formula for Nuclear Reactions}",
    doi = "10.1103/PhysRev.56.750",
    journal = "Phys. Rev.",
    volume = "56",
    number = {8},
    pages = "750--752",
    year = "1939",
    month = {Oct},
    publisher = {American Physical Society}
}

@article{IntroductionGamovVectors,
    author = "de la Madrid, R. and Gadella, M.",
    title = "{A pedestrian introduction to Gamow vectors}",
    eprint = "quant-ph/0201091",
    archivePrefix = "arXiv",
    doi = "10.1119/1.1466817",
    journal = "Am. J. Phys.",
    volume = "70",
    number = {6},
    pages = "626--638",
    year = "2002",
    month = {06}
}

@article{Enomoto:2023cun,
    author = "Enomoto, Seishi and Su, Yu-Hang and Zheng, Man-Zhu and Zhang, Hong-Hao",
    title = "{Boltzmann equation and its cosmological applications}",
    eprint = "2301.11819",
    archivePrefix = "arXiv",
    primaryClass = "hep-ph",
    doi = "10.3390/sym17060921",
    journal = "Symmetry",
    volume = "17",
    number = "6",
    pages = "921",
    year = "2025"
}

@article{Berges:2004yj,
    author = "Berges, Juergen",
    editor = "Bracco, Mirian and Chiapparini, Marcelo and Ferreira, Erasmo and Kodama, Takeshi",
    title = "{Introduction to Nonequilibrium Quantum Field Theory}",
    eprint = "hep-ph/0409233",
    archivePrefix = "arXiv",
    doi = "10.1063/1.1843591",
    journal = "AIP Conf. Proc.",
    volume = "739",
    number = "1",
    pages = "3--62",
    year = "2004"
}

@article{Weinstock:2005jw,
    author = "Weinstock, Steffen",
    title = "{Boltzmann collision term}",
    eprint = "hep-ph/0510417",
    archivePrefix = "arXiv",
    doi = "10.1103/PhysRevD.73.025005",
    journal = "Phys. Rev. D",
    volume = "73",
    number = {2},
    pages = "025005",
    numpages = {9},
    year = "2006",
    month = {Jan}
}

@article{Harvey:1990qw,
    author = "Harvey, Jeffrey A. and Turner, Michael S.",
    title = {Cosmological baryon and lepton number in the presence of electroweak fermion-number violation},
    doi = "10.1103/PhysRevD.42.3344",
    journal = "Phys. Rev. D",
    volume = "42",
    number = {10},
    pages = "3344--3349",
    year = "1990",
    month = {Nov}
}

@book{LeBellac:1996,
    author    = {Le Bellac, Michel},
    title     = {Thermal Field Theory},
    publisher = {Cambridge University Press},
    series    = {Cambridge Monographs on Mathematical Physics},
    year      = {1996},
    doi       = {10.1017/CBO9780511721700},
    isbn      = {978-0-521-65477-7}
}

@article{Bodeker:1999gx,
    author = "B{\"o}deker, D. and Moore, Guy D. and Rummukainen, K.",
    title = "{Chern--Simons number diffusion and hard thermal loops on the lattice}",
    eprint = "hep-ph/9907545",
    archivePrefix = "arXiv",
    doi = "10.1103/PhysRevD.61.056003",
    journal = "Phys. Rev. D",
    volume = "61",
    number = "5",
    pages = "056003",
    numpages = {21},
    year = "2000",
    month = {Feb}
}

@article{Arnold:1996dy,
    author = "Arnold, Peter Brockway and Son, Dam and Yaffe, Laurence G.",
    title = {The hot baryon violation rate is $O\big({\ensuremath{\alpha}}_{\mathrm{w}}^{5}{T}^{4}\big)$},
    eprint = "hep-ph/9609481",
    archivePrefix = "arXiv",
    doi = "10.1103/PhysRevD.55.6264",
    journal = "Phys. Rev. D",
    volume = "55",
    number = {10},
    pages = "6264--6273",
    year = "1997",
    month = {May}
}

@article{Bodeker:1998hm,
    author = "B{\"o}deker, Dietrich",
    title = "{Effective dynamics of soft non-Abelian gauge fields at finite temperature}",
    eprint = "hep-ph/9801430",
    archivePrefix = "arXiv",
    doi = "10.1016/S0370-2693(98)00279-2",
    journal = "Phys. Lett. B",
    volume = "426",
    number = {3},
    pages = "351--360",
    year = "1998"
}

@article{DOnofrio:2014rug,
    author = "D'Onofrio, Michela and Rummukainen, Kari and Tranberg, Anders",
    title = "{Sphaleron Rate in the Minimal Standard Model}",
    eprint = "1404.3565",
    archivePrefix = "arXiv",
    primaryClass = "hep-ph",
    doi = "10.1103/PhysRevLett.113.141602",
    journal = "Phys. Rev. Lett.",
    volume = "113",
    number = "14",
    pages = "141602",
    numpages = {5},
    year = "2014",
    month = {Oct}
}

@article{Kubo1957,
  title={Statistical-Mechanical Theory of Irreversible Processes. I. General Theory and Simple Applications to Magnetic and Conduction Problems},
  author={Ryogo Kubo},
  journal={J. Phys. Soc. Jpn.},
  volume={12},
  number={6},
  pages={570--586},
  year={1957},
  doi={10.1143/JPSJ.12.570}
}

@article{KuboYokotaNakajima1957,
  title={Statistical-Mechanical Theory of Irreversible Processes. II. Response to Thermal Disturbance},
  author={Ryogo Kubo and Mario Yokota and Sadao Nakajima},
  journal={J. Phys. Soc. Jpn.},
  volume={12},
  number={11},
  pages={1203--1211},
  year={1957},
  doi={10.1143/JPSJ.12.1203}
}

@article{Zwanzig:1961zz,
    author = "Zwanzig, Robert",
    title = "{Memory Effects in Irreversible Thermodynamics}",
    doi = "10.1103/PhysRev.124.983",
    journal = "Phys. Rev.",
    volume = "124",
    pages = "983--992",
    year = "1961"
}

@article{Nakajima:1958pnl,
    author = {Nakajima, Sadao},
    title = {On Quantum Theory of Transport Phenomena: Steady Diffusion},
    journal = "Prog. Theor. Phys.",
    volume = {20},
    number = {6},
    pages = {948--959},
    year = {1958},
    doi = {10.1143/PTP.20.948}
}

@article{Kadanoff:1963axw,
    author = "Kadanoff, Leo P. and Martin, Paul C.",
    title = "{Hydrodynamic equations and correlation functions}",
    doi = "10.1016/0003-4916(63)90078-2",
    journal = "Annals Phys.",
    volume = "24",
    pages = "419--469",
    year = "1963"
}

@article{Mori:1965oqj,
    author = "Mori, Hazime",
    title = "{Transport, Collective Motion, and Brownian Motion}",
    doi = "10.1143/PTP.33.423",
    journal = "Prog. Theor. Phys.",
    volume = "33",
    number = "3",
    pages = "423--455",
    year = "1965"
}

@article{Mabillard2023,
  title = {Quantum local-equilibrium approach to dissipative hydrodynamics},
  author = {Mabillard, Jo\"el and Gaspard, Pierre},
  journal = {Phys. Rev. E},
  volume = {107},
  number = {1},
  pages = {014102},
  numpages = {15},
  year = {2023},
  month = {Jan},
  publisher = {American Physical Society},
  doi = {10.1103/PhysRevE.107.014102},
  eprint = "2208.02544",
  archivePrefix = "arXiv",
  primaryClass = "cond-mat.stat-mech"
}

@article{Manton:1983nd,
    author = "Manton, N. S.",
    title = "{Topology in the Weinberg-Salam theory}",
    doi = "10.1103/PhysRevD.28.2019",
    journal = "Phys. Rev. D",
    volume = "28",
    number = {8},
    pages = "2019--2026",
    year = "1983",
    month = {Oct}
}

@article{Klinkhamer:1984di,
    author = "Klinkhamer, Frans R. and Manton, N. S.",
    title = "{A saddle-point solution in the Weinberg-Salam theory}",
    reportNumber = "NSF-ITP-84-57",
    doi = "10.1103/PhysRevD.30.2212",
    journal = "Phys. Rev. D",
    volume = "30",
    number = "10",
    pages = "2212--2220",
    year = "1984",
    month = {Nov}
}

@article{Kuzmin:1985mm,
    author = "Kuzmin, V. A. and Rubakov, V. A. and Shaposhnikov, M. E.",
    title = "{On anomalous electroweak baryon-number non-conservation in the early universe}",
    reportNumber = "IC/85/8",
    doi = "10.1016/0370-2693(85)91028-7",
    journal = "Phys. Lett. B",
    volume = "155",
    number = {1},
    pages = "36--42",
    year = "1985"
}

@article{Burnier:2005hp,
    author = "Burnier, Y. and Laine, M. and Shaposhnikov, M.",
    title = "{Baryon and lepton number violation rates across the electroweak crossover}",
    eprint = "hep-ph/0511246",
    archivePrefix = "arXiv",
    reportNumber = "BI-TP-2005-48",
    doi = "10.1088/1475-7516/2006/02/007",
    journal = "JCAP",
    volume = "02",
    pages = "007",
    year = "2006"
}

@article{Lindner:2005kv,
    author = "Lindner, Manfred and M{\"u}ller, Markus Michael",
    title = "{Comparison of Boltzmann equations with quantum dynamics for scalar fields}",
    eprint = "hep-ph/0512147",
    archivePrefix = "arXiv",
    doi = "10.1103/PhysRevD.73.125002",
    journal = "Phys. Rev. D",
    volume = "73",
    number = {12},
    pages = "125002",
    numpages = {13},
    year = "2006",
    month = {Jun}
}

@article{Grozdanov:2018fic,
    author = "Grozdanov, Sa{\v{s}}o and Lucas, Andrew and Poovuttikul, Napat",
    title = "{Holography and hydrodynamics with weakly broken symmetries}",
    eprint = "1810.10016",
    archivePrefix = "arXiv",
    primaryClass = "hep-th",
    doi = "10.1103/PhysRevD.99.086012",
    journal = "Phys. Rev. D",
    volume = "99",
    number = "8",
    numpages = {36},
    pages = "086012",
    year = "2019"
}

@article{Harutyunyan:2021rmb,
    author = "Harutyunyan, Arus and Sedrakian, Armen and Rischke, Dirk H.",
    title = "{Relativistic second-order dissipative hydrodynamics from Zubarev’s non-equilibrium statistical operator}",
    eprint = "2110.04595",
    archivePrefix = "arXiv",
    primaryClass = "nucl-th",
    doi = "10.1016/j.aop.2022.168755",
    journal = "Annals Phys.",
    volume = "438",
    pages = "168755",
    year = "2022"
}

@article{Moore:1996qs,
    author = "Moore, Guy D.",
    title = "{Motion of Chern--Simons number at high temperatures under a chemical potential}",
    eprint = "hep-ph/9603384",
    archivePrefix = "arXiv",
    doi = "10.1016/S0550-3213(96)00445-2",
    journal = "Nucl. Phys. B",
    volume = "480",
    number = {3},
    pages = "657--688",
    year = "1996"
}

@article{Jeon:1994if,
    author = "Jeon, Sangyong",
    title = "{Hydrodynamic transport coefficients in relativistic scalar field theory}",
    eprint = "hep-ph/9409250",
    archivePrefix = "arXiv",
    doi = "10.1103/PhysRevD.52.3591",
    journal = "Phys. Rev. D",
    volume = "52",
    number = {6},
    pages = "3591--3642",
    year = "1995",
    month = {Sep}
}

@article{Zub61,
  author  = {Zubarev, D. N.},
  title   = {The Statistical Operator for Nonequilibrium Systems},
  journal = {Doklady Akademii Nauk SSSR},
  volume  = {140},
  number  = {1},
  pages   = {92--95},
  year    = {1961},
  url     = {http://mi.mathnet.ru/dan25474},
  mrnumber = {0139434},
  zbl      = {0112.22204}
}

@article{Kuz18,
  author        = {Kuzemsky, A. L.},
  title         = {Nonequilibrium Statistical Operator Method and Generalized Kinetic Equations},
  journal       = {Theor. Math. Phys.},
  volume        = {194},
  number        = {1},
  pages         = {30--56},
  year          = {2018},
  month         = {Jan},
  doi           = {10.1134/S004057791801004X},
  url           = {https://doi.org/10.1134/S004057791801004X}
}

@article{Bodeker:2012gs,
    author = "B{\"o}deker, D. and Laine, M.",
    title = "{Heavy quark chemical equilibration rate as a transport coefficient}",
    eprint = "1205.4987",
    archivePrefix = "arXiv",
    primaryClass = "hep-ph",
    doi = "10.1007/JHEP07(2012)130",
    journal = "JHEP",
    volume = "2012",
    number = "7",
    pages = "130",
    year = "2012",
    month = "Jul"
}

@article{Bodeker:2014hqa,
    author = "B{\"o}deker, D. and Laine, M.",
    title = "{Kubo relations and radiative corrections for lepton number washout}",
    eprint = "1403.2755",
    archivePrefix = "arXiv",
    primaryClass = "hep-ph",
    reportNumber = "BI-TP-2014-03",
    doi = "10.1088/1475-7516/2014/05/041",
    journal = "JCAP",
    volume = "2014",
    number = "05",
    pages = "041",
    year = "2014",
    month = "May"
}

@article{Bamler20215,
  title = {Equilibration and approximate conservation laws: Dipole oscillations and perfect drag of ultracold atoms in a harmonic trap},
  author = {Bamler, Robert and Rosch, Achim},
  journal = {Phys. Rev. A},
  volume = {91},
  number = {6},
  pages = {063604},
  numpages = {13},
  year = {2015},
  publisher = {American Physical Society},
  doi = {10.1103/PhysRevA.91.063604}
}

@article{Armas:2021vku,
    author = "Armas, Jay and Jain, Akash and Lier, Ruben",
    title = "{Approximate symmetries, pseudo-Goldstones, and the second law of thermodynamics}",
    eprint = "2112.14373",
    archivePrefix = "arXiv",
    primaryClass = "hep-th",
    doi = "10.1103/PhysRevD.108.086011",
    journal = "Phys. Rev. D",
    volume = "108",
    number = "8",
    numpages = "13",
    pages = "086011",
    year = "2023",
    month = "Oct"
}

@article{Hongo:2024brb,
    author = "Hongo, Masaru and Sogabe, Noriyuki and Stephanov, Mikhail A. and Yee, Ho-Ung",
    title = "{Schwinger--Keldysh effective action for hydrodynamics with approximate symmetries}",
    eprint = "2411.08016",
    archivePrefix = "arXiv",
    primaryClass = "hep-th",
    doi = "10.1103/d6y8-q8dz",
    journal = "Phys. Rev. D",
    volume = "113",
    number = "1",
    pages = "014020",
    numpages = "20",
    year = "2026",
    month = "Jan"
}

@article{Kovtun:2012rj,
    author = "Kovtun, Pavel",
    title = "{Lectures on hydrodynamic fluctuations in relativistic theories}",
    eprint = "1205.5040",
    archivePrefix = "arXiv",
    primaryClass = "hep-th",
    doi = "10.1088/1751-8113/45/47/473001",
    journal = "J. Phys. A",
    volume = "45",
    number = {47},
    pages = "473001",
    year = "2012",
    month = {Nov},
}

@article{Calzetta:1986cq,
    author = "Calzetta, E. and Hu, B. L.",
    title = "{Nonequilibrium quantum fields: Closed-time-path effective action, Wigner function, and Boltzmann equation}",
    doi = "10.1103/PhysRevD.37.2878",
    journal = "Phys. Rev. D",
    volume = "37",
    issue = {10},
    pages = "2878--2900",
    year = "1988",
    month = {May}
}

@article{Vrugt:2019qhr,
    author = "te Vrugt, Michael and Wittkowski, Raphael",
    title = "{Projection operators in statistical mechanics: a pedagogical approach}",
    eprint = "2001.01572",
    archivePrefix = "arXiv",
    primaryClass = "cond-mat.stat-mech",
    doi = "10.1088/1361-6404/ab8e28",
    journal = "Eur. J. Phys.",
    volume = "41",
    number = {4},
    pages = "045101",
    year = "2020",
    month = {jun}
}

@article{Kubo:1966,
    title = {The fluctuation-dissipation theorem},
    author = {Kubo, R.},
    journal = {Rep. Prog. Phys.}, 
    doi = {10.1088/0034-4885/29/1/306},
    year = {1966},
    month = {Jan},
    volume = {29},
    number = {1},
    pages = {255}
}

@article{ZwanzigProjection1960,
    author = {Zwanzig, Robert},
    title = {Ensemble Method in the Theory of Irreversibility},
    journal = {J. Chem. Phys.},
    volume = {33},
    number = {5},
    pages = {1338--1341},
    year = {1960},
    month = {11},
    issn = {0021-9606},
    doi = {10.1063/1.1731409}
}

@article{Callen:1951,
  title = {Irreversibility and Generalized Noise},
  author = {Callen, Herbert B. and Welton, Theodore A.},
  journal = {Phys. Rev.},
  volume = {83},
  number = {1},
  pages = {34--40},
  year = {1951},
  month = {Jul},
  doi = {10.1103/PhysRev.83.34}
}

@article{Green:1952,
    author = {Green, Melville S.},
    title = {Markoff Random Processes and the Statistical Mechanics of Time‐Dependent Phenomena},
    journal = "J. Chem. Phys.",
    volume = {20},
    number = {8},
    pages = {1281--1295},
    year = {1952},
    doi = {10.1063/1.1700722}
}

@article{Green:1954ubq,
    author = {Green, Melville S.},
    title = {Markoff Random Processes and the Statistical Mechanics of Time‐Dependent Phenomena. II. Irreversible Processes in Fluids},
    journal = "J. Chem. Phys.",
    volume = {22},
    number = {3},
    pages = {398--413},
    year = {1954},
    doi = {10.1063/1.1740082}
}

@article{Robertsen1966,
  title = {Equations of Motion in Nonequilibrium Statistical Mechanics},
  author = {Robertson, Baldwin},
  journal = {Phys. Rev.},
  volume = {144},
  number = {1},
  pages = {151--161},
  numpages = {0},
  year = {1966},
  month = {Apr},
  doi = {10.1103/PhysRev.144.151}
}

@article{Jaynes:1957,
  title = {Information Theory and Statistical Mechanics},
  author = {Jaynes, E. T.},
  journal = {Phys. Rev.},
  volume = {106},
  issue = {4},
  pages = {620--630},
  year = {1957},
  month = {May},
  doi = {10.1103/PhysRev.106.620}
}

@article{Jaynes:1957II,
    author = "Jaynes, E. T.",
    title = "{Information Theory and Statistical Mechanics. II}",
    doi = "10.1103/PhysRev.108.171",
    journal = "Phys. Rev.",
    volume = "108",
    number = "2",
    pages = "171--190",
    year = "1957"
}

@book{Forster:1975,
  title = {Hydrodynamic Fluctuations, Broken Symmetry, And Correlation Functions},
  series = {Frontiers in Physics},
  volume = {47},
  author = {Forster, D.}, 
  year = {1975},
  month = {May},
  publisher={W. A. Benjamin}
}

@book{Grabert:1975,
  title = {Projection Operator Techniques in Nonequilibrium Statistical Mechanics},
  series = {Springer Tracts in Modern Physics},
  volume = {95},
  author = {Hermann Grabert}, 
  year = {1982},
  publisher={Springer Berlin}, 
  doi = {10.1007/BFb0044591}
}

@article{Hosoya:1983id,
    author = "Hosoya, Akio and Sakagami, Masa-aki and Takao, Masaru",
    title = "{Nonequilibrium thermodynamics in field theory: Transport coefficients}",
    doi = "10.1016/0003-4916(84)90144-1",
    journal = "Annals Phys.",
    volume = {154},
    number = {1},
    pages = {229--252},
    year = "1984"
}

@article{vanWeert:1983,
    title = {Maximum entropy principle and relativistic hydrodynamics},
    author = {Ch.G {van Weert}},
    journal = {Annals Phys.},
    volume = {140},
    number = {1},
    pages = {133--162},
    year = {1982},
    doi = {https://doi.org/10.1016/0003-4916(82)90338-4}
}

@article{tHooft:1976rip,
    author = "'t Hooft, Gerard",
    title = "{Symmetry Breaking through Bell-Jackiw Anomalies}",
    doi = "10.1103/PhysRevLett.37.8",
    journal = "Phys. Rev. Lett.",
    volume = "37",
    number = {1},
    pages = "8--11",
    year = "1976",
    month = {Jul}
}

@article{Adler:1969gk,
    author = "Adler, Stephen L.",
    title = "{Axial vector vertex in spinor electrodynamics}",
    doi = "10.1103/PhysRev.177.2426",
    journal = "Phys. Rev.",
    volume = "177",
    number = {5},
    pages = "2426--2438",
    year = "1969",
    month = {Jan}
}

@article{Bell:1969ts,
    author = "Bell, J. S. and Jackiw, R.",
    title = "{A PCAC puzzle: $\pi^0 \to \gamma \gamma$ in the $\sigma$ model}",
    doi = "10.1007/BF02823296",
    journal = "Nuovo Cim. A",
    volume = "60",
    number = {1},
    pages = "47--61",
    year = "1969"
}

@article{Rubakov:1996,
	author = {V. A. Rubakov and M. E. Shaposhnikov},
	title = {Electroweak baryon number non-conservation in the early Universe and in high-energy collisions},
	year = {1996},
	journal = {Phys. Usp.},
	volume = {39},
	number = {5},
	pages = {461--502},
	doi = {10.1070/PU1996v039n05ABEH000145},
    eprint = "hep-ph/9603208",
    archivePrefix = "arXiv",
}

@article{Bodeker:2019ajh,
    author = {B{\"o}deker, Dietrich and Schr{\"o}der, Dennis},
    title = "{Equilibration of right-handed electrons}",
    eprint = "1902.07220",
    archivePrefix = "arXiv",
    primaryClass = "hep-ph",
    doi = "10.1088/1475-7516/2019/05/010",
    journal = "JCAP",
    volume = {2019},
    number = {05},
    pages = "010",
    year = "2019",
    month = {May}
}

@article{Jeon:1995zm,
  title = {From quantum field theory to hydrodynamics: Transport coefficients and effective kinetic theory},
  author = {Jeon, Sangyong and Yaffe, Laurence G.},
  journal = {Phys. Rev. D},
  volume = {53},
  issue = {10},
  pages = {5799--5809},
  year = {1996},
  month = {May},
  doi = {10.1103/PhysRevD.53.5799},
  eprint = "hep-ph/9512263",
  archivePrefix = "arXiv",
}

@article{Bogoliubov:1946,
  title = {Kinetic Equations},
  author = {N. N. Bogoliubov},
  journal = { J. Phys. (USSR)},
  volume = {10},
  issue = {257},
  pages = {265},
  year = {1946}
}

@article{Zub80,
  author  = {Zubarev, D. N.},
  title   = {Contemporary methods of the statistical theory of nonequilibrium processes},
  journal = {J. Sov. Math.},
  year    = {1981},
  volume  = {16},
  number  = {6},
  pages   = {1509--1571},
  doi     = {10.1007/BF01091712},
  note    = {Translated from \emph{Itogi Nauki i Tekhniki. Ser. Sovrem. Probl. Mat.}, vol. \emph{15} (1980), pp. 131--226}
}

@article{Hayata:2015lga,
    author = "Hayata, Tomoya and Hidaka, Yoshimasa and Noumi, Toshifumi and Hongo, Masaru",
    title = "{Relativistic hydrodynamics from quantum field theory on the basis of the generalized Gibbs ensemble method}",
    eprint = "1503.04535",
    archivePrefix = "arXiv",
    primaryClass = "hep-ph",
    doi = "10.1103/PhysRevD.92.065008",
    journal = "Phys. Rev. D",
    volume = "92",
    number = "6",
    pages = "065008",
    year = "2015"
}

@article{Hidaka:2022dmn,
    author = "Hidaka, Yoshimasa and Pu, Shi and Wang, Qun and Yang, Di-Lun",
    title = "{Foundations and applications of quantum kinetic theory}",
    eprint = "2201.07644",
    archivePrefix = "arXiv",
    primaryClass = "hep-ph",
    doi = "10.1016/j.ppnp.2022.103989",
    journal = "Prog. Part. Nucl. Phys.",
    volume = "127",
    pages = "103989",
    year = "2022"
}

@article{Liu:2018kfw,
    author = "Liu, Hong and Glorioso, Paolo",
    title = "{Lectures on non-equilibrium effective field theories and fluctuating hydrodynamics}",
    eprint = "1805.09331",
    archivePrefix = "arXiv",
    primaryClass = "hep-th",
    reportNumber = "MIT-CTP/5018; EFI-18-8, MIT-CTP-5018, EFI-18-8",
    doi = "10.22323/1.305.0008",
    journal = "PoS",
    volume = "TASI2017",
    pages = "008",
    year = "2018"
}

@article{Firat:2025upx,
    author = "Firat, Eren and Gomes, Andrew and Nardi, Filippo and Penco, Riccardo and Rattazzi, Riccardo",
    title = "{Schwinger--Keldysh effective theory of charge transport: redundancies and systematic {\ensuremath{\omega}}/T expansion}",
    eprint = "2508.18346",
    archivePrefix = "arXiv",
    primaryClass = "hep-th",
    doi = "10.1007/JHEP05(2026)079",
    journal = "JHEP",
    volume = "05",
    pages = "079",
    year = "2026"
}

@article{Akyuz:2023lsm,
    author = "Akyuz, Can Onur and Goon, Garrett and Penco, Riccardo",
    title = "{The Schwinger--Keldysh coset construction}",
    eprint = "2306.17232",
    archivePrefix = "arXiv",
    primaryClass = "hep-th",
    doi = "10.1007/JHEP06(2024)004",
    journal = "JHEP",
    volume = "06",
    pages = "004",
    year = "2024"
}

@article{Cline:2000nw,
    author = "Cline, James M. and Joyce, Michael and Kainulainen, Kimmo",
    title = "{Supersymmetric electroweak baryogenesis}",
    eprint = "hep-ph/0006119",
    archivePrefix = "arXiv",
    doi = "10.1088/1126-6708/2000/07/018",
    journal = "JHEP",
    volume = "2000",
    number = "07",
    pages = "018",
    year = "2000"
}

@article{Bochkarev:1987wf,
    author = "Bochkarev, A. I. and Shaposhnikov, M. E.",
    title = "{Electroweak Production of Baryon Asymmetry and Upper Bounds on the Higgs and Top Masses}",
    reportNumber = "NBI-HE-87-18",
    doi = "10.1142/S0217732387000537",
    journal = "Mod. Phys. Lett. A",
    volume = "2",
    pages = "417",
    year = "1987"
}

@article{Figueroa:2017hun,
    author = "Figueroa, Daniel G. and Shaposhnikov, Mikhail",
    title = "{Anomalous non-conservation of fermion/chiral number in Abelian gauge theories at finite temperature}",
    eprint = "1707.09967",
    archivePrefix = "arXiv",
    primaryClass = "hep-ph",
    doi = "10.1007/JHEP04(2018)026",
    journal = "JHEP",
    volume = "2018",
    number = "04",
    pages = "026",
    year = "2018",
    note = "[Erratum: \href{https://doi.org/10.1007/JHEP07(2020)217}{\emph{JHEP}, vol. \emph{2020}\,(07), p.\,217}]"
}

@article{Lucas:2015sya,
    author = "Lucas, Andrew and Crossno, Jesse and Fong, Kin Chung and Kim, Philip and Sachdev, Subir",
    title = "{Transport in inhomogeneous quantum critical fluids and in the Dirac fluid in graphene}",
    eprint = "1510.01738",
    archivePrefix = "arXiv",
    primaryClass = "cond-mat.str-el",
    doi = "10.1103/PhysRevB.93.075426",
    journal = "Phys. Rev. B",
    volume = "93",
    number = "7",
    pages = "075426",
    year = "2016"
}

@article{Bouchoule:2020lwr,
    author = "Bouchoule, Isabelle and Doyon, Benjamin and Dubail, Jerome",
    title = "{The effect of atom losses on the distribution of rapidities in the one-dimensional Bose gas}",
    eprint = "2006.03583",
    archivePrefix = "arXiv",
    primaryClass = "cond-mat.quant-gas",
    doi = "10.21468/SciPostPhys.9.4.044",
    journal = "SciPost Phys.",
    volume = "9",
    pages = "044",
    year = "2020"
}

@article{Mallayya:2019uqw,
    author = "Mallayya, Krishnanand and Rigol, Marcos and Roeck, Wojciech De",
    title = "{Prethermalization and Thermalization in Isolated Quantum Systems}",
    eprint = "1810.12320",
    archivePrefix = "arXiv",
    primaryClass = "cond-mat.stat-mech",
    doi = "10.1103/PhysRevX.9.021027",
    journal = "Phys. Rev. X",
    volume = "9",
    number = "2",
    pages = "021027",
    year = "2019"
}

@article{Surace:2023wqq,
    author = "Surace, Federica Maria and Motrunich, Olexei",
    title = "{Weak integrability breaking perturbations of integrable models}",
    eprint = "2302.12804",
    archivePrefix = "arXiv",
    primaryClass = "cond-mat.stat-mech",
    doi = "10.1103/PhysRevResearch.5.043019",
    journal = "Phys. Rev. Res.",
    volume = "5",
    number = "4",
    pages = "043019",
    year = "2023"
}

@article{Delacretaz:2019brr,
    author = "Delacr{\'e}taz, Luca V. and Hofman, Diego M. and Mathys, Gr{\'e}goire",
    title = "{Superfluids as Higher-form Anomalies}",
    eprint = "1908.06977",
    archivePrefix = "arXiv",
    primaryClass = "hep-th",
    doi = "10.21468/SciPostPhys.8.3.047",
    journal = "SciPost Phys.",
    volume = "8",
    pages = "047",
    year = "2020"
}

@article{Armas:2023tyx,
    author = "Armas, Jay and Jain, Akash",
    title = "{Approximate higher-form symmetries, topological defects, and dynamical phase transitions}",
    eprint = "2301.09628",
    archivePrefix = "arXiv",
    primaryClass = "hep-th",
    doi = "10.1103/PhysRevD.109.045019",
    journal = "Phys. Rev. D",
    volume = "109",
    number = "4",
    pages = "045019",
    year = "2024"
}

@article{Hartnoll:2012rj,
    author = "Hartnoll, Sean A. and Hofman, Diego M.",
    title = "{Locally Critical Resistivities from Umklapp Scattering}",
    eprint = "1201.3917",
    archivePrefix = "arXiv",
    primaryClass = "hep-th",
    doi = "10.1103/PhysRevLett.108.241601",
    journal = "Phys. Rev. Lett.",
    volume = "108",
    pages = "241601",
    year = "2012"
}

@article{Garbrecht:2018mrp,
    author = {Garbrecht, Bj{\"o}rn},
    title = "{Why is there more matter than antimatter? Calculational methods for leptogenesis and electroweak baryogenesis}",
    eprint = "1812.02651",
    archivePrefix = "arXiv",
    primaryClass = "hep-ph",
    reportNumber = "TUM-HEP-1177-18",
    doi = "10.1016/j.ppnp.2019.103727",
    journal = "Prog. Part. Nucl. Phys.",
    volume = "110",
    pages = "103727",
    year = "2020"
}

@article{Morrissey:2012db,
    author = "Morrissey, David E. and Ramsey-Musolf, Michael J.",
    title = "{Electroweak baryogenesis}",
    eprint = "1206.2942",
    archivePrefix = "arXiv",
    primaryClass = "hep-ph",
    reportNumber = "NPAC-12-08",
    doi = "10.1088/1367-2630/14/12/125003",
    journal = "New J. Phys.",
    volume = "14",
    pages = "125003",
    year = "2012"
}

@article{Shaposhnikov:1986jp,
    author = "Shaposhnikov, M. E.",
    title = "{Possible Appearance of the Baryon Asymmetry of the Universe in an Electroweak Theory}",
    journal = "JETP Lett.",
    volume = "44",
    pages = "465--468",
    year = "1986"
}

@article{Carena:1996wj,
    author = "Carena, Marcela and Quiros, M. and Wagner, C. E. M.",
    title = "{Opening the window for electroweak baryogenesis}",
    eprint = "hep-ph/9603420",
    archivePrefix = "arXiv",
    reportNumber = "CERN-TH-96-30, IEM-FT-126-96",
    doi = "10.1016/0370-2693(96)00475-3",
    journal = "Phys. Lett. B",
    volume = "380",
    pages = "81--91",
    year = "1996"
}

@article{Nelson:1991ab,
    author = "Nelson, A. E. and Kaplan, D. B. and Cohen, Andrew G.",
    title = "{Why there is something rather than nothing: Matter from weak interactions}",
    reportNumber = "UCSD-PTH-91-20, BUHEP-91-15",
    doi = "10.1016/0550-3213(92)90440-M",
    journal = "Nucl. Phys. B",
    volume = "373",
    pages = "453--478",
    year = "1992"
}

@article{Benso:2026csa,
    author = {Benso, Cristina and B{\"o}deker, Dietrich and Kamada, Kohei and Mukaida, Kyohei and Sagunski, Laura and Schicho, Philipp and Schmitz, Kai},
    title = "{Baryon number freeze-out in the Standard Model, precisely}",
    eprint = "2609.07467",
    archivePrefix = "arXiv",
    primaryClass = "hep-ph",
    reportNumber = "MS-TP-26-25, KEK-TH-2869",
    month = "9",
    year = "2026"
}

@article{Hindmarsh:2020hop,
    author = {Hindmarsh, Mark B. and L{\"u}ben, Marvin and Lumma, Johannes and Pauly, Martin},
    title = "{Phase transitions in the early universe}",
    eprint = "2008.09136",
    archivePrefix = "arXiv",
    primaryClass = "astro-ph.CO",
    reportNumber = "MPP-2020-163, HIP-2020-27/TH",
    doi = "10.21468/SciPostPhysLectNotes.24",
    journal = "SciPost Phys. Lect. Notes",
    volume = "24",
    pages = "1",
    year = "2021"
}

@article{Moore:1995ua,
    author = "Moore, Guy D. and Prokopec, Tomislav",
    title = "{Bubble wall velocity in a first order electroweak phase transition}",
    eprint = "hep-ph/9503296",
    archivePrefix = "arXiv",
    reportNumber = "PUPT-1531, LANCASTER-TH-9503",
    doi = "10.1103/PhysRevLett.75.777",
    journal = "Phys. Rev. Lett.",
    volume = "75",
    pages = "777--780",
    year = "1995"
}

@article{Moore:1995si,
    author = "Moore, Guy D. and Prokopec, Tomislav",
    title = "{How fast can the wall move? A Study of the electroweak phase transition dynamics}",
    eprint = "hep-ph/9506475",
    archivePrefix = "arXiv",
    reportNumber = "PUPT-1544, LANCS-TH-9517",
    doi = "10.1103/PhysRevD.52.7182",
    journal = "Phys. Rev. D",
    volume = "52",
    pages = "7182--7204",
    year = "1995"
}

@article{Ai:2025bjw,
    author = "Ai, Wen-Yuan and Carosi, Matthias and Garbrecht, Bj{\"o}rn and Tamarit, Carlos and Vanvlasselaer, Miguel",
    title = "{Bubble wall dynamics from nonequilibrium quantum field theory}",
    eprint = "2504.13725",
    archivePrefix = "arXiv",
    primaryClass = "hep-ph",
    reportNumber = "MITP-25-029",
    doi = "10.1007/JHEP08(2025)077",
    journal = "JHEP",
    volume = "08",
    pages = "077",
    year = "2025"
}

@article{Moore:1998swa,
    author = "Moore, Guy D.",
    title = "{Measuring the broken phase sphaleron rate nonperturbatively}",
    eprint = "hep-ph/9805264",
    archivePrefix = "arXiv",
    reportNumber = "MCGILL-98-7",
    doi = "10.1103/PhysRevD.59.014503",
    journal = "Phys. Rev. D",
    volume = "59",
    pages = "014503",
    year = "1999"
}

@article{Joyce:1994zn,
    author = "Joyce, Michael and Prokopec, Tomislav and Turok, Neil",
    title = "{Nonlocal electroweak baryogenesis. Part 1: Thin wall regime}",
    eprint = "hep-ph/9410281",
    archivePrefix = "arXiv",
    reportNumber = "PUPT-1495",
    doi = "10.1103/PhysRevD.53.2930",
    journal = "Phys. Rev. D",
    volume = "53",
    pages = "2930--2957",
    year = "1996"
}

@article{Konstandin:2013caa,
    author = "Konstandin, Thomas",
    title = "{Quantum Transport and Electroweak Baryogenesis}",
    eprint = "1302.6713",
    archivePrefix = "arXiv",
    primaryClass = "hep-ph",
    reportNumber = "DESY-13-036",
    doi = "10.3367/UFNe.0183.201308a.0785",
    journal = "Phys. Usp.",
    volume = "56",
    pages = "747--771",
    year = "2013"
}

@article{Linde:1978px,
    author = "Linde, Andrei D.",
    title = "{Phase Transitions in Gauge Theories and Cosmology}",
    reportNumber = "LEBEDEV-78-166",
    doi = "10.1088/0034-4885/42/3/001",
    journal = "Rept. Prog. Phys.",
    volume = "42",
    pages = "389",
    year = "1979"
}

@article{Kibble:1980mv,
    author = "Kibble, T. W. B.",
    title = "{Some Implications of a Cosmological Phase Transition}",
    reportNumber = "ICTP-79-80-23",
    doi = "10.1016/0370-1573(80)90091-5",
    journal = "Phys. Rept.",
    volume = "67",
    pages = "183",
    year = "1980"
}

@article{Witten:1984rs,
    author = "Witten, Edward",
    title = "{Cosmic Separation of Phases}",
    reportNumber = "PRINT-84-0400",
    doi = "10.1103/PhysRevD.30.272",
    journal = "Phys. Rev. D",
    volume = "30",
    pages = "272--285",
    year = "1984"
}

@article{vandeVis:2025efm,
    author = "van de Vis, Jorinde and de Vries, Jordy and Postma, Marieke",
    title = "{Bubble Trouble: a Review on Electroweak Baryogenesis}",
    eprint = "2508.09989",
    archivePrefix = "arXiv",
    primaryClass = "hep-ph",
    reportNumber = "CERN-TH-2025-161, Nikhef 2025-012",
    doi = "10.1016/j.ppnp.2026.104244",
    journal = "Prog. Part. Nucl. Phys.",
    volume = "150",
    pages = "104244",
    year = "2026"
}

@article{Peletminskii1968,
  author  = {Peletminskii, S. V. and Yatsenko, A. A.},
  title   = {Contribution to the Quantum Theory of Kinetic and Relaxation Processes},
  journal = {Sov. Phys. JETP},
  volume  = {26},
  number  = {4},
  pages   = {773--778},
  year    = {1968}, 
  note = {[Russian original: \emph{Zh. Eksp. Teor. Fiz.}, vol. \emph{53}\,(4), pp. 1327--1339, 1967]}
}

@article{Onsager1931,
  author  = {Onsager, Lars},
  title   = {Reciprocal Relations in Irreversible Processes. II},
  journal = {Physical Review},
  volume  = {38},
  pages   = {2265--2279},
  year    = {1931},
  doi     = {10.1103/PhysRev.38.2265}
}

\end{document}